\documentclass[12pt,preprint]{aastex}

\usepackage{graphicx}
\usepackage{amsmath}
\usepackage{color}

\shorttitle{BH X-ray Polarization in Thermal State}
\shortauthors{Schnittman \& Krolik}

\begin{document}

\title{X-ray Polarization from Accreting Black Holes: Coronal Emission}

\author{Jeremy D.\ Schnittman}
\affil{Department of Physics and Astronomy,
Johns Hopkins University\\
Baltimore, MD 21218}
\email{schnittm@pha.jhu.edu}

\and
\author{Julian H.\ Krolik}
\affil{Department of Physics and Astronomy,
Johns Hopkins University\\
Baltimore, MD 21218}
\email{jhk@pha.jhu.edu}

\begin{abstract}

We present new calculations of X-ray polarization from accreting black
holes (BHs), using a Monte-Carlo ray-tracing code in full general
relativity.  In our model, an optically thick disk in the BH equatorial plane
produces thermal seed photons with polarization oriented parallel to the disk
surface. These seed photons are then inverse-Compton scattered through a hot
(but thermal) corona, producing a hard X-ray power-law spectrum.  We consider
three different models for the corona geometry: a wedge ``sandwich''
with aspect ratio $H/R$ and vertically-integrated
optical depth $\tau_0$ constant throughout the disk; an inhomogeneous
``clumpy'' corona with a finite number of hot clouds
distributed randomly above the disk within a wedge geometry;
and a spherical corona of uniform density, centered on the BH and surrounded by a
truncated thermal disk with inner radius $R_{\rm edge}$. In all cases
we find a characteristic transition from horizontal polarization
at low energies to vertical polarization above the thermal peak;
the vertical direction is defined as the projection of the BH
spin axis on the plane of the sky. We show how the details of the spectropolarization
signal can be used to distinguish between these models and infer
various properties of the corona and BH.  Although the bulk of this
paper focuses on stellar-mass BHs, we also consider the effects of
coronal scattering on the X-ray polarization signal from
supermassive BHs in active galactic nuclei.

\end{abstract}


\section{INTRODUCTION}\label{intro}

A recent flurry of new mission proposals has renewed interest in
measuring
X-ray polarization from a variety of astrophysical sources. The
Gravity and Extreme Magnetism SMEX ({\it GEMS}) mission\footnote{\tt
  heasarc.gsfc.nasa.gov/docs/gems}, which has
recently been approved for funding in
the latest round of NASA Small Explorer proposals, should be able to
detect a degree of polarization $\delta \lesssim 1\%$ for a flux of a few
mCrab \citep{black:03, bellazzini:06, swank:09}. A similar detector for
the International X-ray Observatory ({\it IXO}) could achieve
sensitivity roughly $10\times$ greater ($\lesssim 0.1\%$ degree of
polarization; Jahoda et al.\ 2007, Costa et al.\ 2008). Projects
like these could potentially detect a large number
of galactic and extra-galactic sources
at the $\delta \sim 1\%$ level, including stellar-mass black holes, magnetars,
pulsar wind nebulae, and active galactic nuclei. In this paper, we
focus on accreting black holes (BHs) in the ``Hard'' or ``Steep Power
Law'' state \citep{remillard:06}, which are 
characterized by a broad-band spectrum with a thermal peak around 1
keV and a strong high-energy power-law component extending above 100
keV. Early estimates suggest that the typical level of polarization
from these sources should be a few
percent in the $1-10$ keV range, depending on the geometry of the
accretion system and the inclination to the observer
\citep{connors:80, sunyaev:85}.  We also extend these results to
AGN, whose hard X-ray properties are in many ways qualitatively similar.

Symmetry demands that in the Newtonian limit the observed
polarization from a flat disk must 
be either parallel or perpendicular to its rotation axis.
However, the effects of relativistic beaming,
gravitational lensing, and gravito-magnetic frame-dragging can break that
symmetry and give
a non-trivial net rotation to the integrated polarization
vector.  Because the temperature in an accretion disk should
increase closer to the BH, where these relativistic effects are
strongest, it was predicted long ago that the observed angle and degree of
polarization of thermal disk emission should depend
on photon energy \citep{stark:77,connors:77,connors:80}.
In more recent years, discussion of polarization in
accreting black holes has expanded to include a number of other
aspects of thermal disk emission, such as UV and X-ray
emission from AGN disks \citep{laor:90,matt:93} and ``lamp post''
models for
irradiating the accretion disk with a non-thermal source above the
plane \citep{dovciak:04}.  Quite recently, \citet{dovciak:08}
investigated the effect of atmospheric optical depth on the disk's
polarization signal, and \citet{li:08} applied the original
calculations of thermal X-ray polarization to the problem of measuring
the inclination of the inner accretion disk. \citet{davis:09}, using
a Monte Carlo ray-tracing code on data from shearing-box simulations, estimated
the effects of magnetic fields on the polarization of thermal emission.
\citet{schnittman:09a} (hereafter SK09) showed the importance of including
returning radiation \citep{cunningham:76} when calculating the polarization of
BHs in the thermal state, an effect previously noted by
\citet{agol:00}. 

In addition to the relatively well-understood thermal state, it has also
long been known that most AGN and stellar-mass BHs can produce
significant levels of harder X-rays with energies well above the
thermal peak. In both cases, the hard flux is thought to be produced via
inverse Compton scattering of the disk photons in a corona of hot (yet
thermal) electrons (e.g.\ \citet{haardt:93}). From the shape of this
hard spectrum, the basic geometry and optical depth of the corona
may be constrained \citep{haardt:94,PK95,stern:95,poutanen:97}.

Despite these constraints, there remain sizable uncertainties
about the nature and geometry of the hard X-ray emitting region.
Popular models for the hard state of stellar-mass BHs include a
cool disk truncated at large radius ($\sim 100M$)
surrounding a hot, radiatively inefficient flow
\citep{gierlinski:97,mcclintock:01,esin:01,done:09}, or
alternatively a more extended disk
surrounded by an optically thick hot corona, possibly in the form of a
hot wind \citep{blandford:04,miller:06,reis:09}.  Even less is known
about the steep power-law (SPL) state.  As for
the hard state, most popular models are based on the inverse Compton
scattering of thermal seed photons from a thin disk surrounded by a
hot corona \citep{zdziarski:04}, but bulk Comptonization of a converging
accretion flow has also been suggested \citep{titarchuk:02,turolla:02}.
Although the geometry of the corona could be as simple as
a uniform slab in the SPL state \citep{zdziarski:05}, the hard state
of galactic BH binaries,
as well as the X-ray emission from AGN, are more likely caused by
clumpy, inhomogeneous coronae, possibly caused by magnetic flares
\citep{haardt:94, poutanen:99}.

Here we explore the X-ray polarization signatures of three simple
models for the corona geometry: a smooth sandwich with uniform optical
depth in the vertical direction; an inhomogeneous model made of a
finite number of spherical clouds, randomly distributed above the
disk; and a truncated thin disk surrounding a spherical corona. In all
cases, seed photons are emitted from the thermal, optically thick
disk and are up-scattered in the hot corona (in the truncated disk
case, we also include low-energy seed photons embedded in the
corona). We find that in all
cases, the spectrum comprises a thermal peak ($\sim 1-3$ keV for
stellar-mass BHs; $\sim 30-100$ eV for AGN) and a
power-law component dominating above that, which typically makes
up $\sim 50-80\%$ of the total flux. The gross features of the
polarization spectra are quite robust: horizontal polarization with a
few percent amplitude at low energies, with a transition to vertical
orientation above the thermal peak, where the polarization
amplitude can be as large as $\sim 10\%$. The specific details of the
polarization spectrum, i.e., the amplitude of polarization at low and
high energies, and the shape and location of the transition, provide
constraints on the global geometry, temperature, and optical depth of
the corona, as well as the BH mass, spin, accretion rate, and the observer
inclination angle.

In this paper, we begin in Section
\ref{methods} with a brief overview of the computational methods used
in the calculations. In Sections \ref{wedge}--\ref{sphere} we present
our results for three different corona geometries: sandwich, hot
spots/clumps, and a sphere embedded in a truncated disk, all in the
context of galactic X-ray binaries. In Section \ref{AGN}, we apply the
results to AGN. In Section
\ref{discussion} we discuss possible applications to
observations. 

\section{METHODOLOGY}\label{methods}

Although a detailed description of the ray-tracing code will be given in a
companion paper \citep{schnittman:09c}, we give a brief summary of the
relevant physics here. The basic geometry is described by a
thin disk with inner radius $R_{\rm edge}$, typically placed at the
inner-most stable circular orbit (ISCO), but possibly at smaller radii
so long as $R_{\rm edge}$ is outside the horizon.  In the latter case,
the gas inside the ISCO follows
plunging trajectories along geodesics with energy and angular momentum
determined by the ISCO values. The orbital angular momentum of the
disk is prograde and aligned with the BH spin axis.
Seed photons are emitted from
the thin disk with a diluted black-body spectrum characterized by a
hardening factor $f=1.8$ \citep{shimura:95}. The local flux is given
by the Novikov-Thorne emissivity profile \citep{novikov:73}, but can
be modified to include emission inside the ISCO, as described in
SK09. 

The seed photons have initial polarization parallel to the disk
surface in the local fluid frame, as determined by the classical result for
scattering-dominated atmospheres \citep{chandra:60}. The degree of
polarization varies from
zero for photons emitted normal to the disk surface up to $\sim 12$\%
for an inclination angle of $90^\circ$. In addition to the
polarization effects, the scattering of the outgoing flux causes
limb-darkening, effectively focusing the emitted radiation in the
direction normal to the disk surface. We define our seed photons in
accordance with the polarization and limb-darkening factors tabulated
as a function of emission angle in Table XXIV in \citet{chandra:60}.

After leaving the disk, the photon packets follow geodesic
trajectories around the black hole, eventually reaching a distant
observer, returning to the disk via gravitational deflection, or
getting captured by the 
horizon. En route, many photons scatter in the corona.  We
describe these
events with the classical Thomson electron cross section.
The differential probability for scattering along a path length $dl$
is simply 
\begin{equation}\label{diff_prob}
P_{\rm scat}= 1-e^{-d\tau} = 1-e^{-\kappa\rho\, dl}\, ,
\end{equation}
where $\kappa=0.4\, {\rm cm}^2\, {\rm g}^{-1}$ is the opacity to
electron scattering, $\rho$ is the local mass density of the corona,
and $\tau$ is the
optical depth. Unlike some recent Monte Carlo ray-tracing codes (e.g.\
\citet{davis:09,dolence:09}) that use the more physically accurate,
energy-dependent Klein-Nishina cross section, the Thomson
cross section allows for the use of photon packets that include
the entire broad-band spectrum, as described in
\citet{schnittman:09c}. This simplification gives improved
computational efficiency, albeit at the loss of physical accuracy for
above $\sim 200$ keV, where the classical and relativistic electron
cross sections begin to diverge. For most stellar-mass BH 
sources, and for most X-ray polarimetry missions 
in the foreseeable future, it should be quite safe to focus on photons
with energies less than $100$ keV. 

At each step along the photon geodesic, the probability for scattering
is calculated according to equation (\ref{diff_prob}), and then a
uniform random number in $[0,1)$ determines whether that photon
scatters at that location\footnote{Our adaptive-step Cash-Karp
  geodesic integrator takes smaller steps where the
  electron density is high, ensuring that $d\tau<<1$.}.
When the photon {\it does} scatter, we first transform the photon
4-momentum $\mathbf{k}$ and polarization vector $\mathbf{f}$ into the local inertial frame of
the corona, then do a special-relativistic boost into the
electron's rest frame, where the electron velocity is taken from an
isotropic thermal distribution. In the electron's rest frame, the
scattering event is treated completely
classically, with a differential cross section given by \citep{rybicki:79}
\begin{equation}\label{dsigma_dOmega}
\frac{d\sigma}{d\Omega} = \frac{1}{2}r_0^2 [(1-\delta)(\cos^2\Theta+1) +
  2\delta\cos^2\Theta \cos^2\psi + 2\delta\sin^2\psi]\, ,
\end{equation}
where $r_0=2.82\times 10^{-13}$ cm is the classical electron radius,
$\delta$ is the degree of polarization, $\Theta$
is the angle between incident and outgoing photon directions, and
$\psi$ is the polarization angle, measured with respect to the
scattering plane. The angles $\Theta$ and $\psi$ are selected
appropriately from the probability distribution corresponding to
(\ref{dsigma_dOmega}). The outgoing Stokes parameters are given by 
\begin{subequations}
\begin{eqnarray}\label{stokes_f}
I' &=& \frac{3}{2}( I_\parallel\cos^2\Theta+I_\perp) \\
Q' &=& \frac{3}{2}(I_\parallel\cos^2\Theta -I_\perp) \\
U' &=& \frac{3}{2}U\cos\Theta \, ,
\end{eqnarray}
\end{subequations}
where $I_\parallel$ and $I_\perp$ are the components of the intensity
with polarization parallel to and perpendicular to the scattering
plane, respectively.
These Stokes parameters are then used to reconstruct the new
polarization degree and angle through 
\begin{subequations}\label{X_Y}
\begin{eqnarray}
X &=& Q/I, \\
Y &=& U/I,
\end{eqnarray}
\end{subequations}
and
\begin{subequations}\label{delta_psi}
\begin{eqnarray}
\delta &=& (X^2+Y^2)^{1/2}, \\
\psi &=& \frac{1}{2}\tan^{-1}(Y/X)\, .
\end{eqnarray}
\end{subequations}

The scattered photon packet is boosted back to the coronal fluid
frame, then transformed back into the
coordinate basis and continues propagating along its new
geodesic. This ``transform--boost--scatter--boost--transform'' algorithm
automatically carries out the inverse Compton change in
photon energy, giving the outgoing photon an average energy increase
of $\gamma^2$, where $\gamma$ is the electron Lorentz factor ($\gamma
\approx 1+kT/(m_e c^2)$ for thermal electrons). The
polarization 4-vector is parallel-transported simply by satisfying the
constraints $\mathbf{k}\cdot\mathbf{f}=0$ and
$\mathbf{f}\cdot\mathbf{f}=1$ and using the complex-valued
Walker-Penrose integral of motion $\kappa_{\rm WP}$ \citep{walker:70}.
Analogous to the way that Carter's constant \citep{carter:68}
can be used to constrain 4-velocity components,
$\kappa_{\rm WP}$ can be used to reconstruct the polarization vector at
any point along the geodesic.  The two orthonormality conditions
stated above, along with the real and imaginary parts of $\kappa_{\rm WP}$,
give a total of four equations for the four components of the
polarization vector $\mathbf{f}$.

Some photons encounter the disk before reaching their final
destination.  When we treat models relevant to AGN, photons striking
the disk are absorbed because our primary interest is in photons $< 10$~keV,
and AGN disks are generally thought to have enough photo-ionization
opacity to have little albedo at these energies.  In the stellar-mass
black hole case, disks are expected to be highly reflective.  For
these models, we use the results given in Section
70.3 of \citet{chandra:60} for diffuse reflection, i.e.\ multiple
scattering events in a semi-infinite plane.

When a ray reaches an observer at infinity, the polarization vector is
projected onto the detector plane. In our convention, the y-axis of
the detector is parallel to the projected symmetry axis of the black
hole and accretion disk, so $\psi=0^\circ$ corresponds to ``horizontal''
polarization and $\psi = \pm 90^\circ$ is ``vertical.'' Finally, we
integrate over all photon bundles
to obtain the energy-dependent Stokes parameters $I_\nu$, $Q_\nu$, and
$U_\nu$, which in turn give $\delta_\nu$ and $\psi_\nu$, our preferred
observables for most of the results presented below. 

\section{WEDGE GEOMETRY}\label{wedge}

We begin by considering the simplest corona geometry, an isothermal
layer of uniform vertical optical depth forming a
wedge with constant opening angle $\tan\theta_c = H/R$. At each point in the
disk, the atmosphere has an exponential profile in the vertical
direction with $\rho(z,R)=\rho_0\exp[-z/H(R)]$. The
vertically-integrated optical depth $\tau_0$ is constant, giving
\begin{equation}
\rho_0(R) = \frac{\tau_0}{\kappa H(R)}\, .
\end{equation}
We set the local rest frame of the corona to be corotating with the
underlying disk: $u^r=u^\theta=0$, $u^\phi/u^t=\Omega(R)$, and $u^\mu
u_\mu=-1$. $\Omega(R)$ is the orbital frequency (measured at infinity)
of a planar circular orbit at radius $R$, which in Boyer-Lindquist
coordinates is 
\begin{equation}
\Omega(R) = \frac{1}{(R/M)^{3/2}+a/M}\, .
\end{equation}

\begin{figure}
\caption{\label{schem_sandwich} Schematic diagram of the disk and
  corona for the sandwich geometry. Thermal seeds are emitted from a
  thin disk in the midplane, then scatter off hot electrons in a
  corona with a wedge geometry and constant scale height $H/R$. The
  optical depth to electron scattering in the vertical is constant
  throughout the disk. For observers at high inclination (edge-on),
  the scattered photons will likely have high polarization, oriented
  perpendicular to the scattering plane, as indicated by the small
  black lines. Some photons emitted from the inner disk are deflected
  by the BH and then scatter off the disk on the far side, also
  leading to large amplitude polarization.}
\begin{center}
\rotatebox{90}{\includegraphics{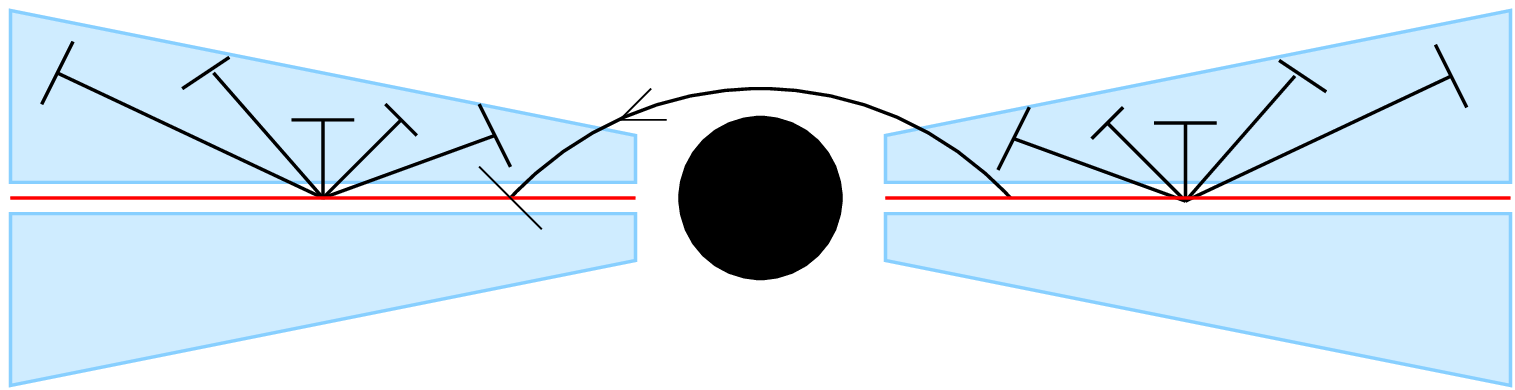}}
\end{center}
\end{figure}

Figure \ref{schem_sandwich} gives a schematic view of the accretion
geometry, showing how the seed photons originate in the midplane, then
scatter through the hot corona before reaching a distant
observer. The coronal scattering has two major effects on the observed
polarization signature. First, it changes the underlying spectrum by
inverse Compton scattering a portion of the photons, boosting them to
higher energies, and thus producing a harder spectrum. Second, it
changes the amplitude and orientation of the net polarization
at high energies, rotating it from horizontal to vertical,
similar to the way returning radiation gets scattered into a vertical
orientation by a thermal disk (SK09).  The coronal polarization
rotation effect
is essentially the same as that described in \citet{sunyaev:85}, who
considered the up-scattering of low-energy photons embedded in a
hot electron corona 
with planar geometry, finding strong vertical polarization for
high-energy photons and observers at high inclinations.

The cause of this rotation in systems with moderate optical depth
($\tau_0 \lesssim 2$) can be understood easily from the scattering geometry
in Figure \ref{schem_sandwich}.  Photons initially emitted
in directions near the disk plane are almost certain to scatter
because they face a large optical depth.  To reach an observer
who views the disk nearly edge-on, they must stay in the disk
plane even after scattering, but this requires a vertical polarization
direction. Additionally, those photons that scatter
multiple times in the corona---and are thus boosted to higher
energies---are geometrically more likely to move in the plane of the
corona, parallel to the disk. This further increases the amplitude of
their vertical polarization and preferentially scatters them to
infinity with large emission angle, leading to a limb-{\it brightening}
effect at high energies with respect to the classical Chandrasekhar
result \citep{sunyaev:85}. 

For hot coronae with high optical depth, the multiple scatters
cause the polarization and limb-darkening to tend towards the
Chandrasekhar limit, but with the photon energy dependence given by a Wien
spectrum. In the limit of a cold corona,
where there is no transfer of energy via inverse Compton scattering,
there is no way to distinguish between photons coming directly from the
disk atmosphere and those that are scattered in the corona, and the
classical result is reproduced regardless of the coronal optical
depth.

\begin{figure}
\caption{\label{image_wedge} Ray-traced images of polarized flux from
  an accretion disk with a sandwich corona of scale height $H/R=0.1$.
  The observer is located at an inclination of
  $75^\circ$ relative to the rotation axis, with the gas
  on the left side of the disk moving towards the observer, which
  causes the characteristic increase in intensity due to relativistic
  beaming and boosting.
  The black hole has spin $a/M=0.9$, mass $M=10 M_\odot$.
  The observed intensity is color-coded on a
  logarithmic scale (normalized to the {\it net} intensity $\int
  d\nu\, I_\nu$), and the
  energy-integrated polarization vectors are
  projected onto the image plane with lengths proportional to the
  local degree of polarization. The four panels correspond to the
  contributions to the observed flux (a) directly from the thermal
  disk; (b) return radiation scattered once off the disk; (c) photons
  scattered once in the corona; and (d) photons scattered multiple
  times in the corona.}
\begin{center}
\scalebox{0.75}{\includegraphics*[52,400][360,660]{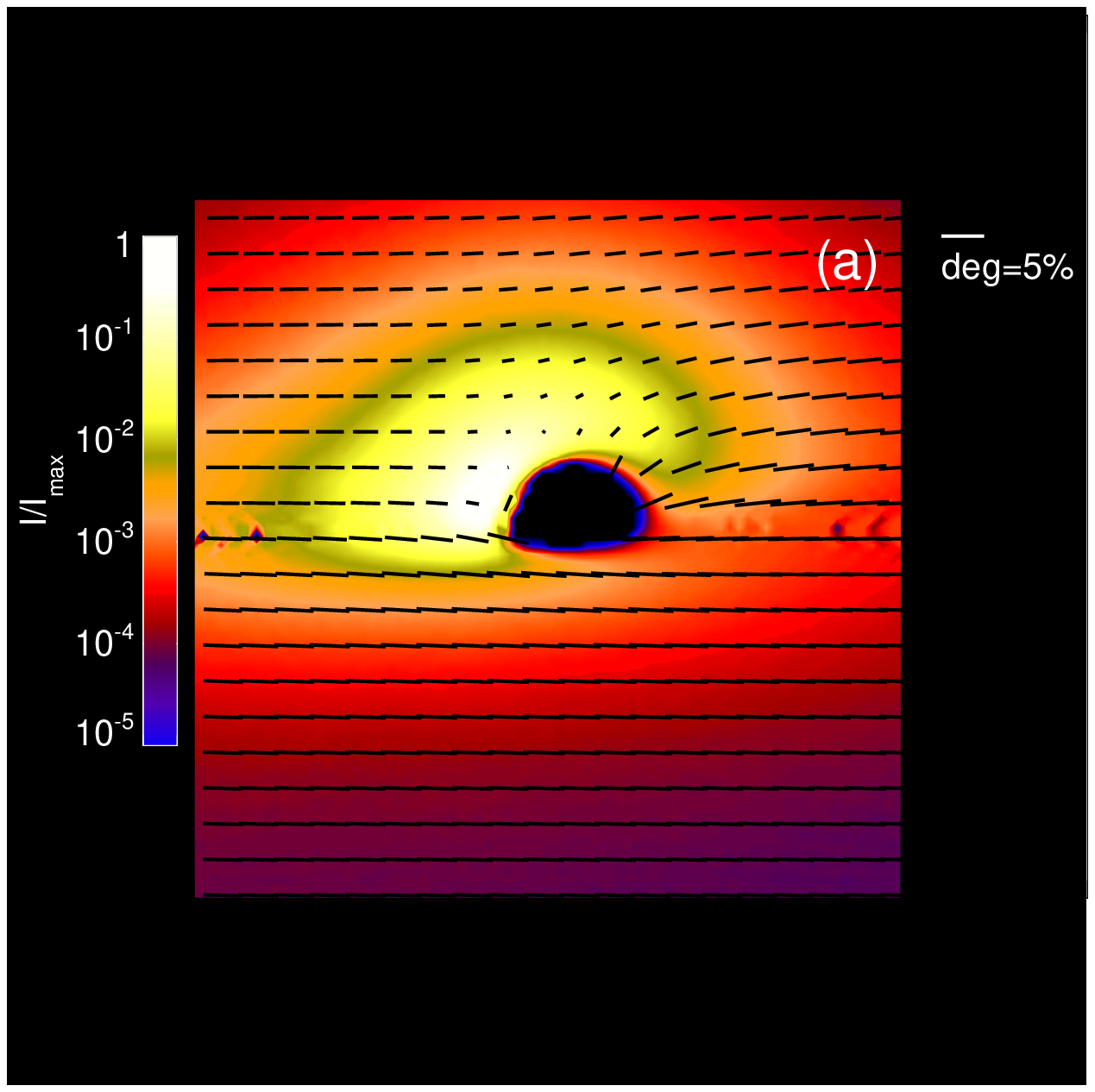}}
\scalebox{0.75}{\includegraphics*[112,400][410,660]{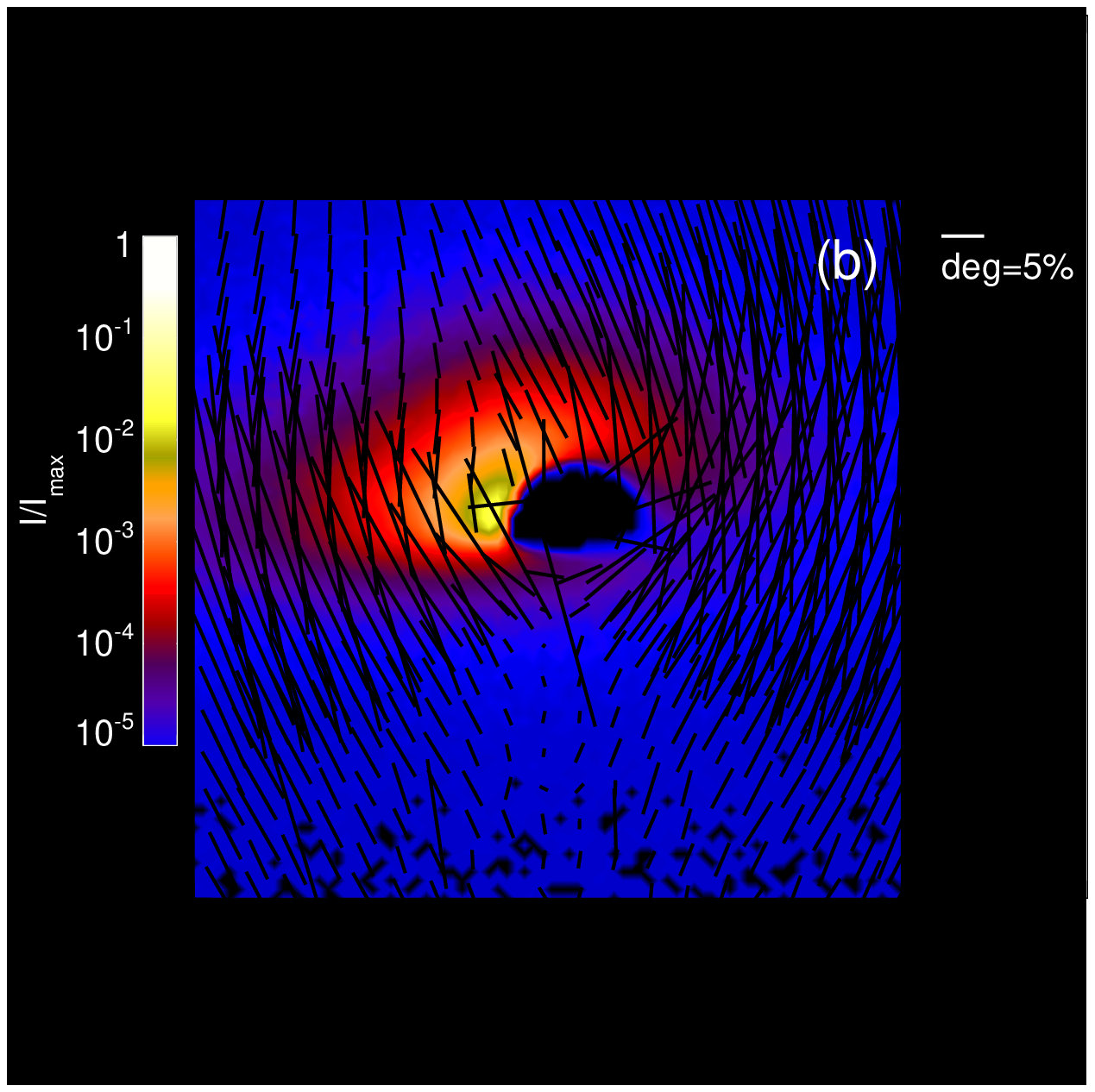}}\\
\vspace{0.1cm}
\scalebox{0.75}{\includegraphics*[52,400][360,660]{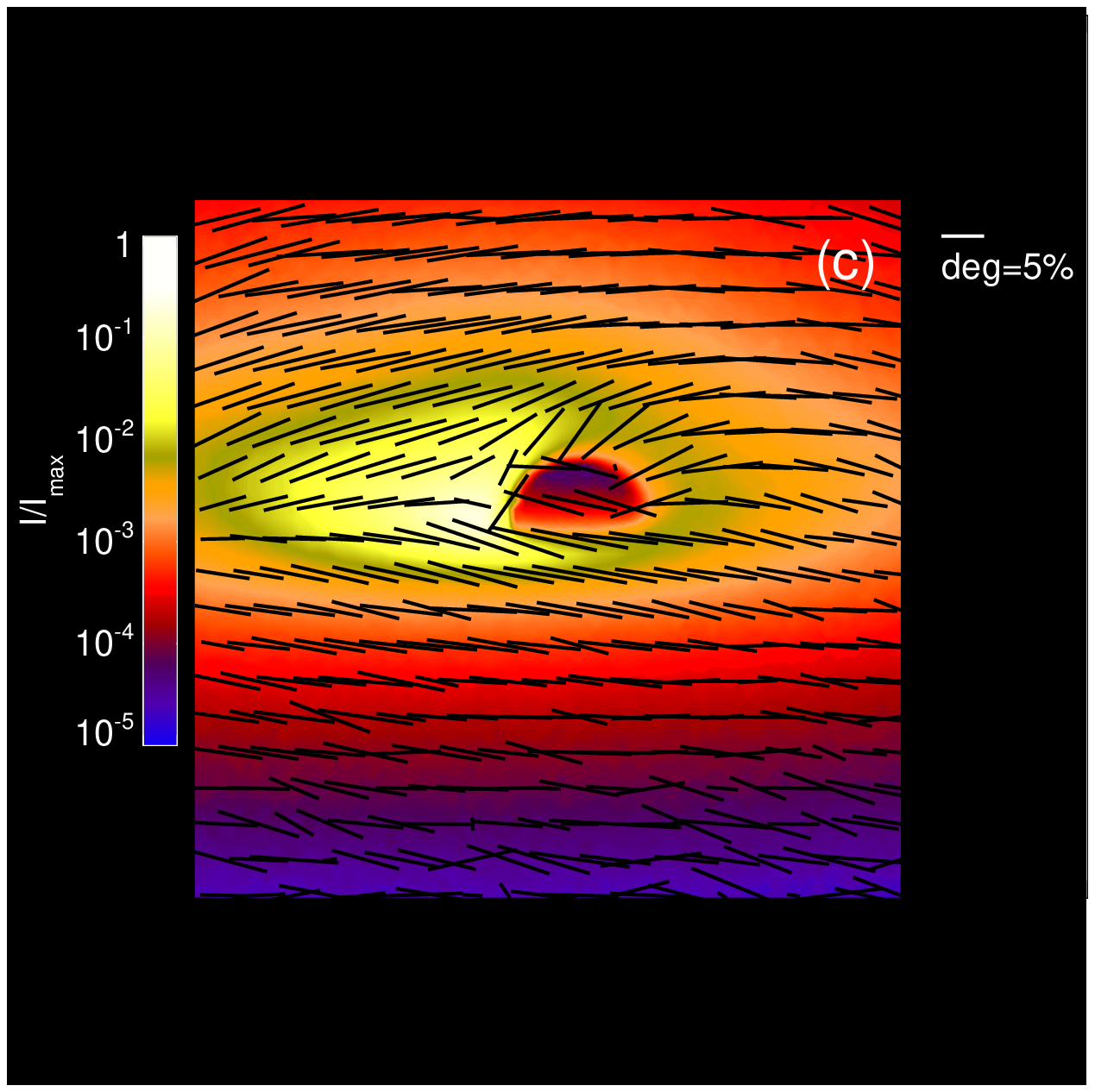}}
\scalebox{0.75}{\includegraphics*[112,400][410,660]{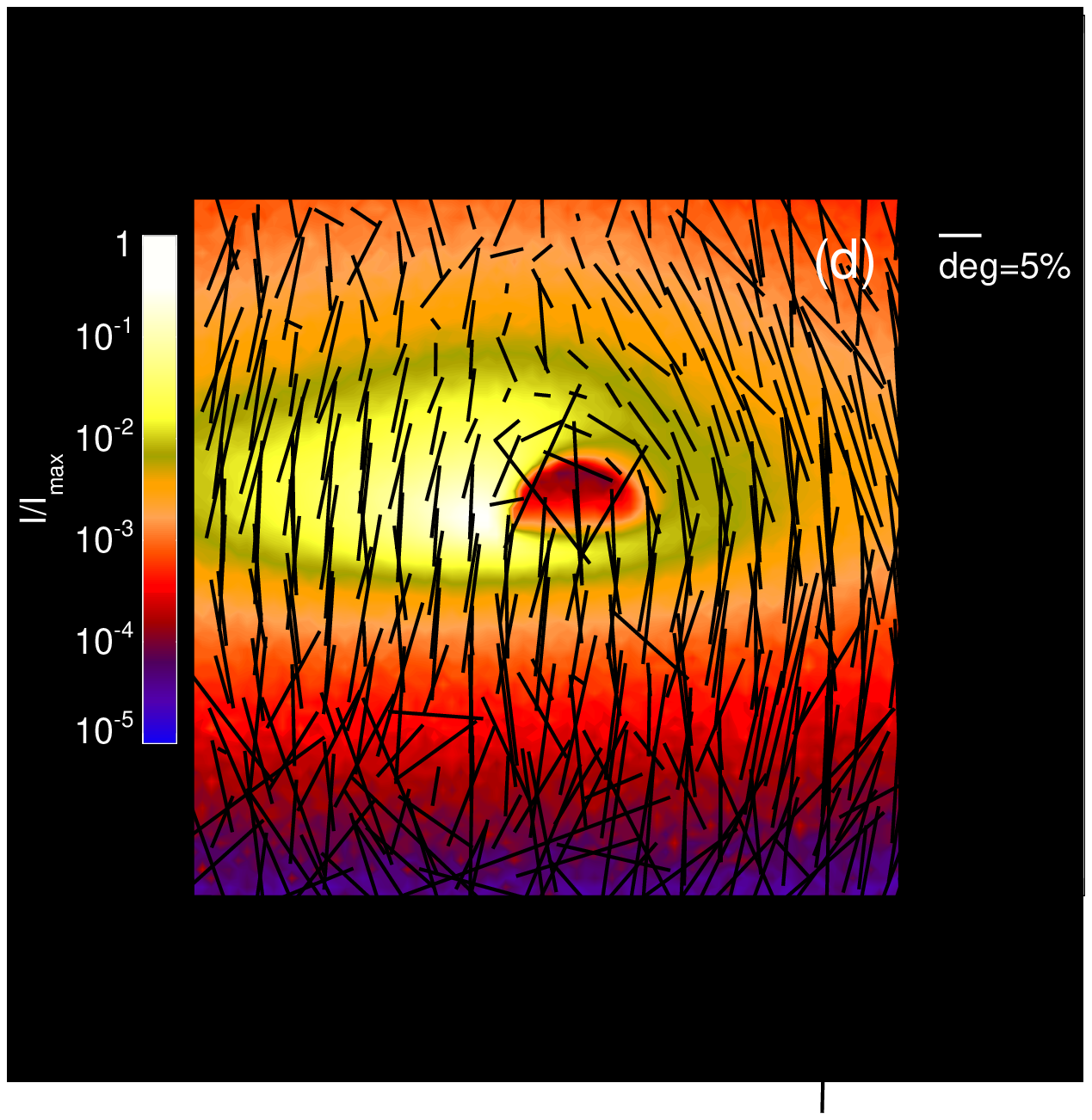}}
\end{center}
\end{figure}

To improve our qualitative understanding of the polarization effects
of the global corona geometry, we plot in Figure \ref{image_wedge} a
series of images of the BH accretion disk plus corona, sorted
by the scattering history of each photon. Figure \ref{image_wedge}a shows
only photons that travel directly from the disk to the observer;
Figure \ref{image_wedge}b shows flux from return radiation---photons that
are bent by gravitational lensing and scatter off the disk to the
observer; Figure \ref{image_wedge}c is the flux from photons that scatter exactly
once in the corona; and Figure \ref{image_wedge}d is made of photons that
scatter multiple times in the corona. The observed intensity is
color-coded on a logarithmic scale (normalized to the peak intensity
from all photons)
and the energy-integrated polarization vectors are projected onto the
image plane with lengths proportional to the local degree of
polarization. The black hole has a spin of
$a/M=0.9$, mass $M=10M_\odot$, and accretion rate such that the
thermal flux alone totals $10\%$ of the Eddington luminosity. The
corona has a temperature of $100$ keV, scale height $H/R=0.1$, and
vertical optical depth $\tau_0=1$. These parameters lead to a coronal luminosity
$2-3$ times that of the thermal luminosity. The disk is rotating in
the counter-clockwise direction, and the observer is located at an
inclination of $75^\circ$ to the rotation axis.  The intensity
maximum at the left of each image is caused by relativistic beaming and
Doppler boosting.  Photons from the far side of the disk are bent by gravitational
lensing, giving the appearance of a warped disk with the far side
having a smaller effective inclination angle.

Not surprisingly, the direct image looks quite similar to that for the
direct radiation from a simple thermal disk with no corona
(SK09). The polarization is modest and predominantly horizontal,
with relativistic effects most noticeable in the inner disk, where
beaming and lensing combine to rotate the observed angle of
polarization at higher energies. The degree of polarization is reduced
above and to the left of center, where the effective inclination of
the disk is smaller, while the opposite occurs in the receding section
of the disk on the right side of the image. However, when carefully
compared to the non-corona system (Fig.\ 1 of SK09),
we see a clockwise shift in the peak flux distribution because
the level of direct flux from a given patch of the disk is dependent
on the optical depth to scattering along the geodesic connecting that
patch with the observer. For the wedge geometry, this means that
where the effective inclination is smallest, the observed flux is
greatest, leading to additional limb-darkening in the direct component.

The return radiation image is also similar to that for the thermal
disk alone, although now the presence of a scattering corona strongly
suppresses any return flux from large radii. Since most returning
photons pass very close to the light orbit around $R=3M$, they would
necessarily intersect the outer disk at very high angles of incidence,
and thus face very large optical depth through the wedge corona. As shown
in SK09, the flux from returning radiation, while
small compared to the total flux, is very highly polarized, and in a
vertical direction, perpendicular to that of the direct flux. Since
the majority of the returning radiation originates in the hot inner
regions of the disk, there is a dependence of the polarization
direction on photon energy. As can be seen from Figure
\ref{image_wedge}b, the angle of polarization for the return radiation
is {\it negative}, as measured in the image plane, with $\psi=0$
defined along the $x$-axis. This can also be understood from the
global geometry of the system, as shown in Figure
\ref{schem_sandwich}. Photons emitted from the right side
of the disk, deflected by the BH and incident on the left side at some
finite angle, will scatter towards a high-inclination observer with
large polarization in the $\psi<0$ orientation.
Of course, the
opposite occurs for photons emitted from the left side and scattered on
the right side, but they get
preferentially scattered away from the observer by relativistic
beaming. 
 
In Figure \ref{image_wedge}c we show the flux from
photons experiencing a single coronal scatter. Since {\it every}
photon in this image scatters exactly once, there is no preference
for oblique
emission, and photons emitted at all inclination angles are equally
weighted. If anything, for $\tau_0=1$, the seed photons emitted normal
to the disk surface are more likely to scatter exactly once and then
escape to an observer at high inclination, thus explaining the
prevailing horizontal polarization seen here. Lastly,
Figure \ref{image_wedge}d shows the multiply-scattered
photons.  These have strong net vertical polarization due to the geometric
effects described above and shown schematically in Figure
\ref{schem_sandwich}. 

Unlike the returning radiation, the coronal flux has a net {\it
positive} polarization angle $\psi>0$ when viewed edge-on. This can
again be understood from the schematic image of the wedge corona in
Figure \ref{schem_sandwich}. It is clear from this geometry that for
two photons, both emitted in the plane of the sky at the same angle
$i_{\rm em}$ to the disk normal, the photon emitted in the direction
{\it away} from the black hole will have a greater path length within
the wedge corona. However, due to the increasing scale height and
constant optical depth in the vertical direction, it will pass through
lower-density gas. Combining these two effects---path length and
coronal density as a function of emission angle---we find that the
optical depths for ingoing and outgoing photons are
\begin{subequations}\label{tau_in_out}
\begin{equation}
\tau_{\rm in}(i_{\rm em}) = \frac{\tau_0}{2}\left[
  \frac{\cos\theta_c}{\cos(i_{\rm em}-\theta_c)}+\frac{1}{\cos i_{\rm
  em}}\right]
\end{equation}
and
\begin{equation}
\tau_{\rm out}(i_{\rm em}) = \frac{\tau_0}{2}\left[
  \frac{\cos\theta_c}{\cos(i_{\rm em}+\theta_c)}+\frac{1}{\cos i_{\rm
  em}}\right] \, .
\end{equation}
Here $\theta_c$ is the opening angle of the corona and
we restrict $i_{\rm em}<\pi/2-\theta_c$ so that the outgoing optical
depth is finite.
\end{subequations}

Since $\tau_{\rm out}>\tau_{\rm in}$ for each $i_{\rm em}$, the
``tilt'' of the polarization vector for corona scattering is opposite
that of return radiation. For the region of the disk beamed
towards the observer (left side of images in Fig.\ \ref{image_wedge}),
this means that photons emitted in the direction away from the BH
are more likely to be scattered in the wedge corona, giving a net
polarization oriented with $\psi>0$, as can be seen
in the lower 
panels of Figure \ref{image_wedge}. In practice, we can't
ever really know whether a given photon came directly from the disk or
was scattered in the corona or off the disk, but the images do provide
important qualitative understanding of the geometric effects
involved. 

For a more quantitative picture, we plot in Figure \ref{wedge_grid}
the observed flux, polarization degree, and polarization angle as a
function of energy for observer inclinations of $i=45^\circ$,
$60^\circ$, and $75^\circ$. We sort the photons into direct
(dotted curves), scattered (dot-dashed curves), and total flux (solid
curves). For a BH mass of $10 M_\odot$ accreting at a rate $\sim
0.1L_{\rm Edd}$, the thermal peak is near 1 keV, with
essentially no flux from the disk above $\sim 10$ keV. With $\tau_0=1$
and $T_c=100$ keV, the Compton $y$ parameter is of order
unity, so that the up-scattered spectrum has $F_\nu \propto
\nu^{-1}$ \citep{rybicki:79}. As the inclination of the observer
increases, the optical depth to the disk increases, thus lowering the
relative contribution from the direct radiation. In all cases, the
up-scattered radiation dominates the flux above a few keV. 

The direct radiation behaves quite similarly to that described in
\citet{connors:80}, reproducing the Chandrasekhar result at low
energy, and then decreasing in fractional polarization at higher energies as the
relativistic effects of the inner disk begin to dominate. As discussed
in SK09, the angular rotation in $\psi$ is greatest for small
observer inclinations simply because there is a smaller degree of
``classical'' polarization to overcome. 

When considering only the scattered photons, the polarization is still
given by the
Chandrasekhar limit at low energies. The scattered
photons below the thermal peak have typically
scattered only once in the corona, thus preserving their
initial horizontal orientation (see Fig.\ \ref{image_wedge}c).
At higher energies, we begin
to sample the multiply-scattered photons that make up Figure
\ref{image_wedge}d, giving strong vertical polarization. Thus there is
a transition from horizontal to vertical orientation even when
considering the scattered photons alone. When including all the flux
(solid curves), this transition point is shifted to somewhat higher
energies than for the scattered photons alone due to the horizontal
contribution of the direct thermal radiation.

As mentioned above, the effects of return radiation and coronal
scattering on the {\it sign} of the polarization angle are opposite:
return radiation leads to a negative angle and the coronal scattering
in a wedge geometry
leads to a positive angle in the transition from horizontal to
vertical polarization. However, some photons that scatter in the
corona are also deflected by gravitational lensing and then scattered in
the corona on the far side of the disk, leading to some ambiguity as
to whether they should be classified as return photons or coronal
photons. From Figure \ref{wedge_grid} we see that the relative
contributions from these two effects are also dependent on the
observer inclination, with low-inclination systems giving a negative
tilt and high-inclination leading to a positive tilt. This distinction
may be observable in practice by comparing the polarization in
the hard state to that of the soft state, where the transition
direction is independent of inclination and therefore can be used to
define the negative orientation (SK09).

Using spectropolarimetry to infer physical properties of the source is
certainly more complicated in the hard state than in the thermal state,
for the simple reason that there are more free parameters in the
underlying model, creating degeneracies in fitting the
data. Instead of attempting to quantify our ability to
measure any particular property of the BH system, we present here the
effects of separately changing individual parameters with respect to
a fiducial model with $M=10M_\odot$, $a/M=0.9$, $L_{\rm
therm}=0.1L_{\rm Edd}$, $i=75^\circ$, $\tau_0=1.0$, $T_c=100$ keV, and $H/R=0.1$. 

\begin{figure}
\caption{\label{wedge_Ledd} Degree and angle of polarization for a
  sandwich corona, varying the luminosity in the thermal flux.}
\begin{center}
\scalebox{0.8}{\includegraphics{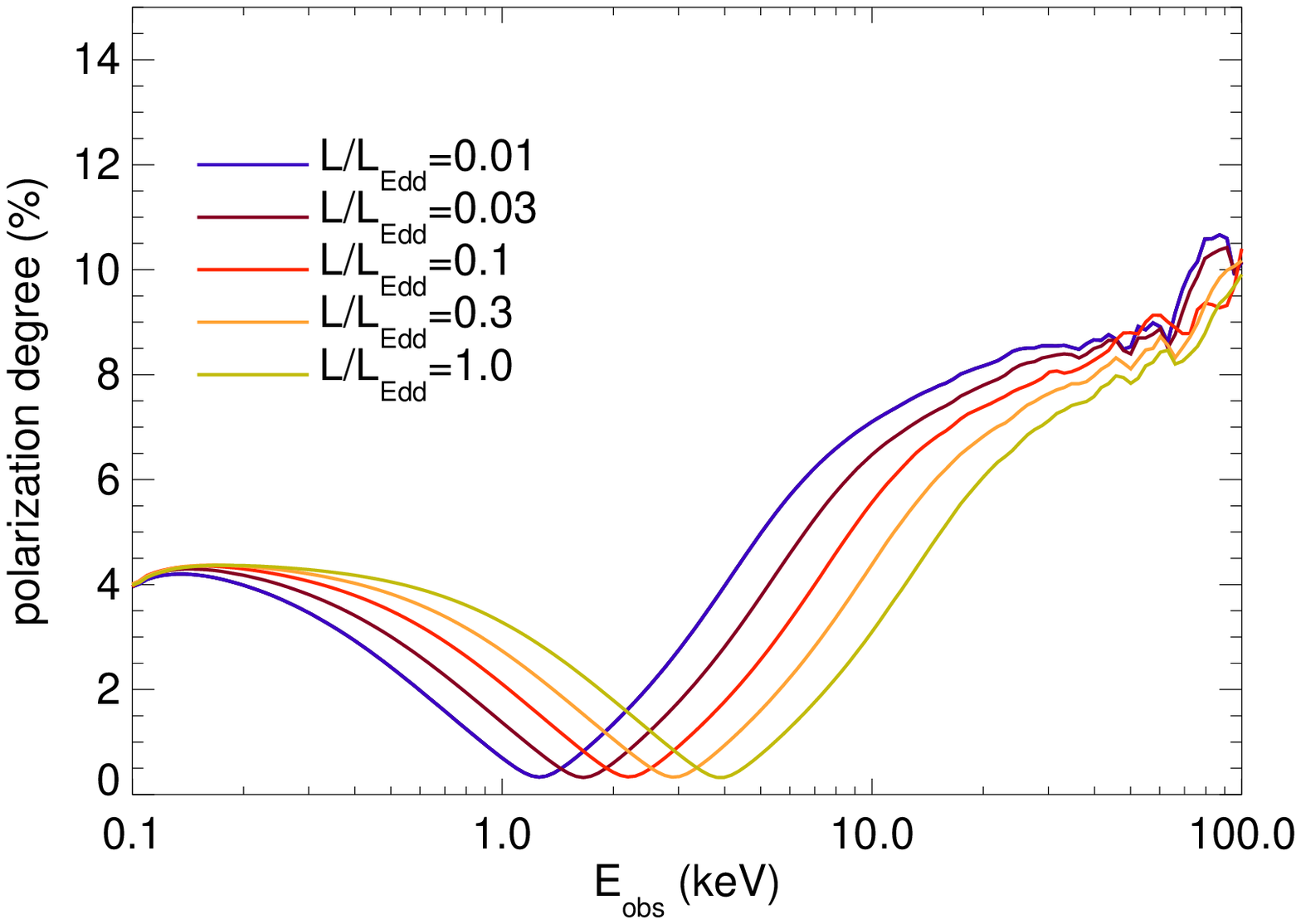}}\\
\scalebox{0.8}{\includegraphics{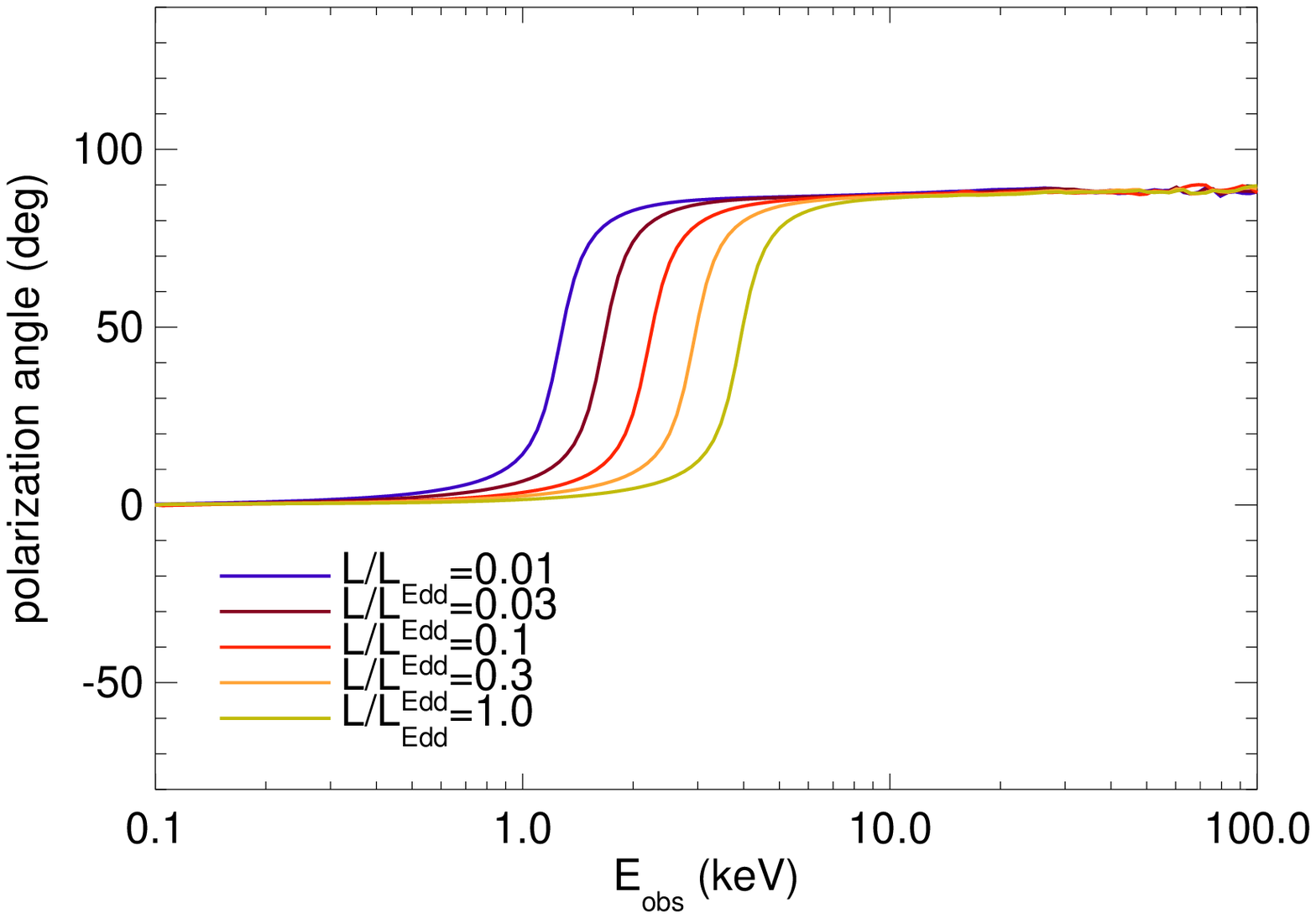}}
\end{center}
\end{figure}

In Figure \ref{wedge_Ledd} we show the polarization degree and angle
for the fiducial model, varying the luminosity in the thermal flux
component from $0.01L_{\rm Edd}$ to $L_{\rm Edd}$. Since in all cases
the thermal peak is well below the corona temperature, the transition
energy scales linearly with the disk temperature, which is in turn
proportional to the quarter power of the luminosity. With multiple
observations of a single source accreting at various rates as in
Figure \ref{wedge_Ledd}, one could confirm that the thermal part of the
spectrum is indeed coming from a thin disk with constant $R_{\rm
  edge}$ \citep{gierlinski:05}. 

Holding the
net thermal flux fixed, but varying the spin, changes the emissivity
profile of the disk, concentrating more of the emission close
to the BH for larger spins (and thus smaller $R_{\rm ISCO}$). As in
the thermal state (SK09), this leads to a greater fraction 
of return radiation, and thus the transition from
horizontal to vertical polarization occurs at lower energy,
as shown in Figure \ref{wedge_spin}.
In the hard/SPL states, the presence of a corona provides
another, complementary mechanism for effecting the 
polarization transition above the thermal peak, regardless of the spin
parameter. Thus the dependence of the polarization signal on BH spin
is weaker than in the thermal state.
In this way, the sandwich corona acts
as a veil around the inner disk, obstructing our view of the plunging
region where the orbital dynamics are most sensitive to the spin.

In Figure
\ref{wedge_tauT}, we simultaneously vary the coronal optical depth and
temperature, maintaining a roughly constant Compton $y$-parameter. In the
wedge corona model, the transition from horizontal to vertical
polarization is largely due to local geometric
effects (as opposed to more global effects like returning
  radiation). In this case, the optical depth and opening angle of the corona
largely determine the scattering history of each photon, and thus the
polarization degree and angle of the outgoing radiation.
Holding the size and shape of the corona constant, two photon
packets experiencing the same number of scattering events will emerge
on average with the same polarization signature. Thus the energy of
the transition is primarily a function of the scattering electron
temperature, and hotter coronae lead to a higher-energy
transition. Similarly, to reach a given energy above the thermal peak,
photons have to scatter more times in a cooler corona,
resulting in a more constrained scattering history, and thus
higher polarization. At the same time, when
changing the optical depth of the corona, the overall scattering geometry {\it
  does} change slightly, so the shape of the transition also varies
in Figure \ref{wedge_tauT}. Furthermore, for small optical depths, a
larger fraction of the photons can return to the disk, giving the
negative polarization angle characteristic of returning radiation.

Lastly, we show in Figure \ref{wedge_HR} the effect of varying the
scale height of the corona.  Here, the fact that return radiation and
coronal scattering rotate the polarization angle in opposite senses
away from horizontal leads to a complicated dependence.
The geometrically thinnest coronae behave similarly to the razor-thin thermal
disks described in SK09, giving a more gradual transition
from horizontal to vertical with $\psi<0$\footnote{From eqn.\
(\ref{tau_in_out}) we see that when $\theta_c\to 0$, there is no
difference between photons emitted towards or away from the BH, leaving
only the global effects of returning radiation.}. As the opening angle of the
corona increases, the transition flips to $\psi>0$ as the coronal scattering
effects of equation (\ref{tau_in_out}) becomes more important. For very large
$H/R$, the lower density of the corona allows longer path lengths and
photons can sample a larger volume of the accretion flow. At
low energies this loosening of the geometrical constraints leads to a
deviation from the classical planar
scattering atmosphere. At higher energies, it leads to a higher
effective temperature as the bulk velocity of the corona
(which can be a substantial fraction of $c$) is added to the thermal
velocities of the scattering electrons. The combination of these
various effects eliminates the existence of any simple trends in Figure
\ref{wedge_HR}. 

\section{INHOMOGENEOUS GEOMETRY}\label{hotspot}

Motivated by spectral models of AGN \cite{stern:95}, whose reasoning
also applies to galactic BHs in the hard state \cite{poutanen:97},
we also consider an inhomogeneous geometry where the hot coronal
plasma is clumped into a large number of small, dense regions.  These clouds are
distributed randomly above the disk, roughly following a wedge
geometry.  For precisely the same reasons that
originally motivated the clumpy corona model, it is not generally
possible to reproduce the same spectrum as in the sandwich geometry.
Yet we find that by fixing the electron temperature and conserving the
total mass in the corona, the net flux in hard X-rays $(\gtrsim
10\mbox{ keV})$ is roughly the same. 

We describe the inhomogeneous corona with five parameters: aspect
ratio $H/R$, temperature $T_c$, mean vertical optical depth
$\tau_0$, number of clouds per unit radius $2n_c$ ($n_c$ above the
plane and $n_c$ below the plane; $n_c \equiv dN_c/dR$ is constant
throughout the disk, giving a larger number of clouds per unit area in
the inner disk), and an overdensity
factor $\rho_c/\rho_0$, where $\rho_0(R)$ is the mean density of a
wedge corona with the same scale height and optical depth. The
distribution of clouds is chosen so as
to reproduce the wedge corona in the limit of $n_c\to \infty$ and
$\rho_c/\rho_0\to 1$. To do this, we require that the average mass in the
inhomogeneous corona is equal to that of a wedge corona at
each annulus $(R,R+dR)$. For spherical clouds of radius $R_c(R)$, the
coronal mass above (or below) the disk at each annulus is given by:
\begin{equation}
2\pi \rho_0 H\, R\, dR = n_c \frac{4}{3}\pi \rho_c R_c^3 \,dR\, ,
\end{equation}
which gives a clump radius
\begin{equation}\label{R_c}
R_c(R) = \left[\frac{3}{2} \left(\frac{H}{R}\right) \frac{R^2}{n_c (\rho_c/\rho_0)}
  \right]^{1/3} \,.
\end{equation}
The covering fraction can be estimated by calculating the probability
that a given region of the disk is {\it not} covered by a coronal
cloud.  Consider an annulus of area $2\pi R\, \Delta R$.  The expectation
value for the number of clouds over it is $N=n_c\, \Delta R$. If each
cloud has cross sectional area $\pi R_c^2 \ll 2\pi R\, \Delta R$, this
probability is given by
\begin{equation}\label{f_c}
p=\left(1-\frac{\pi R_c^2}{2\pi R\, \Delta R}\right)^N {\to \atop {}^{\small N\to\infty}} \exp\left(-\frac{n_c
  R_c^2}{2R}\right) = \exp\left[-\frac{1}{2}n_c^{1/3}R^{1/3}
  \left(\frac{3}{2}\frac{H}{R}\frac{\rho_0}{\rho_c}\right)^{2/3}\right].
\end{equation}
The covering fraction is simply $f_c = 1-p$. 

\begin{figure}
\caption{\label{image_hotspot} Same as Figure \ref{image_wedge}, but
  for a hot spot corona with an overdensity factor of 10. There are
  roughly 100 hot spots distributed randomly within radius $R \le
  100M$ with scale height $H/R = 0.1$.}
\begin{center}
\scalebox{0.75}{\includegraphics*[52,400][360,660]{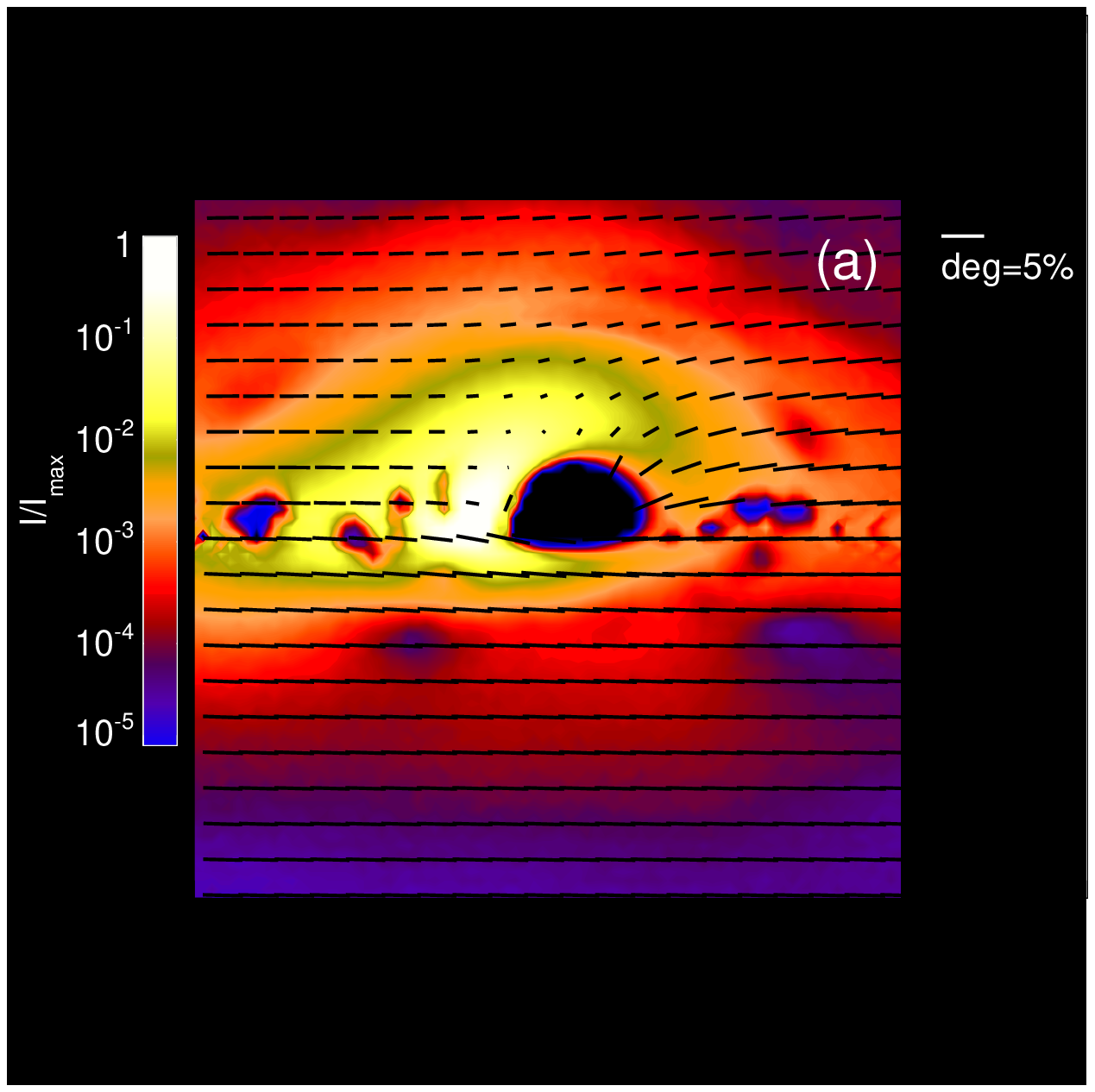}}
\scalebox{0.75}{\includegraphics*[112,400][410,660]{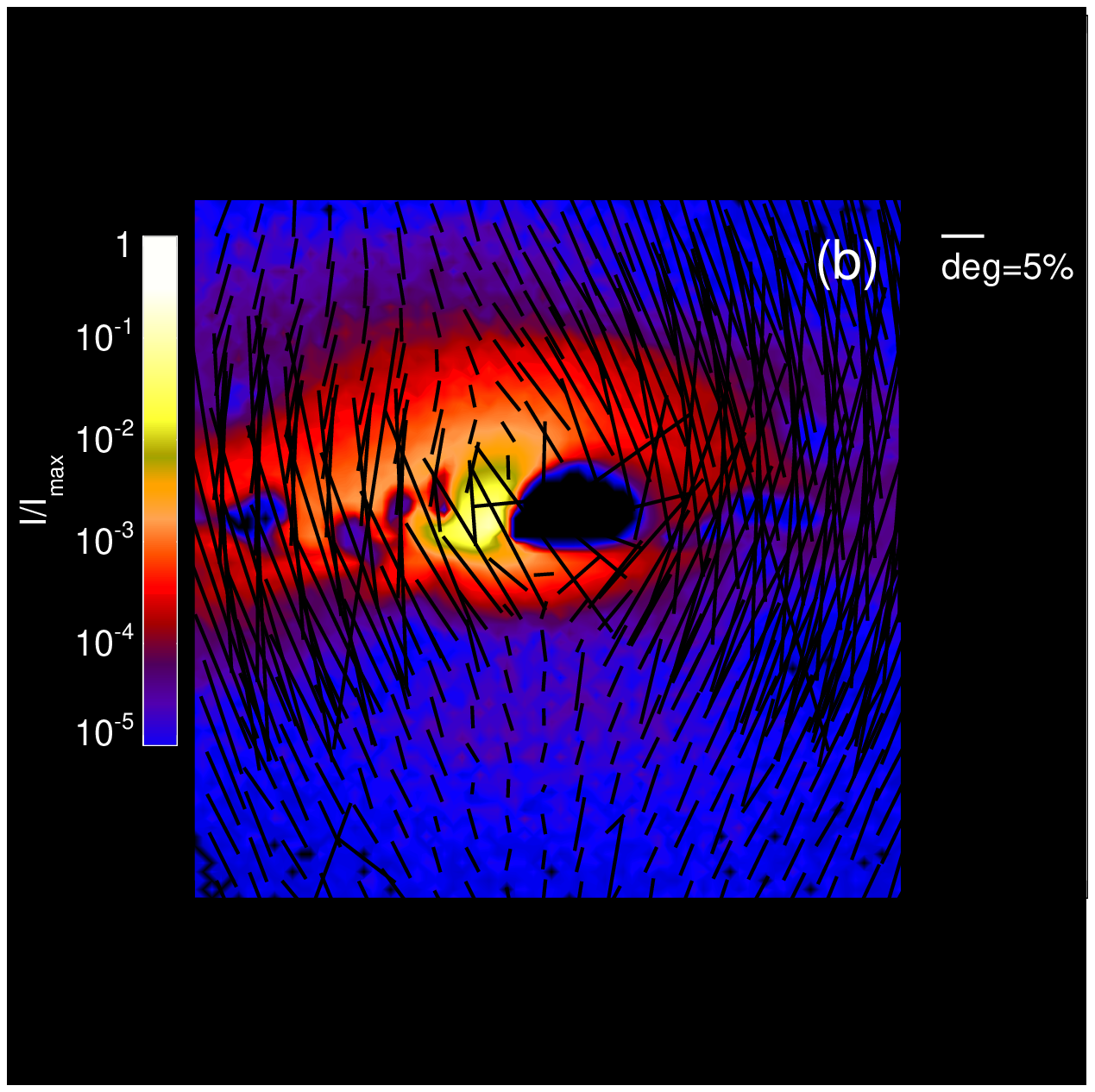}}\\
\vspace{0.1cm}
\scalebox{0.75}{\includegraphics*[52,400][360,660]{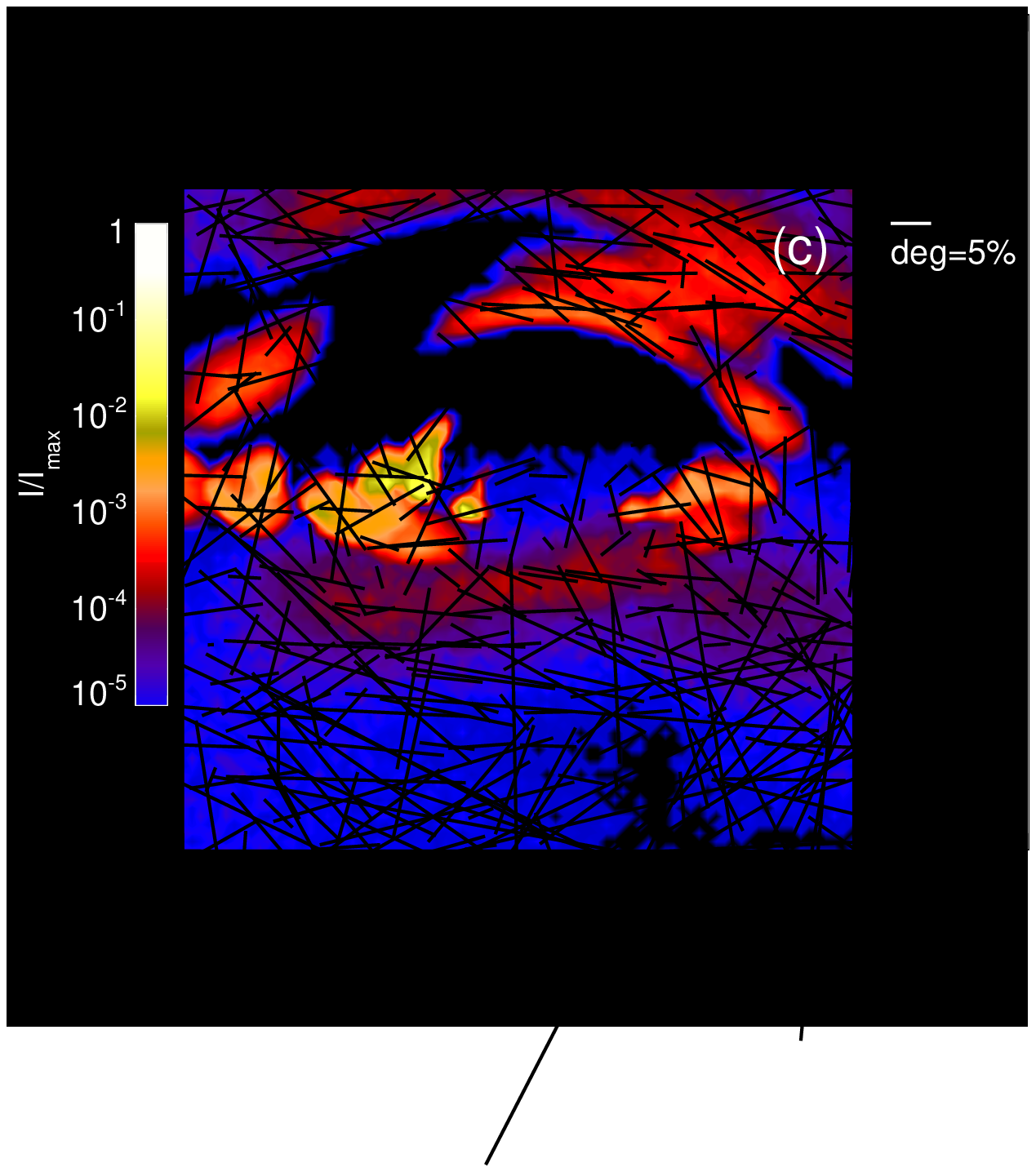}}
\scalebox{0.75}{\includegraphics*[112,400][410,660]{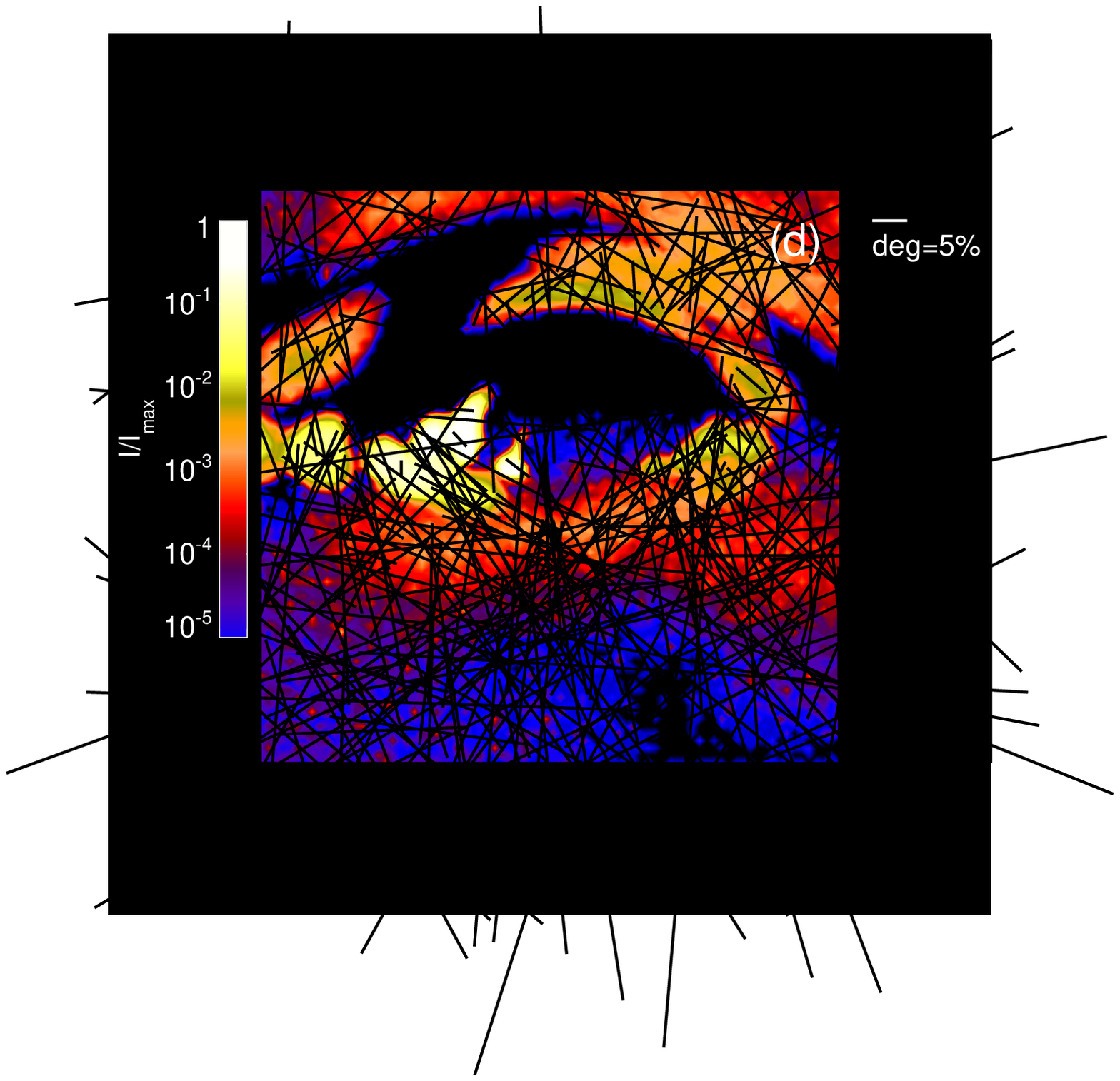}}
\end{center}
\end{figure}

In Figure \ref{image_hotspot} we show a series of images of the BH
accretion disk plus a clumpy corona, again sorted by photon history as
in Figure \ref{image_wedge}. The corona is made up of 100 clouds
distributed randomly with scale height ratio $H/R=0.1$ with constant
probability per unit radius (i.e., $dN/dR = n_c = 1$) inside radius
$R=100M$.  The mean local optical depth
$\tau_0=1$, and the density contrast $\rho_c/\rho_0=10$. For the clumping
algorithm described above, these parameters correspond to a covering
fraction in the inner disk of $f_c\sim 5\%$, comparable to that
inferred for the hard state of Cyg X-1 by \citet{poutanen:97}.
As in Figure \ref{image_wedge}, the
BH has spin $a/M=0.9$, mass $10M_\odot$, thermal flux $L_{\rm
  therm}=0.1L_{\rm Edd}$, corona temperature $T_c=100$ keV,
and observer inclination angle $75^\circ$. The direct flux from
the disk, plotted in Figure \ref{image_hotspot}a, clearly shows a number of
distinct shadows where optically thick clouds block the disk. 

The reduced covering fraction of the corona leads to a greater flux
from return radiation (Fig.\ \ref{image_hotspot}b), which otherwise behaves
much the same as in the sandwich geometry, contributing a strong
vertical component to the polarization signal. As in the direct image,
the return radiation that eventually reaches the observer is also
blocked in places by
intervening coronal patches. In the lower panels of Figure
\ref{image_hotspot}, we show the scattered photons (Fig.\
\ref{image_hotspot}c: one scatter; Fig.\ \ref{image_hotspot}d: many
scatters). Now the polarization of the hard
flux is significantly less coherent than in the sandwich
geometry. Many photons scatter multiple times within a single cloud,
eventually emerging with no particular polarization
direction. At the same time, the global geometry of the
system is still essentially the same as the sandwich model, and photons
are more likely to experience multiple scattering events by
propagating from one coronal clump to another, constrained to move roughly
parallel to the disk surface. As in the homogeneous model, this
leads to a net vertical polarization in the observed flux at high
energies. 

In Figure \ref{hotspot_grid} we plot the flux, polarization degree,
and polarization angle as a function of energy for the inhomogeneous
corona and observer angles of $45^\circ$, $60^\circ$, and
$75^\circ$. 
As in the sandwich geometry, here too we see that with increasing
inclination, a greater fraction of the observed flux comes from
scattered photons. Due to the smaller covering factor, a greater
fraction of the thermal flux can escape at all inclinations, but the
spectrum is still
dominated by Comptonized photons above $7-8$ keV. From equation
(\ref{R_c}), we see that the individual coronal clumps have typical
optical depths of $\tau\approx 10$ in the inner disk. This naturally
leads to a harder power-law spectrum relative to that of the wedge
geometry.

The polarization signature for the inhomogeneous corona is similar to
the sandwich geometry at low energies, but noticeably different
above a few keV. Below the thermal peak, a sandwich corona
merely extends the classic Chandrasekhar atmosphere, leaving the
polarization unchanged. The scattering
contribution (dot-dashed curves) is more weakly polarized when the gas
is clumped because it is easier for a
photon emitted at a large angle relative to the disk normal to scatter
once off a coronal cloud and then reach an observer\footnote{The
  clumpy corona can be thought of as a collection of hard scattering
  spheres where photons are unable to forward-scatter, unlike the
  homogeneous atmosphere of diffuse electrons.}, contributing a
vertical component to the polarization, slightly diluting the dominant
horizontal signal at low energy. On the other hand,  
the degree of polarization at high energies is significantly
lower for a clumpy corona than for the sandwich corona. This is
because many high-energy photons scatter
multiple times within a single cloud, ultimately emerging with little or
no net polarization due to the spherical symmetry of the cloud.

Lastly, we note that when the corona is clumped, 
the {\it shape} of the transition from horizontal to
vertical polarization more closely resembles that
of the thermal state than that of a homogeneous sandwich corona:
relatively smooth and in the $\psi<0$ direction. Because of the
small covering factor of the corona, thermal seeds emitted in the
inner disk have a greater chance of returning to the disk or even
scattering off a
coronal clump on the far side of the BH, thus leading to a $\psi(E)$
distribution quite similar to that of the thermal disk with no
corona (SK09). Not surprisingly, the polarization
signature is sensitive to the compactness of the coronal hot spots, as
shown in Figure \ref{hotspot_rho}. In
the limit of $\rho_c/\rho_0=1$, the signal more closely resembles the
homogeneous sandwich result: a sharp transition at $2-3$ keV, with
$\psi>0$ and the
polarization degree rising above $6\%$ at high energies (this case
does not identically reproduce the homogeneous sandwich when $N$ is
finite because, although the centers of the clouds must lie within
$H$ of the plane, their outer portions may extend farther, resulting
in an inhomogeneous geometry even when $\rho_c/\rho_0=1$). As the
compactness increases, the covering fraction and the overall symmetry
decrease, giving a smoother transition and smaller amplitude above a
few keV. 
In Figure \ref{hotspot_Nc} we show the dependence of the polarization
signal on the number of clumps in the corona, holding the overdensity
fixed at $\rho_c/\rho_0=10$. In the limit
of $N_{\rm clump}\to \infty$, the homogeneous result is
reproduced, and in the limit of $N_{\rm clump}\to 0$ we get the
pure thermal result. 

\section{SPHERE GEOMETRY}\label{sphere}

The last model we consider is a spherical corona immediately
surrounded by a truncated disk, a geometry motivated by a variety of
spectral and timing observations (e.g.\
\citet{gierlinski:97,done:99,makishima:08,ingram:09,done:09}).
The disk extends in to radius $R_{\rm edge}$ with a Novikov-Thorne
emissivity profile, and the corona is
defined by its temperature $T_c$ and optical depth $\tau_0$, as
measured from the horizon out to $R_{\rm edge}$. The electron density in
the corona is taken to be constant, and in keeping with its spherical
shape, the fluid is at rest in the frame
of a zero-angular momentum observer \citep{bardeen:72}.
In order to most closely match the
spectral properties of the sandwich corona, we also embed thermal seed
photons with a uniform distribution inside the corona. The net
flux and spectrum as a function of radius is determined by matching
the shell-integrated flux to that of a standard Novikov-Thorne disk
down to the ISCO. These 
coronal seed photons are emitted isotropically with zero
polarization.

\begin{figure}
\caption{\label{image_sphere} Same as Figure \ref{image_wedge}, but
  for a spherical corona surrounded by a truncated thermal disk with
  an inner edge at $R_{\rm edge}=15M$.} 
\begin{center}
\scalebox{0.75}{\includegraphics*[52,400][360,660]{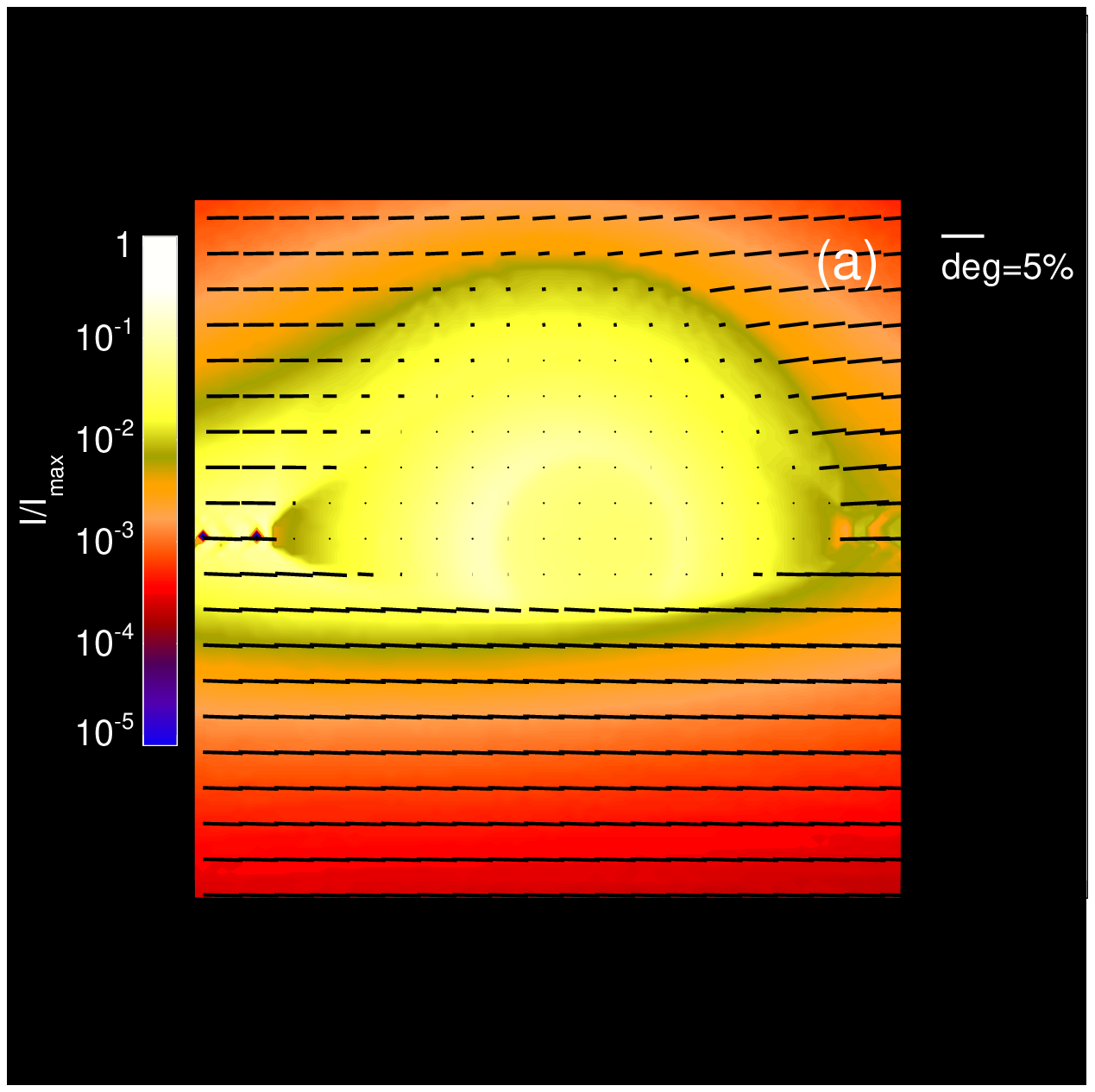}}
\scalebox{0.75}{\includegraphics*[112,400][410,660]{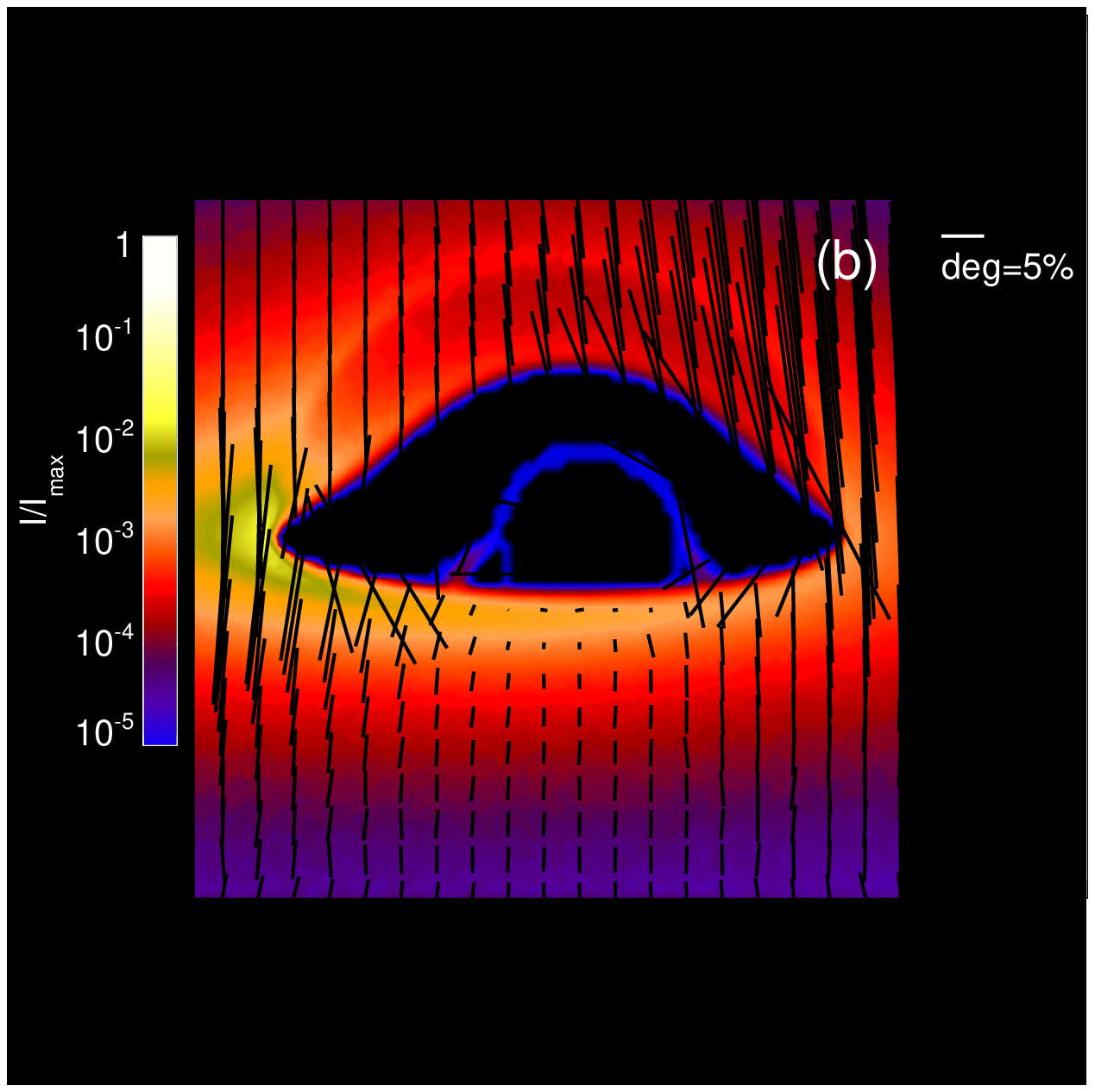}}\\
\vspace{0.1cm}
\scalebox{0.75}{\includegraphics*[52,400][360,660]{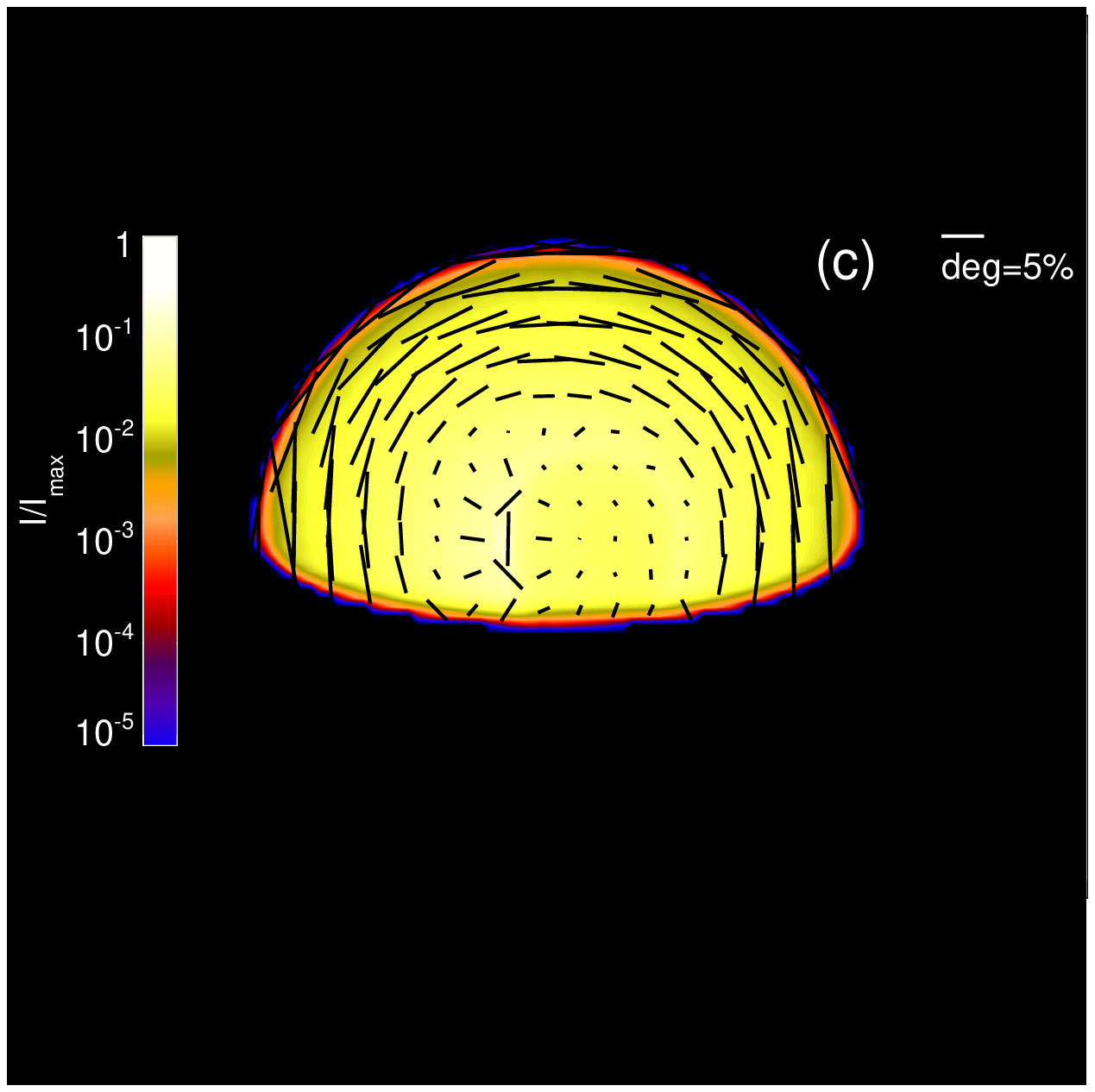}}
\scalebox{0.75}{\includegraphics*[112,400][410,660]{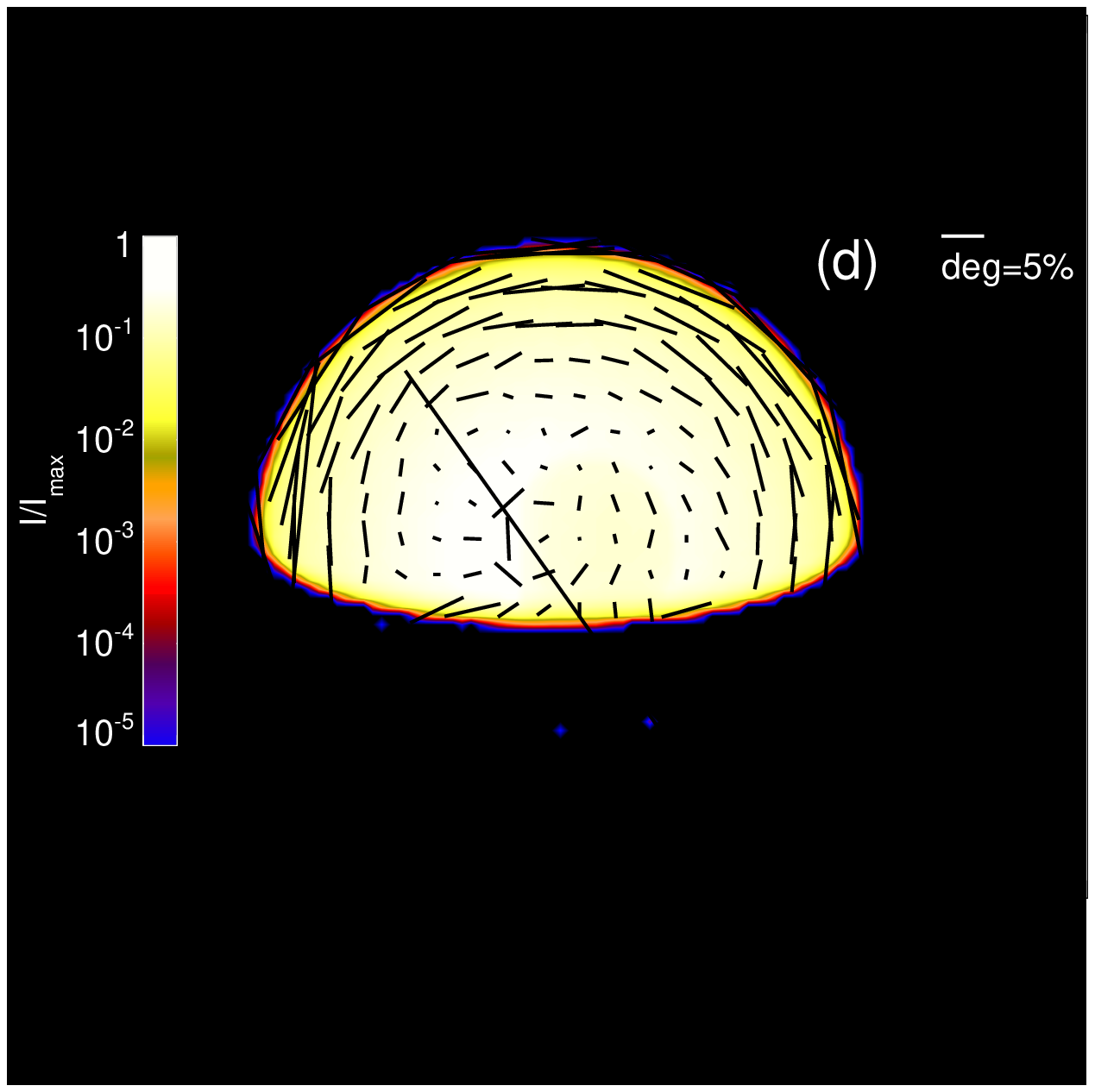}}
\end{center}
\end{figure}

In Figure \ref{image_sphere} we show the same photon-sorted images as
in Figure \ref{image_wedge}, now for the spherical coronal model with
$R_{\rm edge}=15M$. The direct disk flux behaves as expected:
because it is emitted from regions that are almost entirely in the
non-relativistic regime, its polarization is very close to the
Chandrasekhar value.  There is now also a
significant amount of unpolarized thermal flux coming directly from the coronal
seeds. These seed photons can also scatter off the surrounding
disk, in which case they
are classified here as return radiation, again strongly polarized in
the vertical direction. The corona-scattered flux shown in Figures
\ref{image_sphere}c,d closely resembles that of a scattering-dominated atmosphere
around a star or planet: zero polarization from surfaces normal to the
observer, and strongly-polarized (but limb-darkened) flux from the
edges. However, the disk blocks a portion of the bottom half of the
sphere, with the very bottom the most likely to be blocked.  Consequently,
the net polarization from a spherical corona is
non-zero and oriented in the vertical direction. We have not shown
here the contribution from photons that scatter in the corona and 
{\it then} off the disk, but these also are strongly polarized in the
vertical direction, much like the image in the upper-right panel. 

Just as in the analogous figures for the sandwich and clumped models,
Figure \ref{sphere_grid} shows that the total flux spectrum is dominated at
low energies by the thermal contribution (here the sum of the disk proper
and the thermal seeds embedded in the corona), while it is dominated at
high energies by inverse Compton scattered photons.  Because the coronal density
and temperature were specifically chosen to give the same Compton
$y$-parameter as in the sandwich geometry (see Fig.\
\ref{wedge_grid}), it is not surprising that the spectra agree so
closely.

The polarization signature is relatively straightforward to
understand, especially in the context of Figure \ref{image_sphere}:
the polarization from the direct radiation is almost exactly
horizontal, decreasing steadily in amplitude with energy as a greater
proportion of the observed flux comes from the unpolarized and
higher-energy seeds in
the corona. The scattered flux is consistently vertical, since even
the lower-energy photons that have scattered only once have a net
vertical orientation (see the lower-left panel of Fig.\
\ref{image_sphere}). Because the unobscured coronal hemisphere looks
much the same from each of the three viewing angles, the polarization amplitude at
high energies is only weakly dependent on observer
inclination. Combining these effects gives the now-familiar behavior
of horizontal polarization at low energies governed by the
Chandrasekhar limit transitioning to vertical
orientation above the thermal peak. The spherical geometry leads to a
particularly sharp transition, at least for $R_{\rm edge} \gtrsim
10M$, but it is still clear that a negative $\psi$ is preferred in the
transition region. Again this can be understood by the geometry of the
return radiation scattering off the region of the disk most strongly
beamed towards the observer. In this case, the coronal photons
originating from the central region above the disk plane replace the return
radiation that similarly had passed near the photon orbit, just
above the BH.

\begin{figure}
\caption{\label{sphere_Redge} Degree and angle of polarization for a
  spherical corona, varying the radius of the corona (inner edge of
  the disk).}
\begin{center}
\scalebox{0.8}{\includegraphics{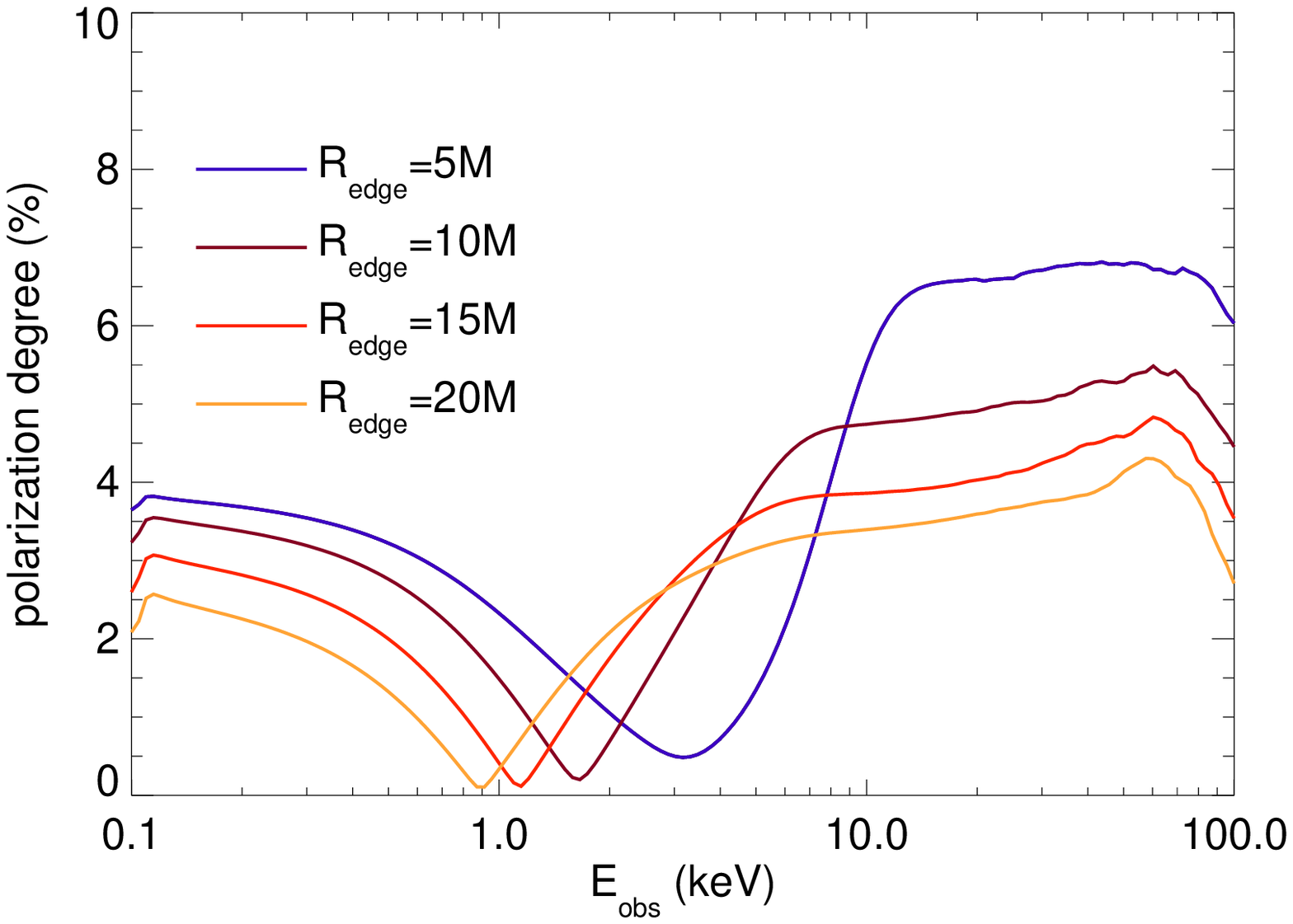}}\\
\scalebox{0.8}{\includegraphics{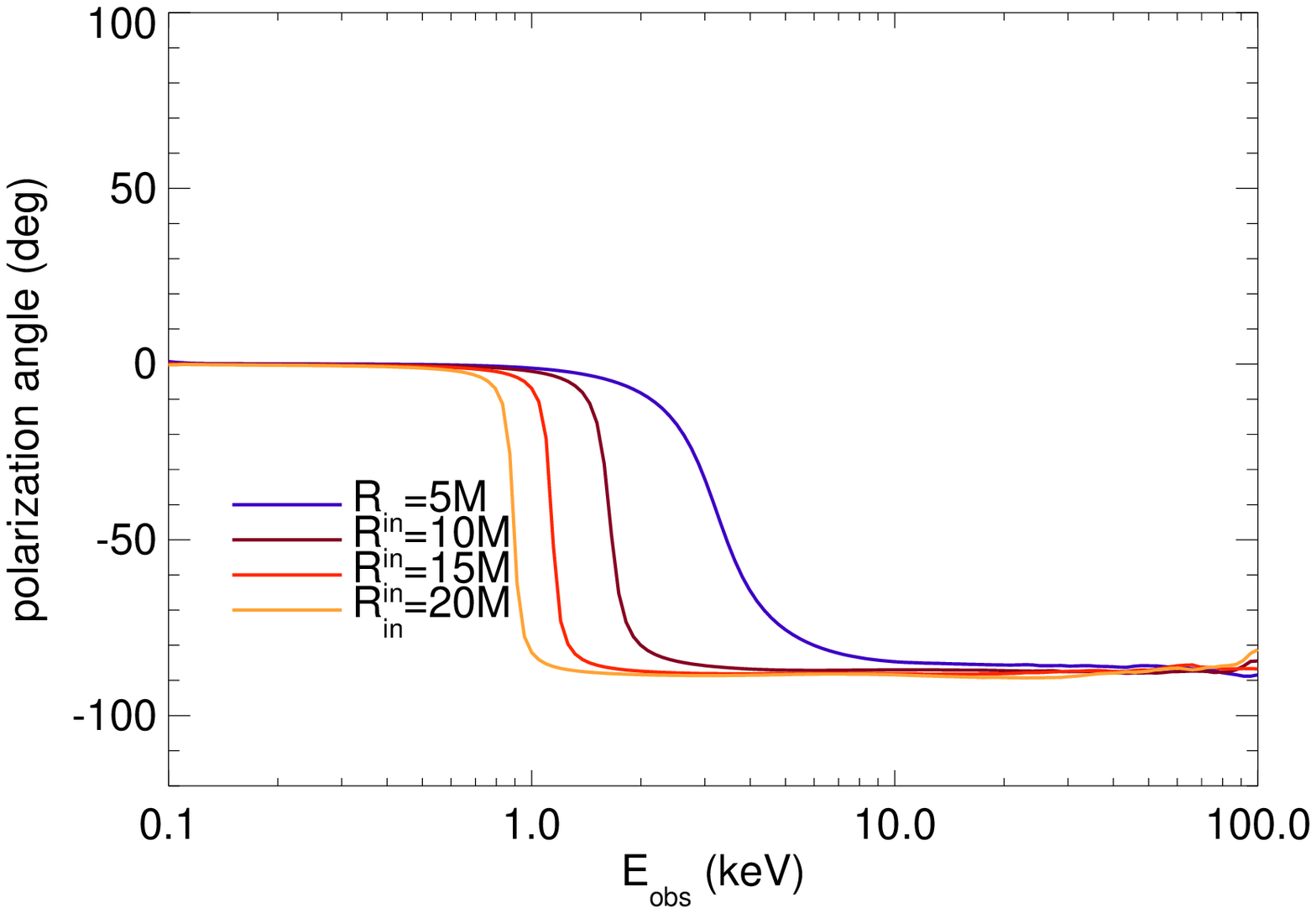}}
\end{center}
\end{figure}

In Figure \ref{sphere_Redge} we investigate the effects of varying
$R_{\rm edge}$, while keeping the total optical depth of the
corona constant. Since the thermal disk moves in with decreasing
$R_{\rm edge}$, we find the transition point moves to higher
energies, as a greater fraction of the flux is coming directly from
the disk and is therefore horizontally polarized. The shape of the
transition grows broader with decreasing $R_{\rm edge}$ as
relativistic effects become more important, rotating the polarization
angle as in the thermal state. In the limit of large $R_{\rm edge}$,
where Newtonian physics dominates the problem, the system becomes
scale-invariant and symmetry constrains the polarization at any given
energy to be exactly horizontal or exactly vertical. In the limit of
very small $R_{\rm edge}$, we reproduce the thermal result, plus a
small power-law contribution to the spectrum at high energy due to the
ultra-compact corona. 

\begin{figure}
\caption{\label{sphere_tauT} Degree and angle of polarization for a
  spherical corona, fixing $R_{\rm edge}=10M$ and varying the optical
  depth and electron temperature.}
\begin{center}
\scalebox{0.8}{\includegraphics{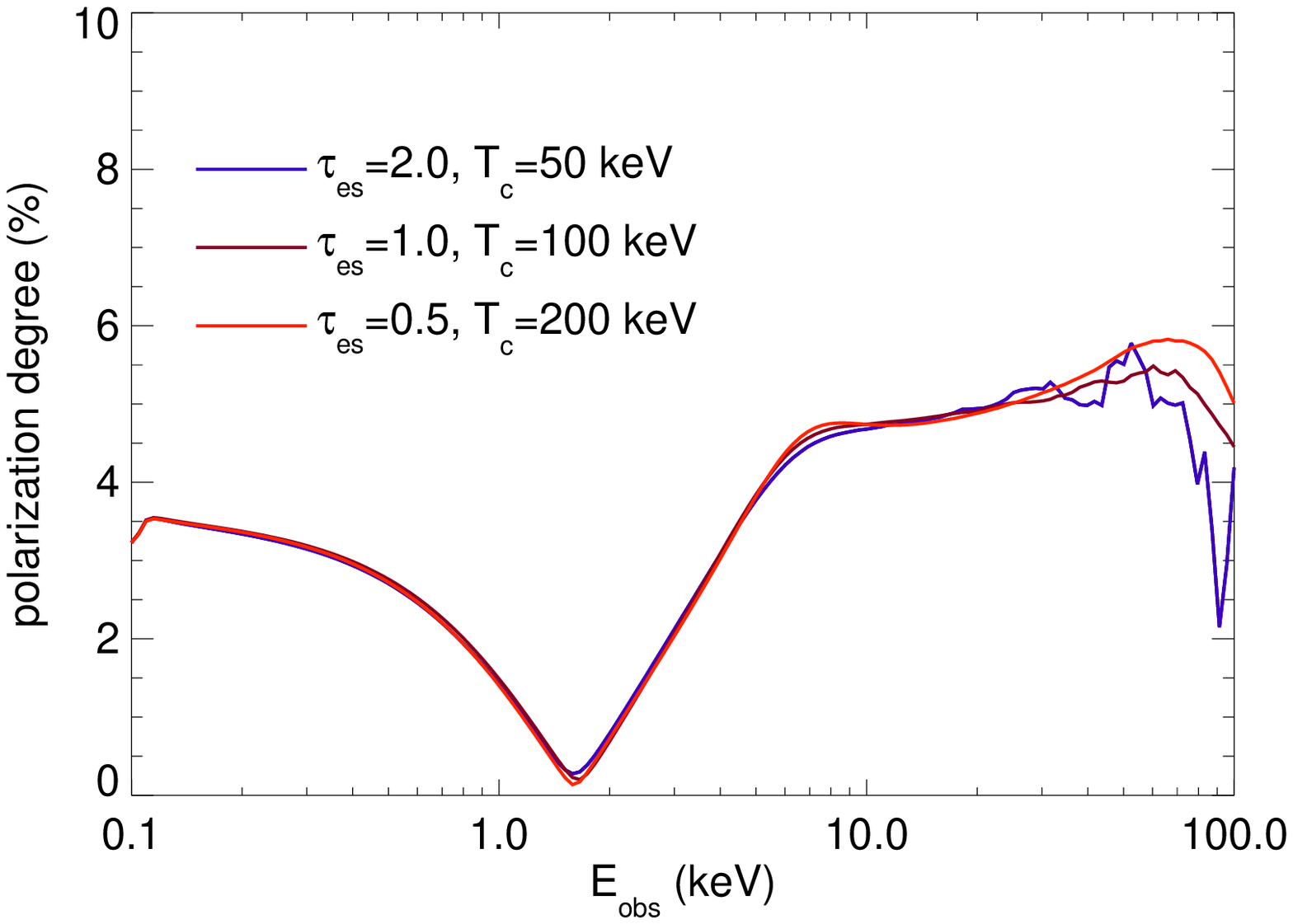}}\\
\scalebox{0.8}{\includegraphics{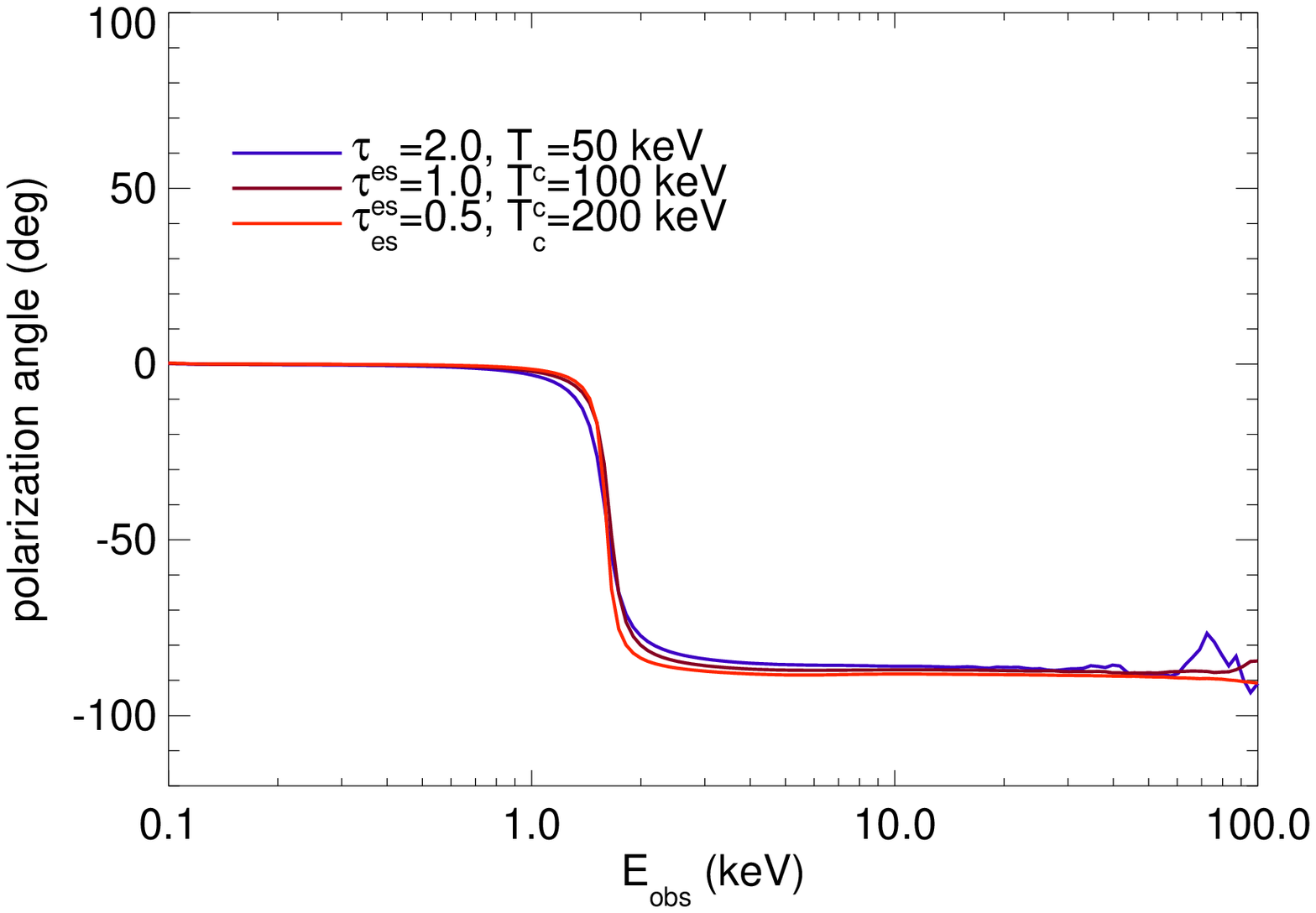}}
\end{center}
\end{figure}

Holding the corona radius constant at $R_{\rm edge}=10M$, in Figure
\ref{sphere_tauT} we show the effects of varying the optical depth and
temperature of the corona. As in Figure \ref{wedge_tauT} for the
sandwich geometry, here too we keep the Compton $y$-parameter fixed so
that the resulting spectra are nearly identical. However, unlike the
sandwich model, for the spherical corona we find essentially no
dependence on the coronal temperature or density. The reason is the
essential simplicity of the scattering geometry: the polarization
signature from the direct radiation alone is nearly
the same, regardless of the coronal optical depth, and from Figure
\ref{sphere_grid} we see that all scattered radiation is polarized
exactly the same over the entire spectrum. Thus the behavior of the
net signal is simply a function of the relative flux in each
component, and this is held constant by fixing the Compton
$y$-parameter.

\section{ACTIVE GALACTIC NUCLEI}\label{AGN}

It has long been known that a significant fraction of the flux
from AGN is emitted in the X-ray band
\citep{elvis:78}. As in stellar-mass BHs, this high-energy flux likely
comes from lower-energy seed photons inverse Compton scattered in a
corona of hot electrons with $T_c \sim 100$ keV. Similar to the
stellar-mass case, this leads to a relatively hard power-law spectrum
with index $\alpha \sim 0.5-1$
\citep{nandra:91,mushotzky:93}. However, unlike the stellar-mass case,
the temperature of the inner disk for an AGN will be well below a keV,
leading to a thermal peak in the UV band. Furthermore, even when the
disk is dominated by radiation pressure and electron scattering
opacity, there should still be a substantial fraction of metals that
are not fully ionized, producing a large opacity for absorption
above $\sim$ 1 keV. 

Both of these AGN features---lower energy seed photons and an X-ray
absorbing disk---lead to important differences in the polarization
signature as compared to the stellar-mass models described
above in Sections \ref{wedge}-\ref{sphere}. Because the seed photons
start off with lower energies, they
must scatter more times in the corona in order
to reach the $\sim 1-10$ keV band. For a thin sandwich corona, this
means that the scattering geometry is even more constrained than in
the stellar-mass case, forcing the photons to move in a plane parallel
to the disk surface, leading to a stronger vertical
polarization. The AGN disk absorbs much of the incident X-ray flux
from the corona, so the Compton $y$-parameter is effectively smaller than
that of a stellar-mass system with the same coronal properties because
scattering sequences are halted once a photon strikes the disk (we do
not re-radiate the X-ray flux, but rather treat it as completely absorbed),
thereby reducing the average path length the photons that escapes to
infinity.  An absorbing disk boundary condition with a sandwich corona
therefore leads to an even higher degree of
X-ray polarization because the photons are forced to scatter in a more
constrained geometry before escaping the corona. 

\begin{figure}
\caption{\label{agn_flux} Observed broad-band flux from a supermassive
  BH with $M=10^7M_\odot$, $a/M=0.9$, $L_{\rm therm}=0.1L_{\rm Edd}$,
  and $i=45^\circ$. The different curves correspond to different
  covering fractions, as described in the text.}
\begin{center}
\scalebox{0.8}{\includegraphics{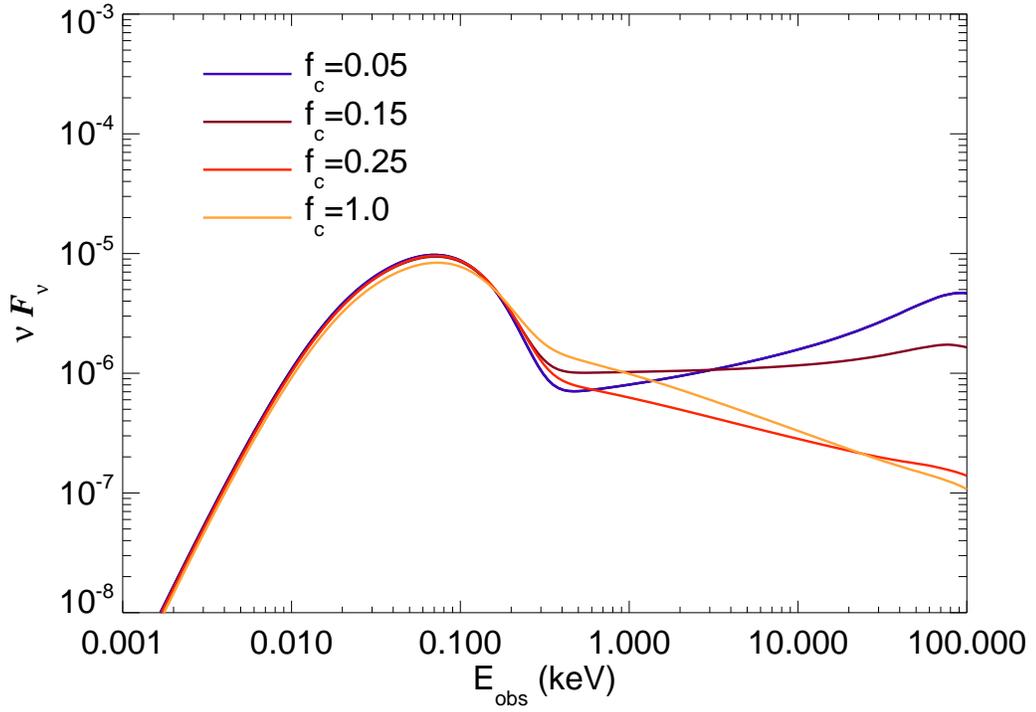}}
\end{center}
\end{figure}

In Figure \ref{agn_flux} we show broad-band spectra from an AGN
with central mass $M=10^7 M_\odot$, spin parameter $a/M=0.9$,
thermal luminosity $L_{\rm therm}=0.1 L_{\rm Edd}$, and observer
inclination angle $i=45^\circ$. Four different corona models are
considered: a homogeneous wedge with $H/R=0.1$, $\tau_0=1$, and
$T_c=100$ keV, and three
clumpy models with the same scale height, mean optical depth, and temperature,
but with $n_c=1$ and overdensities of $\rho_c/\rho_0=1$, $3$, and $10$. The
inhomogeneous models can be characterized by their covering fractions as
measured in the inner disk at $R=10M$; these overdensities correspond to
$f_c=0.25$, $0.15$, and $0.05$, respectively.  As can be seen in
Figure \ref{agn_flux}, a
greater covering fraction leads to a softer spectrum, while an
inhomogeneous corona with smaller yet denser clumps gives a larger
optical depth for those photons that do not escape directly from the
disk, in turn giving a larger Compton $y$-parameter and harder
spectrum. 

\begin{figure}
\caption{\label{agn_C} Degree and angle of polarization for a
  supermassive BH with the same parameters as in Figure
  \ref{agn_flux}.}
\begin{center}
\scalebox{0.8}{\includegraphics{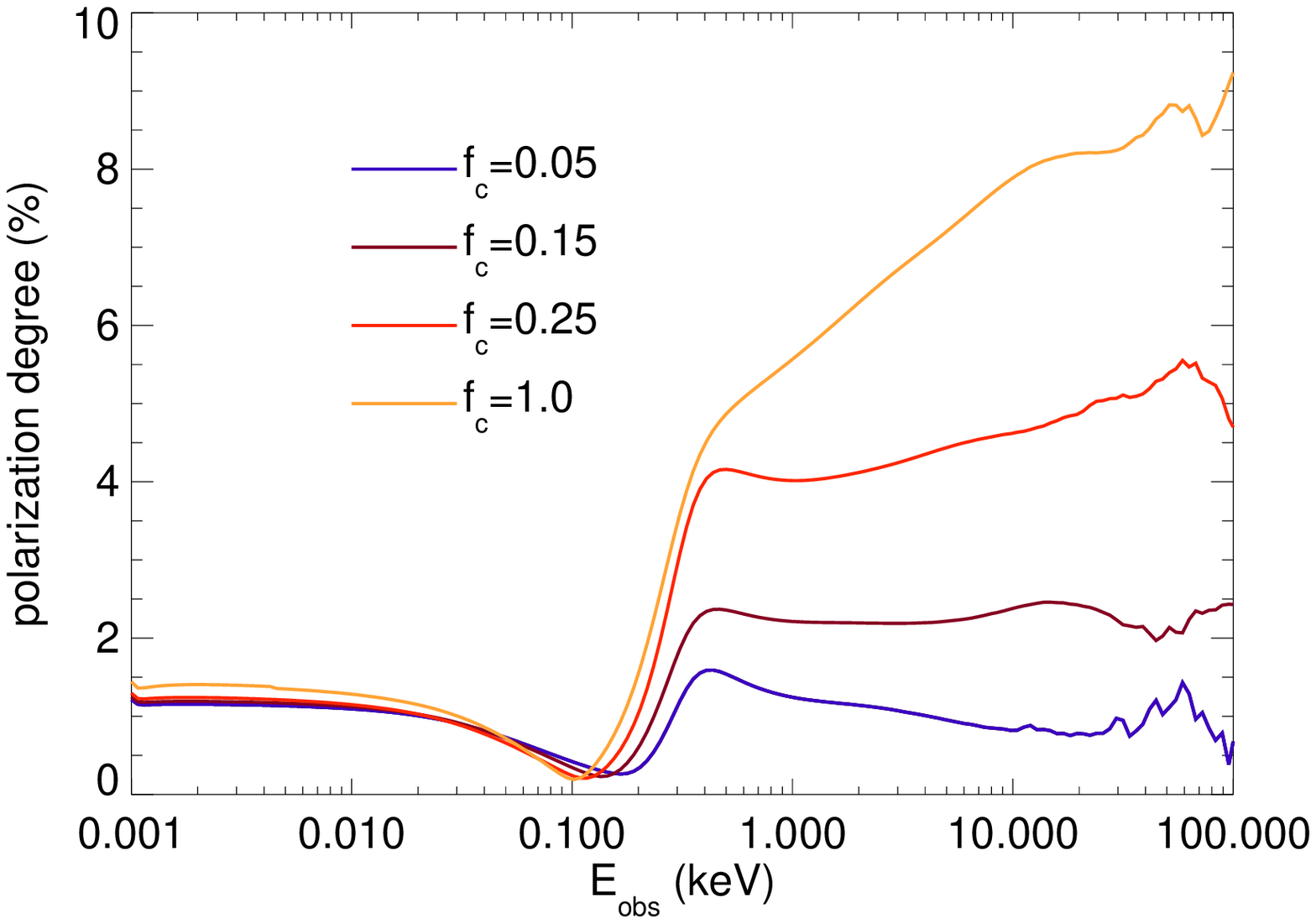}}\\
\scalebox{0.8}{\includegraphics{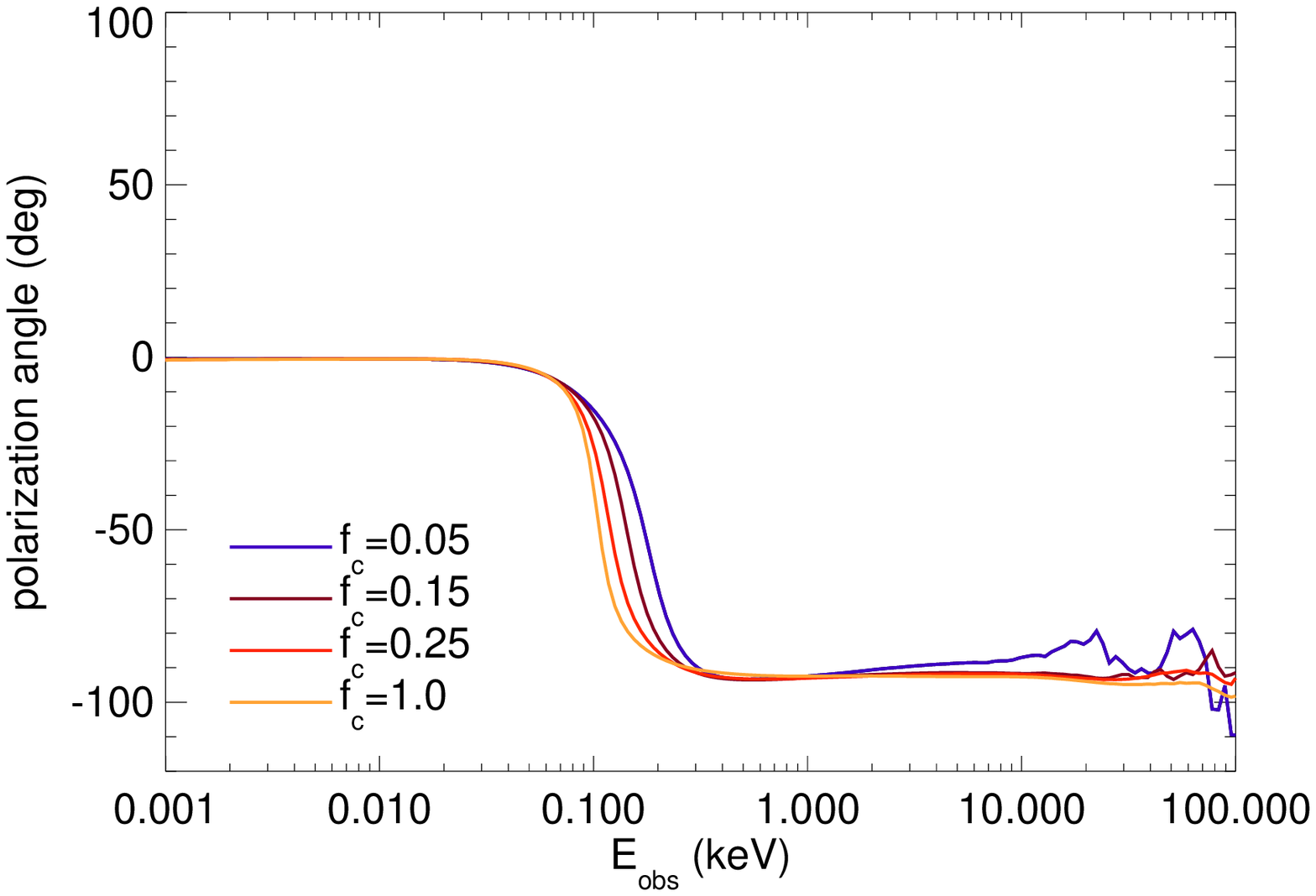}}
\end{center}
\end{figure}

As mentioned above, the disk absorption in the AGN models results
in a softer spectrum than stellar-mass BHs with the same coronal
parameters. This can be seen by comparing Figure \ref{agn_flux} with
the upper-left panels in Figures \ref{wedge_grid} (wedge geometry;
$f_c=1$) and \ref{hotspot_grid} (clumpy geometry; $f_c=0.05$). In both
cases, the spectral index $\alpha$ is increased (i.e., softer) in the
AGN case by about 0.5. The effect of this absorption on the X-ray polarization
can be seen in
Figure \ref{agn_C}, which plots the degree and angle of polarization
as a function of energy for the same model parameters used in Figure
\ref{agn_flux}. As expected, for the wedge geometry with $f_c=1$, the
absorbing disk leads to a larger degree of polarization than the
reflecting boundary condition used in stellar-mass BHs ($\sim 6-8\%$
for AGN compared with the $\sim 4\%$ shown in Fig.\ \ref{wedge_grid} for
$M=10M_\odot$ and $i=45^\circ$).

For the clumpy coronae, however, we find the opposite effect:
including absorption actually {\it reduces} the degree of polarization. Recall from
Section \ref{hotspot} that inhomogeneous coronae generally produce
a weaker polarization signal than the uniform-density sandwich
geometry. This is because photons scattering multiple times in a
single spherical cloud will ultimately escape with no net
polarization. It is only through the {\it global} scattering geometry
that a net vertical polarization is acquired, due to the small number
of photons that scatter from one cloud to another, often
reflecting off the disk surface at grazing incidence along the
way. Since many of these
photons are lost to absorption in the AGN case, global geometric effects
play a smaller role, and thus the degree of polarization is
diminished ($\sim 1\%$ for AGN versus $\sim 2\%$ for the stellar-mass
case in Fig.\ \ref{hotspot_grid}). Thus we find that for AGN, X-ray
polarization is even more sensitive to the inhomogeneity of the corona
than for stellar-mass BHs.

\begin{figure}
\caption{\label{agn_i} Degree of polarization for a
  supermassive BH with covering fractions $f_c=0.15$ ({\it dashed
  curves}) and $f_c=1$ ({\it solid curves}), for a range
  of observer inclination angles.}
\begin{center}
\scalebox{0.8}{\includegraphics{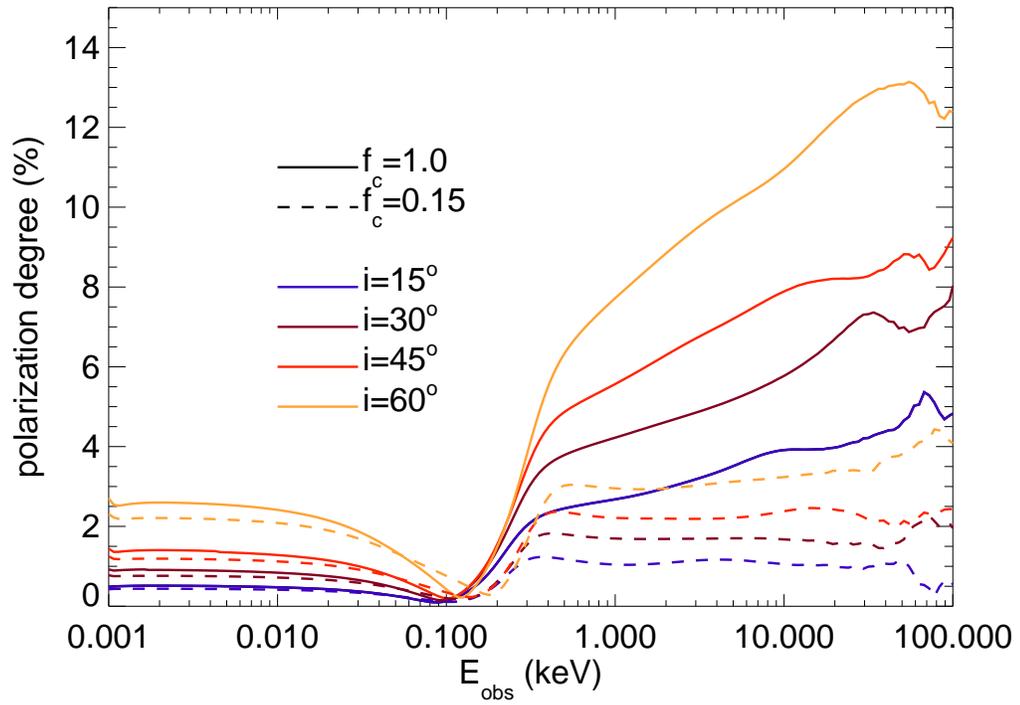}}\\
\end{center}
\end{figure}

From Figure \ref{agn_flux}, we see that it should be possible to
determine the AGN covering fraction from the X-ray spectrum alone,
without polarization information. If so, then the degree of
polarization could rather be used to constrain the inclination of the
AGN disk, a parameter that can be quite difficult to constrain in many
cases. In Figure \ref{agn_i} we compare the degree of polarization for a
range of inclination angles and two different corona models: wedge
geometry with $f_c=1$ (solid curves) and a clumpy geometry with
$f_c=0.15$ (dashed curves). Clearly the inclination will be easier to
measure for a smooth corona, but even for a clumpy corona with
$f_c=0.15$, a first-generation X-ray polarimeter should be able to
distinguish between $i=15^\circ$ and $45^\circ$. 

\begin{figure}
\caption{\label{agn_Nc} Degree of polarization for a
  supermassive BH with covering fraction $f_c=0.15$, inclination
  $i=45^\circ$, and a range of clump sizes. From equation (\ref{f_c})
  we see that, for a constant covering fraction, the number density of
  clouds $n_c$ is a function of the overdensity parameter $\rho_c/\rho_0$.}
\begin{center}
\scalebox{0.8}{\includegraphics{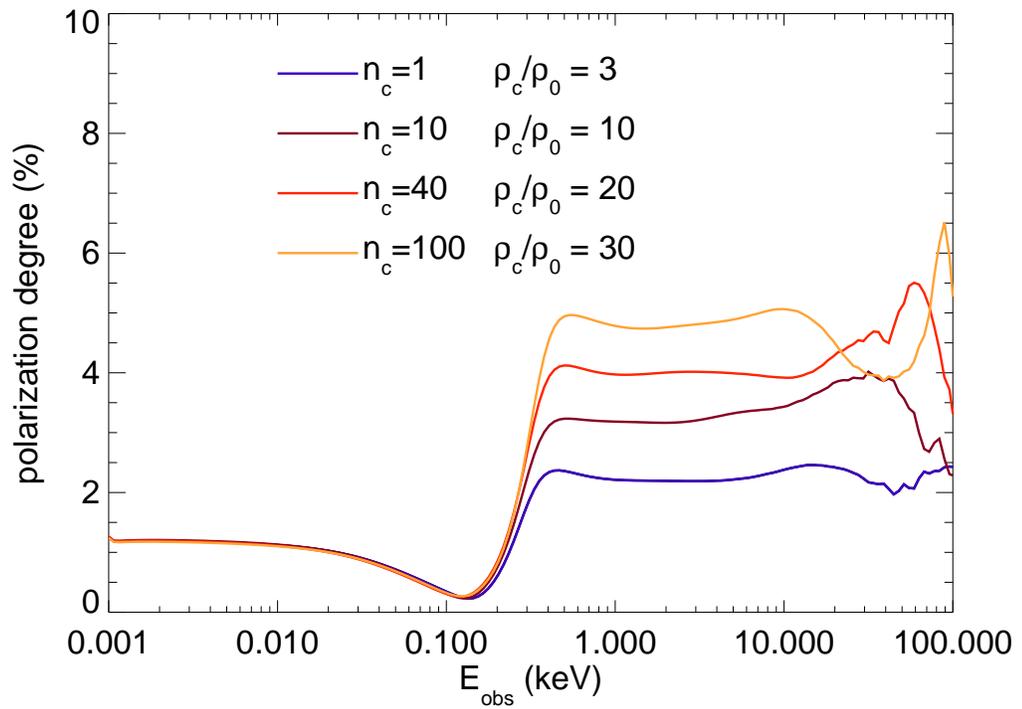}}\\
\end{center}
\end{figure}

In the event that we
can measure the covering fraction {\it and} inclination with other
observations, the polarization can constrain other corona
parameters like the number of clumps $n_c$ and their overdensity
$\rho_c/\rho_0$. From equation (\ref{f_c}), we see that for a constant
covering fraction, $n_c \propto (\rho_c/\rho_0)^2$, so increasing the
number of clouds both decreases their characteristic size and
increases their density. Yet when averaged over a photon's entire
path, a corona with larger $n_c$ will appear more homogeneous, despite
the increased density perturbations on very small scales. This in turn
leads to a greater degree of polarization, as shown in Figure
\ref{agn_Nc}. Holding the inclination and covering fraction fixed at
$i=45^\circ$ and $f_c=0.15$, respectively, we vary the number density
of coronal clumps. While all four cases plotted in Figure \ref{agn_Nc}
have nearly identical spectra, we see that polarization can
distinguish them and thus give improved constraints on the corona
geometry.  The different models in Figure \ref{agn_Nc} would also likely
have observably different timing properties: if individual clumps evolve
coherently, but are independent of one another,
systems with smaller $n_c$ should produce greater amplitude flux
variations in the X-ray band.

In practice, AGN polarization measurements will have some additional
observational challenges not present in stellar-mass BH
systems. First, the typical X-ray flux from nearby Seyfert galaxies and
quasars is at least an magnitude smaller than the
brightest galactic BHs. This means that much longer observation times
will be required to reach a comparable level of polarization
sensitivity. In addition, type 1 AGN are expected to have only
modest inclinations ($i \lesssim 45^\circ$) because at higher inclinations
our view of the inner disk is likely
blocked by the surrounding dusty torus (assuming the torus and disk are
oriented in the same direction).  
There are also two potential sources of dilution.  Radio-loud AGN
generically have somewhat larger ratios of X-ray flux to optical flux
than radio-quiet \cite{shen:06}, a fact plausibly interpreted as due to
a jet contribution; in blazars, the jet contribution is almost certainly
substantial.  In addition, Fe K$\alpha$ emission
often accounts for a few percent of the 2--10~keV flux.
Lastly, the much longer variability timescales for supermassive
BHs make it unlikely that we could use the strategy discussed
in Section~\ref{wedge}, in which observations of multiple spectral
states can jointly determine the orientation of the spin on the sky.
This advantage would be absent in AGN systems, which for the most part do not undergo
significant state transitions on an observable timescale.

\section{DISCUSSION}\label{discussion}

Using a Monte Carlo ray-tracing code in the Kerr metric, we have explored a
variety of models for X-ray-emitting coronae attached to accreting stellar-mass
black holes in the hard/SPL states or to radio-quiet AGN.  Over a wide
range of model parameters, the polarization swings from parallel to the
disk plane (``horizontal") at low energy to perpendicular (``vertical")
at high energy.  The location of this transition is generically at an
energy a few times that of the highest temperature found in the disk;
in the case of stellar-mass BHs, that means $\sim 1--3$~keV; in AGN,
$\sim 100$~eV.  The detailed
properties of this transition, and the polarization amplitude at the
low- and high-energy limits, provide the observer with information
about the BH's mass, spin, inclination, and accretion rate, as well as
the coronal properties: degree of (in)homogeneity, vertical scale
height, temperature, optical depth, and covering fraction.

We do not claim that a single or even multiple
polarization observations of a given source will unambiguously measure
all these model parameters. What does seem possible, even with
relatively low precision data, is to rule out large
regions of parameter space, and in so doing challenge a number of the
traditional paradigms and toy models for BH accretion geometry.
For example, if we were to observe an upper limit of
$\delta \le 2\%$ at $\sim 10$ keV from a stellar-mass system with known binary
inclination $i\gtrsim 60^\circ$, we would have only two choices: to
abandon any form of smooth corona geometry in favor of a highly asymmetric,
inhomogeneous accretion flow, or to posit a misalignment between the
binary orbit and the inner accretion disk (thus giving lower
inclination and polarization than expected). For AGN sources, where
the expected disk inclination is lower ($i \lesssim 45^\circ$), a null
polarization measurement would rule out a smooth sandwich corona for
anything but a nearly face-on system.

\begin{figure}
\caption{\label{wedge_spin} Degree and angle of polarization for a
  sandwich corona, varying the spin of the BH, while holding fixed the
  total thermal flux.}
\begin{center}
\scalebox{0.8}{\includegraphics{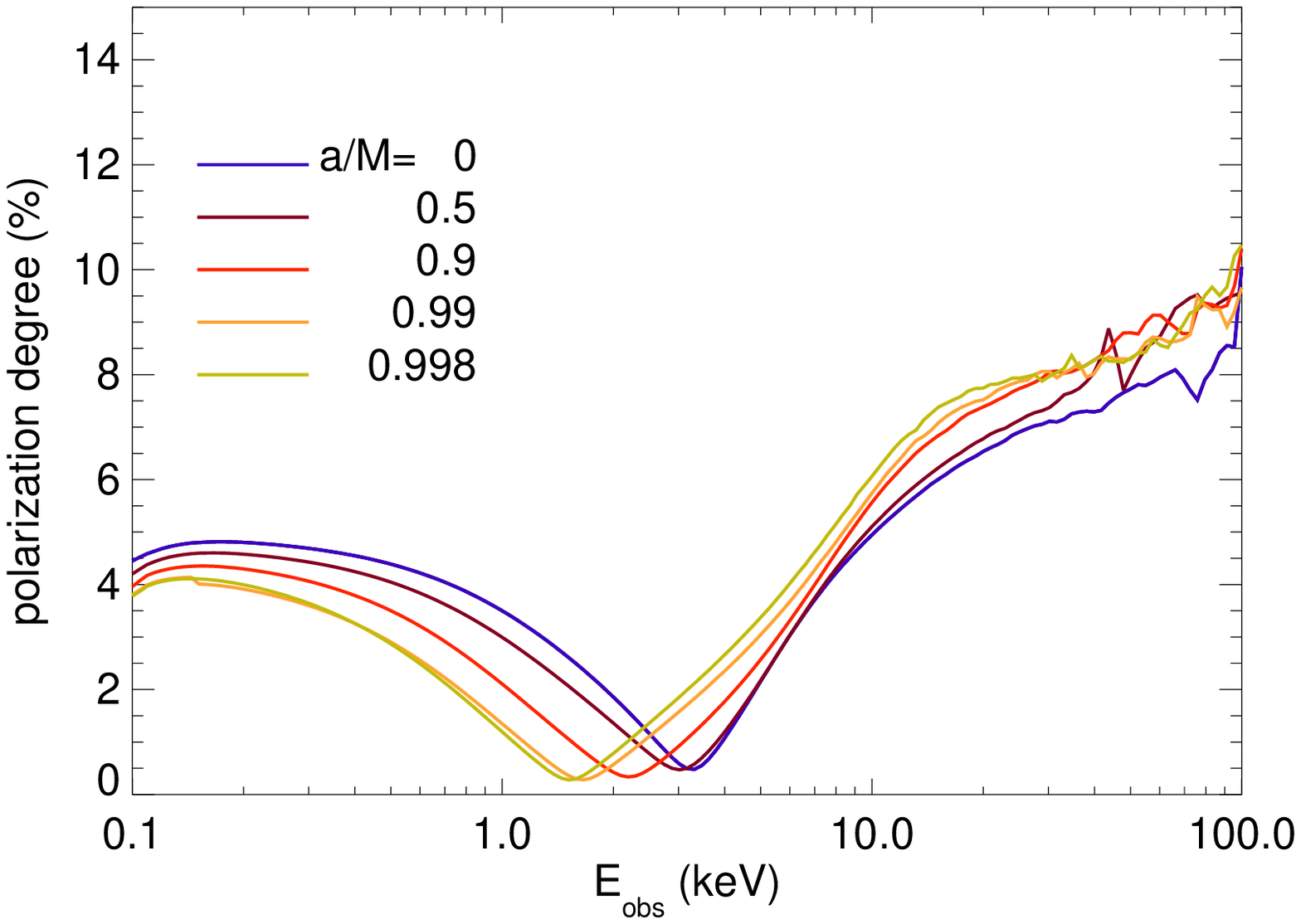}}\\
\scalebox{0.8}{\includegraphics{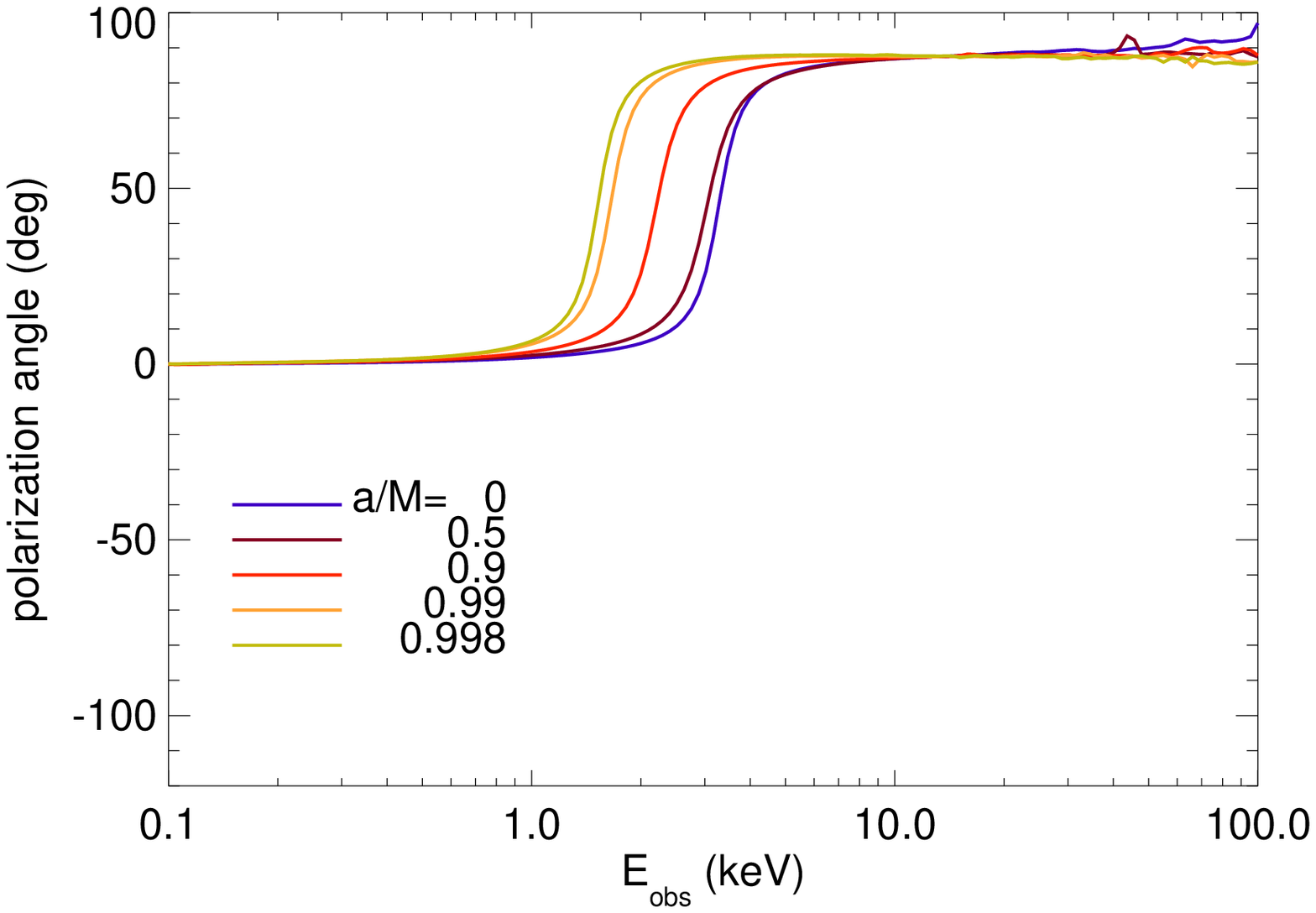}}
\end{center}
\end{figure}

\begin{figure}
\caption{\label{wedge_tauT} Degree and angle of polarization for a
  sandwich corona, varying the optical depth and electron
  temperature, maintaining a roughly constant Compton-$y$ parameter.}
\begin{center}
\scalebox{0.8}{\includegraphics{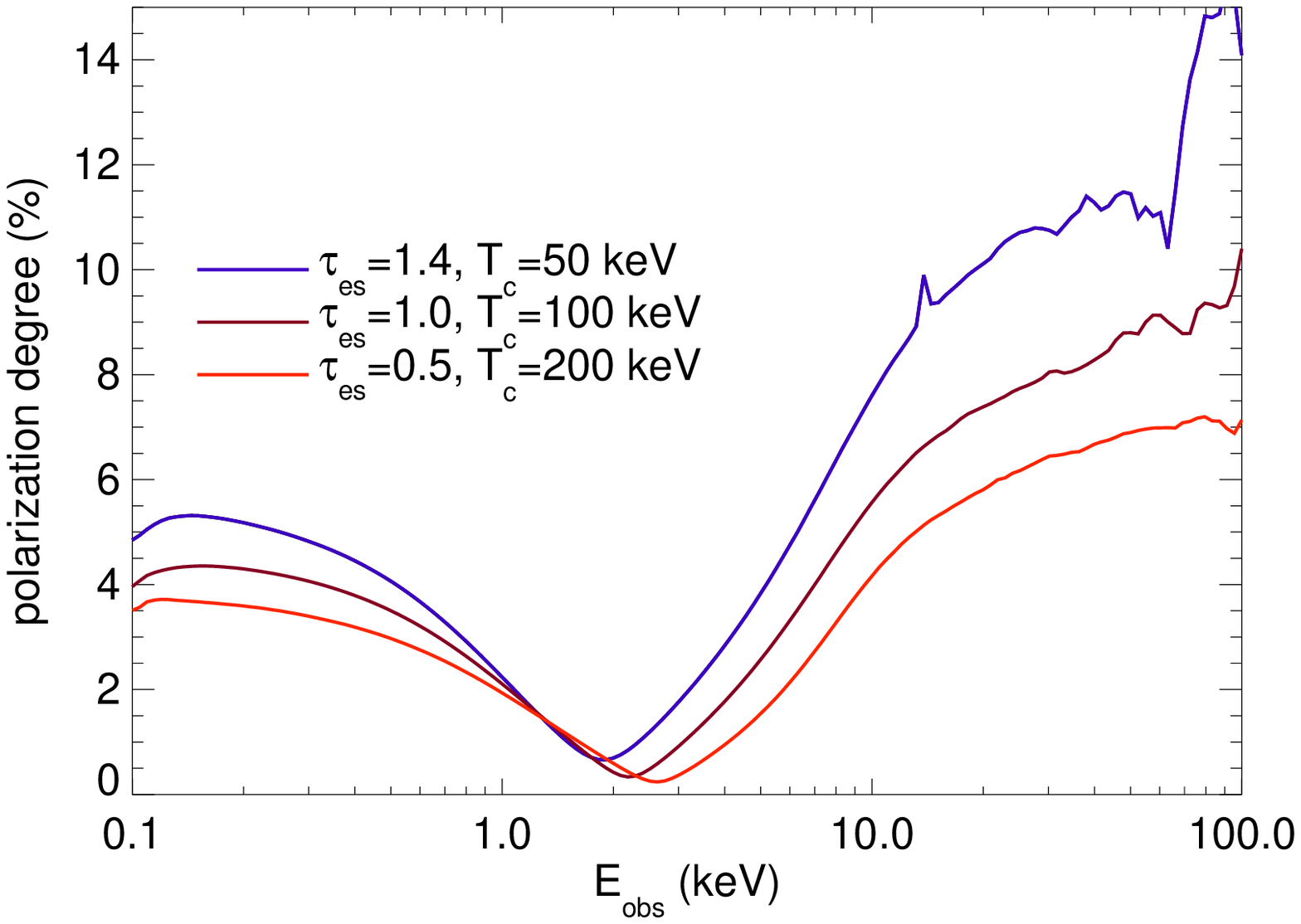}}\\
\scalebox{0.8}{\includegraphics{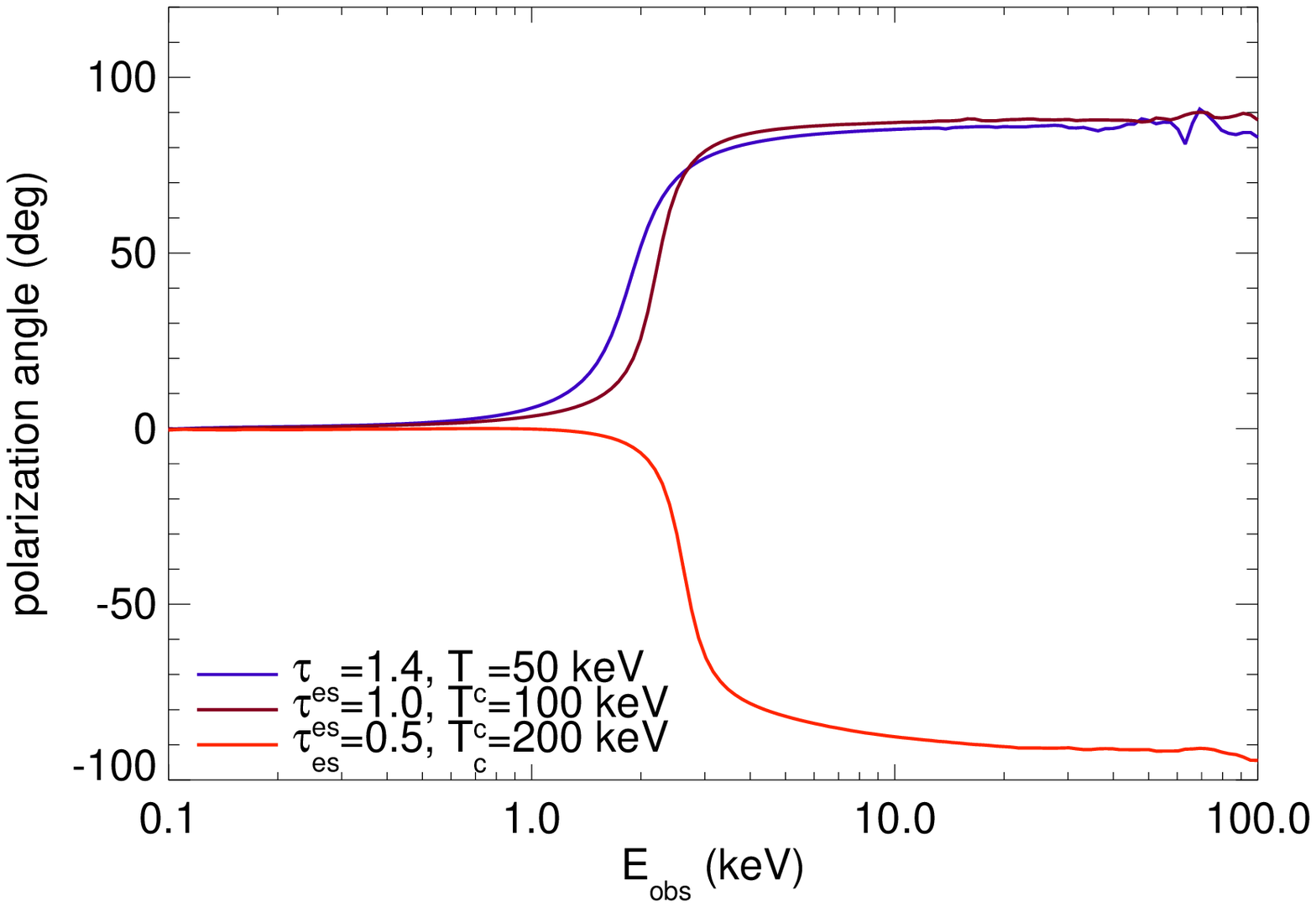}}
\end{center}
\end{figure}

As with any theoretical model based on a large number of
parameters, we expect that certain physical properties of the BH
system will be more easily
constrained than others. For example, even if we assume prior
knowledge of the observer inclination, BH mass and accretion rate, as
well as the temperature, density, and geometry of the corona, Figure
\ref{wedge_spin} suggests that, with a polarization
sensitivity of $\delta \lesssim 1\%$, one could
only just distinguish between a Schwarzschild and extreme Kerr BH in
the hard state. However, this also means that even if we have relatively
poor constraints on the spin parameter (which is generally the case
for most galactic BHs), we should still be able to measure other
parameters robustly. By measuring the power-law slope of the spectrum
around 10 keV, one could estimate the Compton $y$-parameter with
reasonable accuracy, but still face a degeneracy between the corona
temperature and optical depth. Again assuming a known inclination and
corona geometry, we see from Figure \ref{wedge_tauT} that this
degeneracy could be broken with a polarization observation.

\begin{figure}
\caption{\label{wedge_grid} Observed flux and polarization from an
  accretion disk with a sandwich corona geometry. The plots show the
  flux (left; arbitrary units of $\nu F_\nu$), polarization degree (center), and polarization angle
  (right) as a function of observed energy, for inclinations of
  $45^\circ$, $60^\circ$, and $75^\circ$ (top, center, bottom,
  respectively). The dotted lines represent contributions directly
  from the thermal disk, the dot-dashed curves are corona-scattered
  photons, and the solid curves are the total observed flux. }
\begin{center}
\scalebox{0.3}{\includegraphics{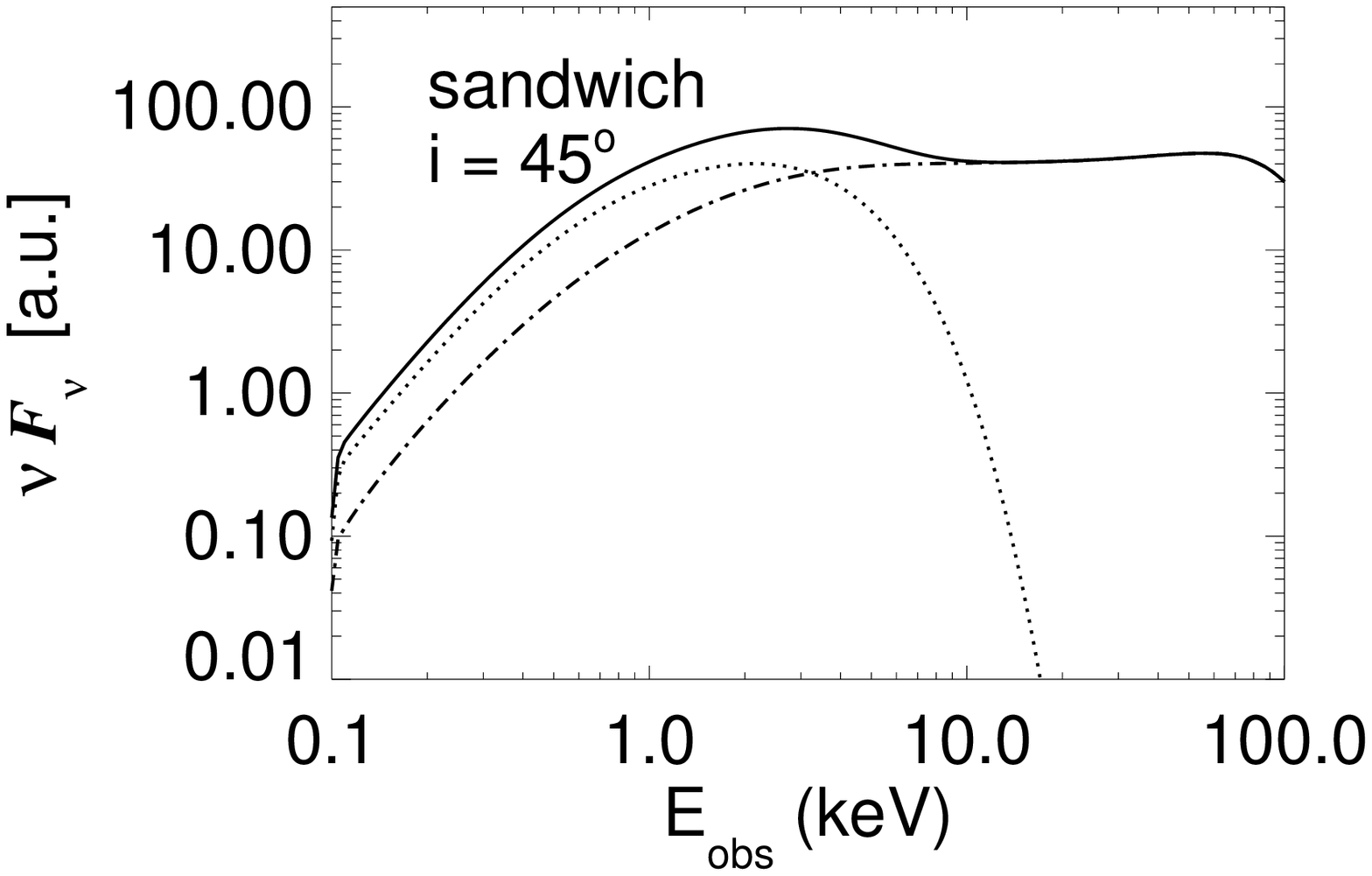}}
\scalebox{0.3}{\includegraphics{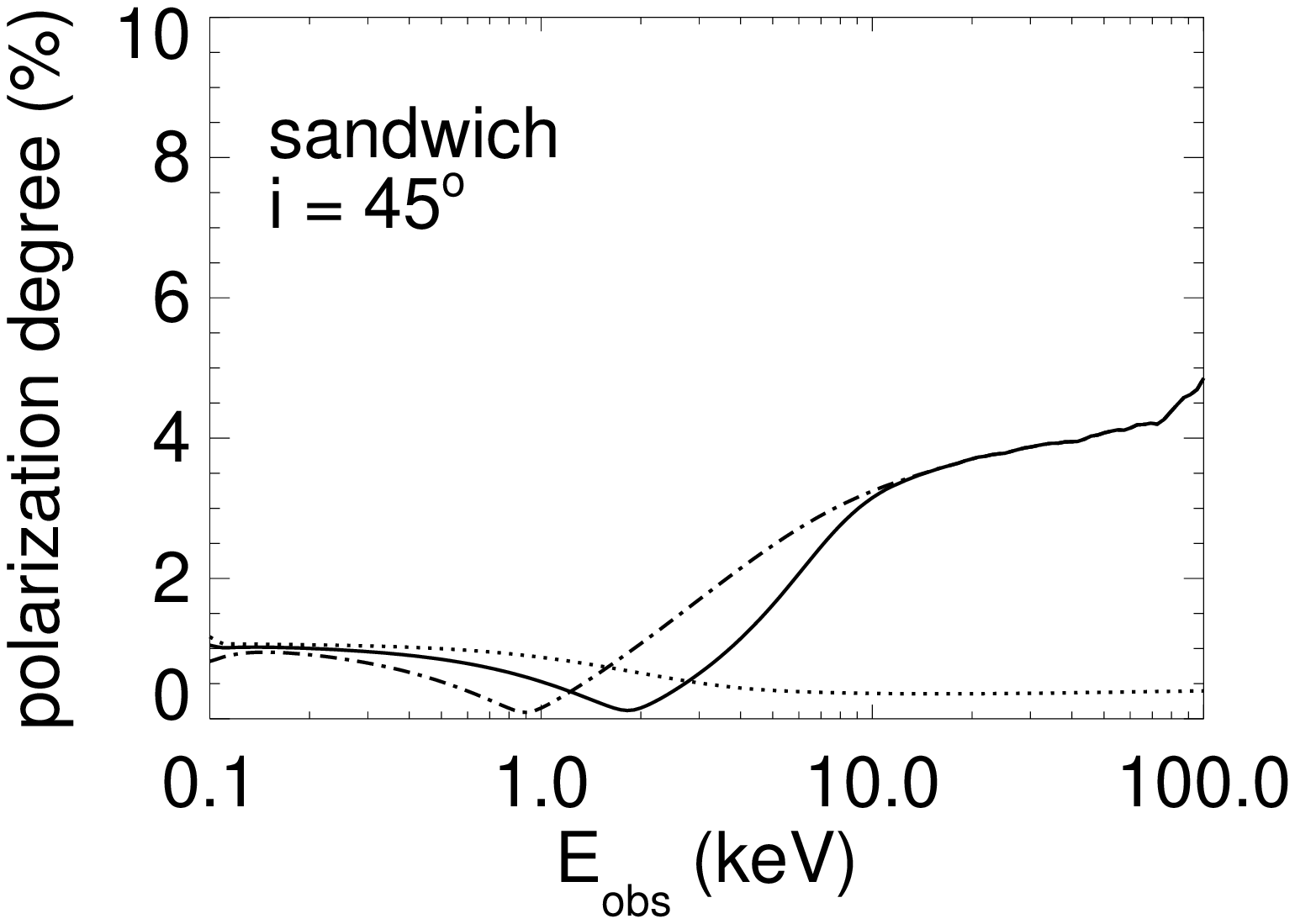}}
\scalebox{0.3}{\includegraphics{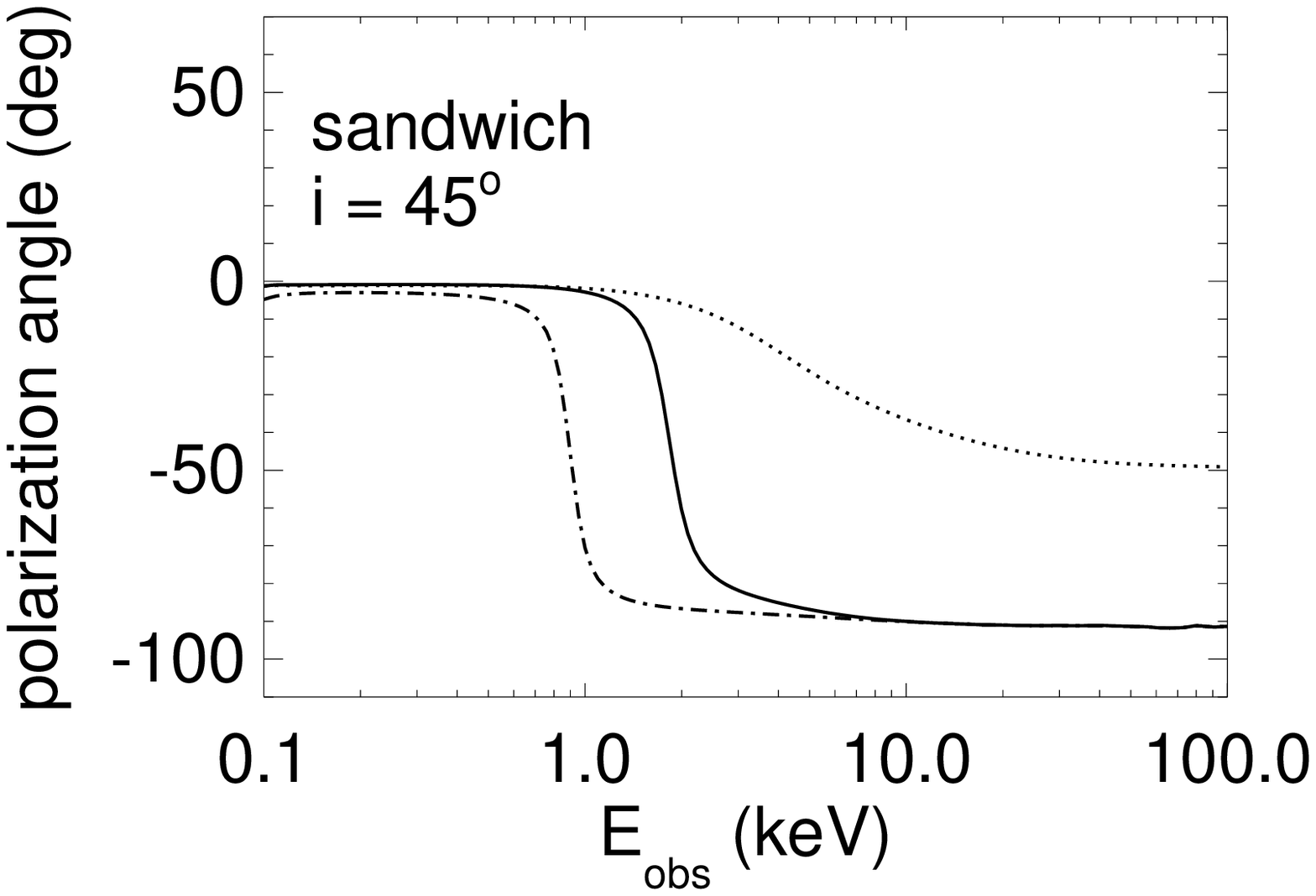}}\\
\scalebox{0.3}{\includegraphics{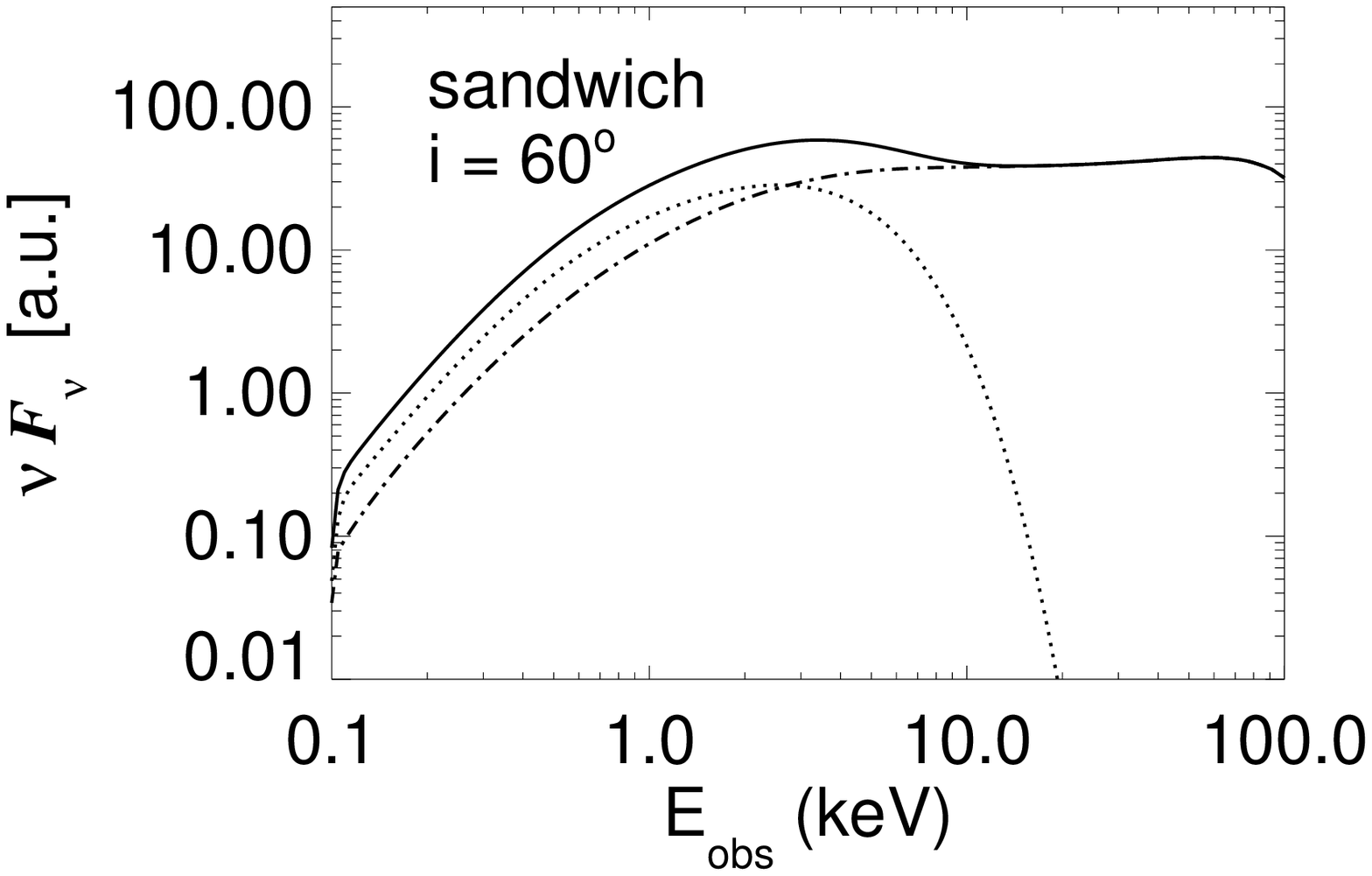}}
\scalebox{0.3}{\includegraphics{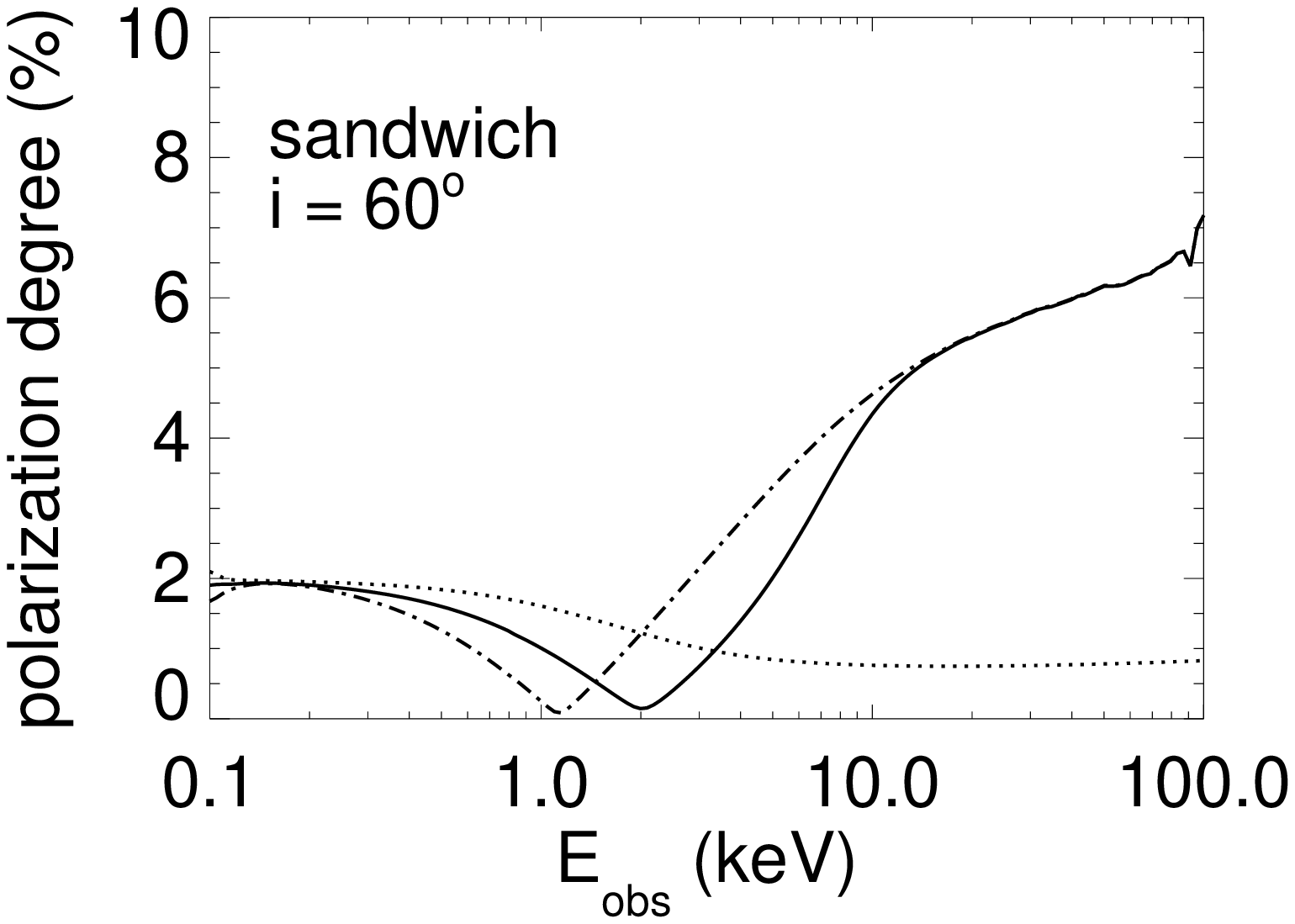}}
\scalebox{0.3}{\includegraphics{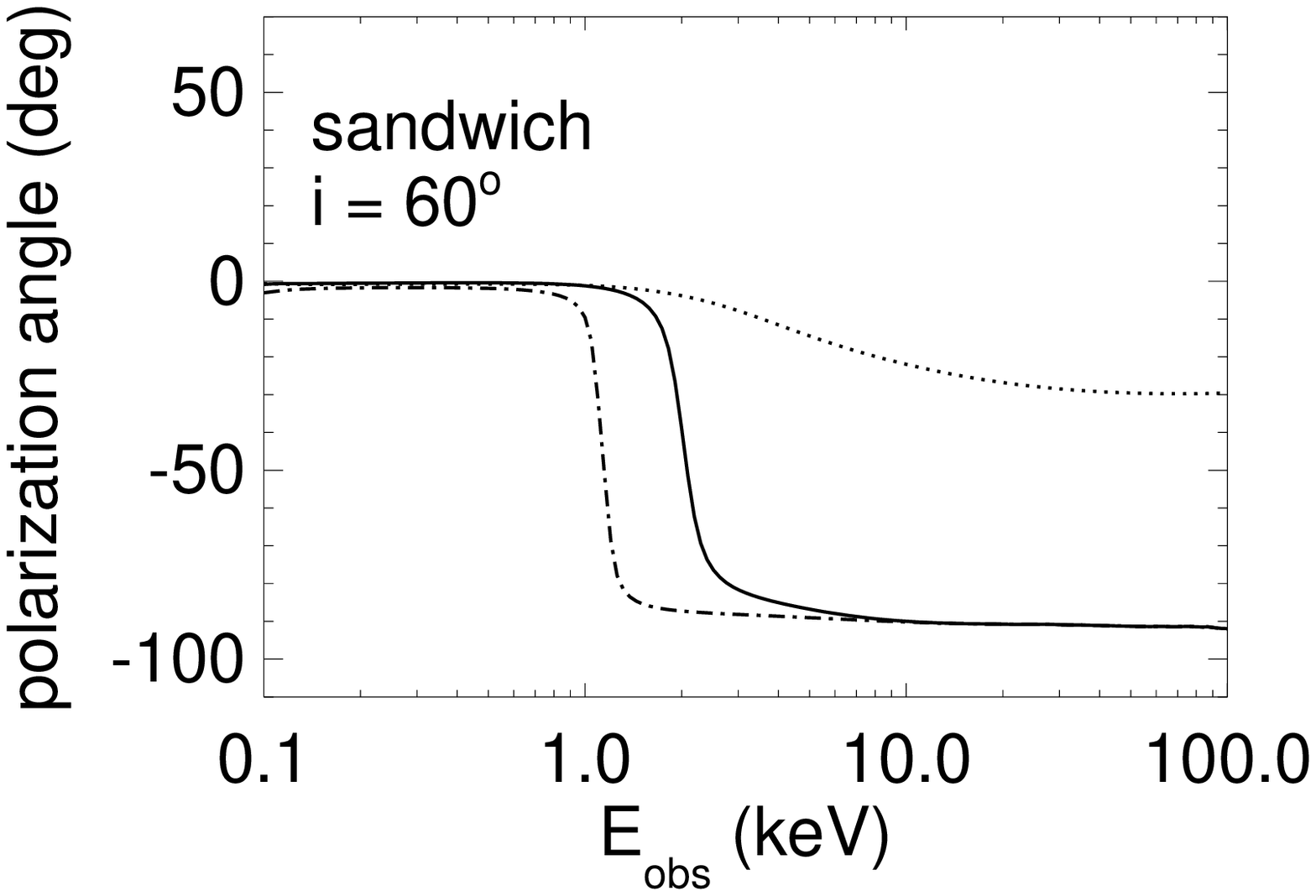}}\\
\scalebox{0.3}{\includegraphics{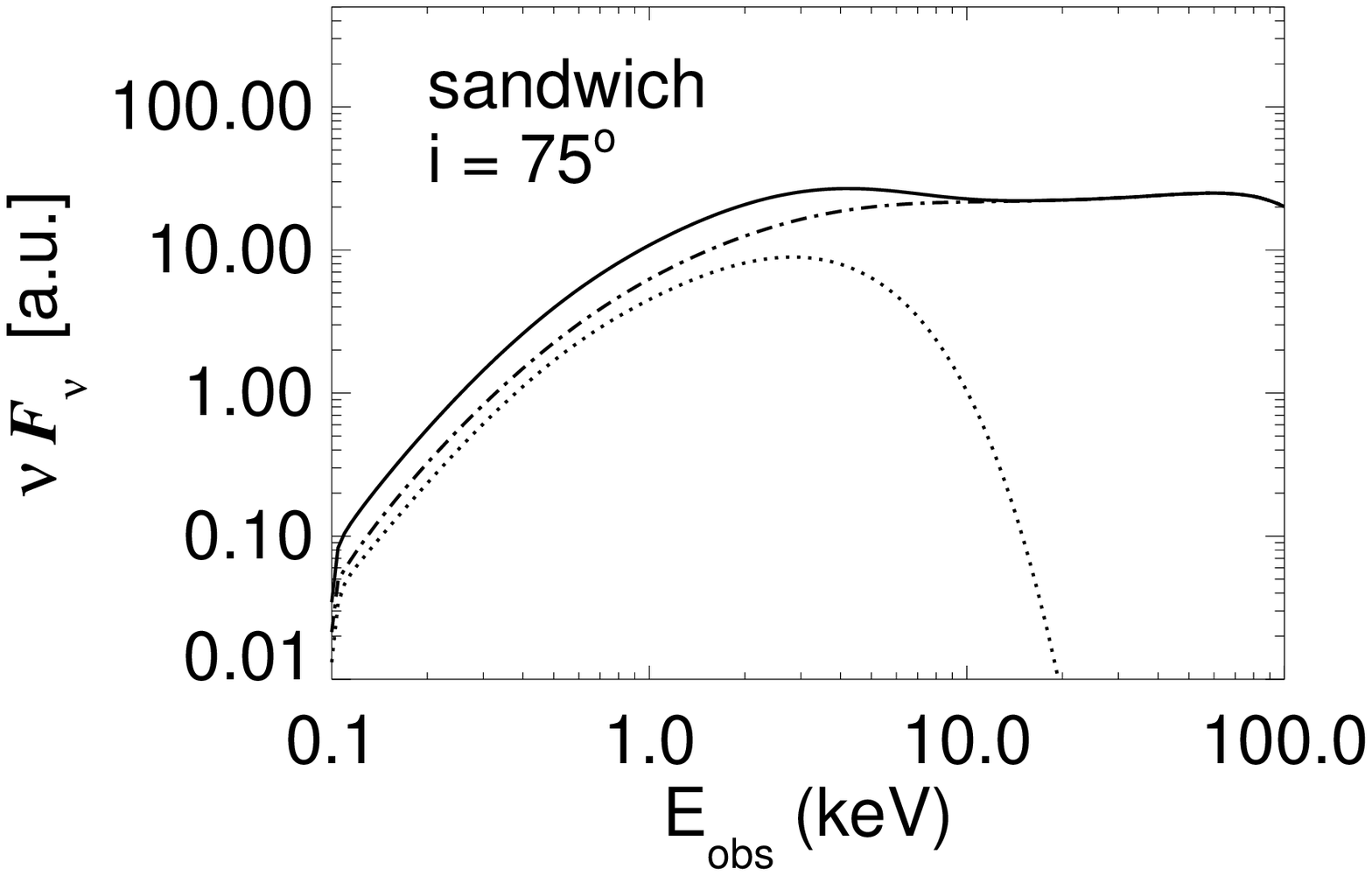}}
\scalebox{0.3}{\includegraphics{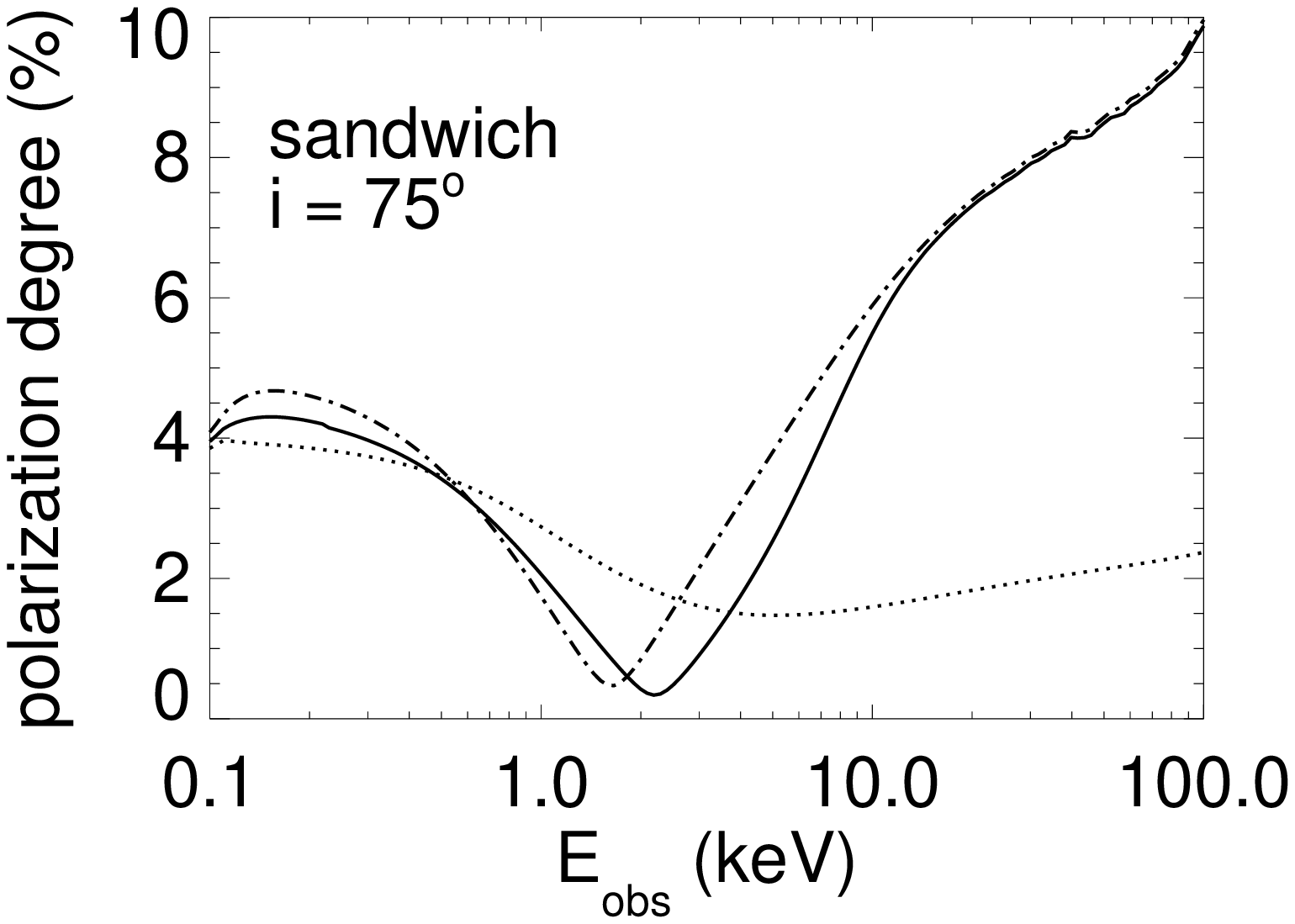}}
\scalebox{0.3}{\includegraphics{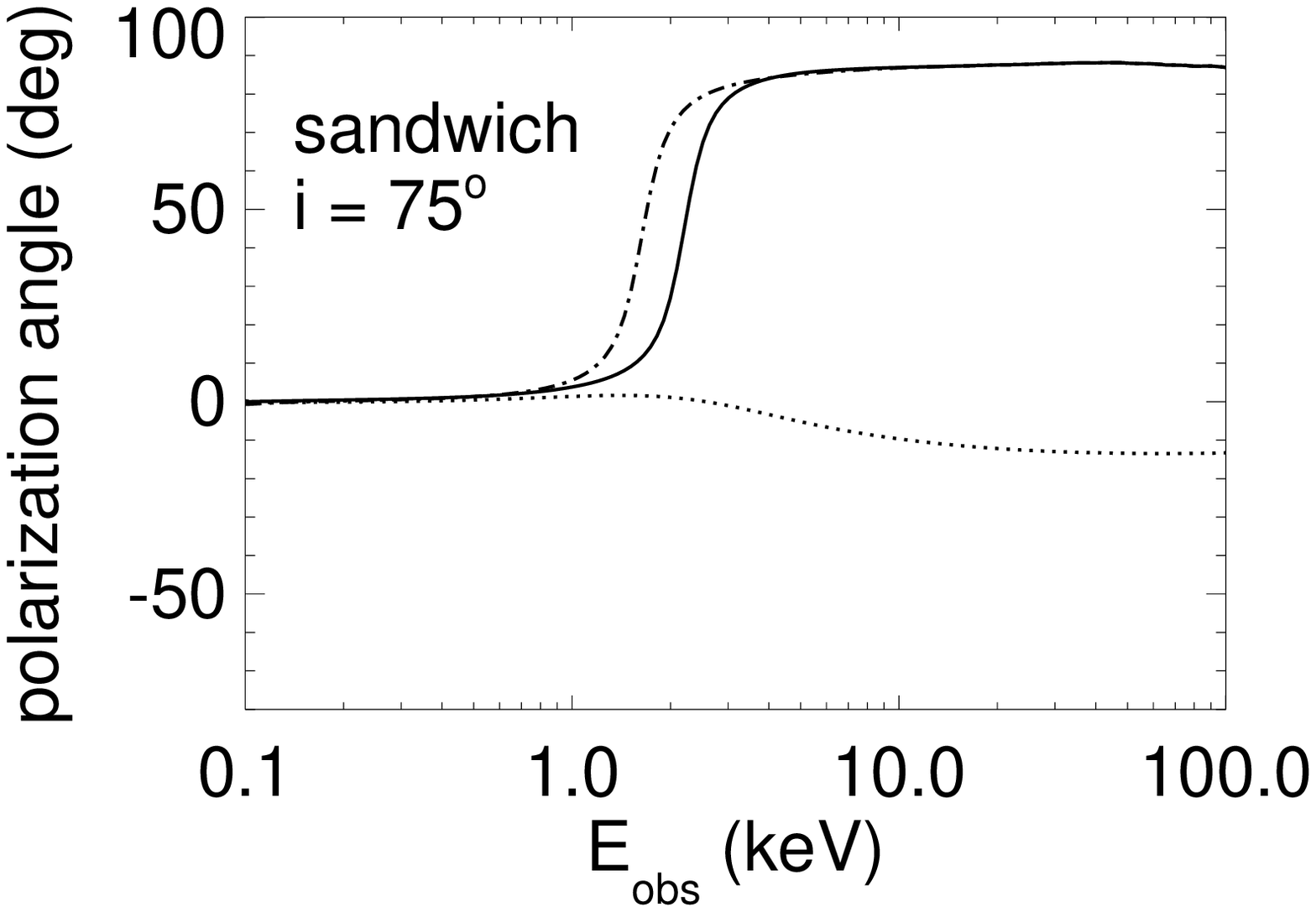}}
\end{center}
\end{figure}

\begin{figure}
\caption{\label{hotspot_grid} Same as Figure \ref{wedge_grid}, but for the
  hot spot corona geometry shown in Figure \ref{image_hotspot}.}
\begin{center}
\scalebox{0.3}{\includegraphics{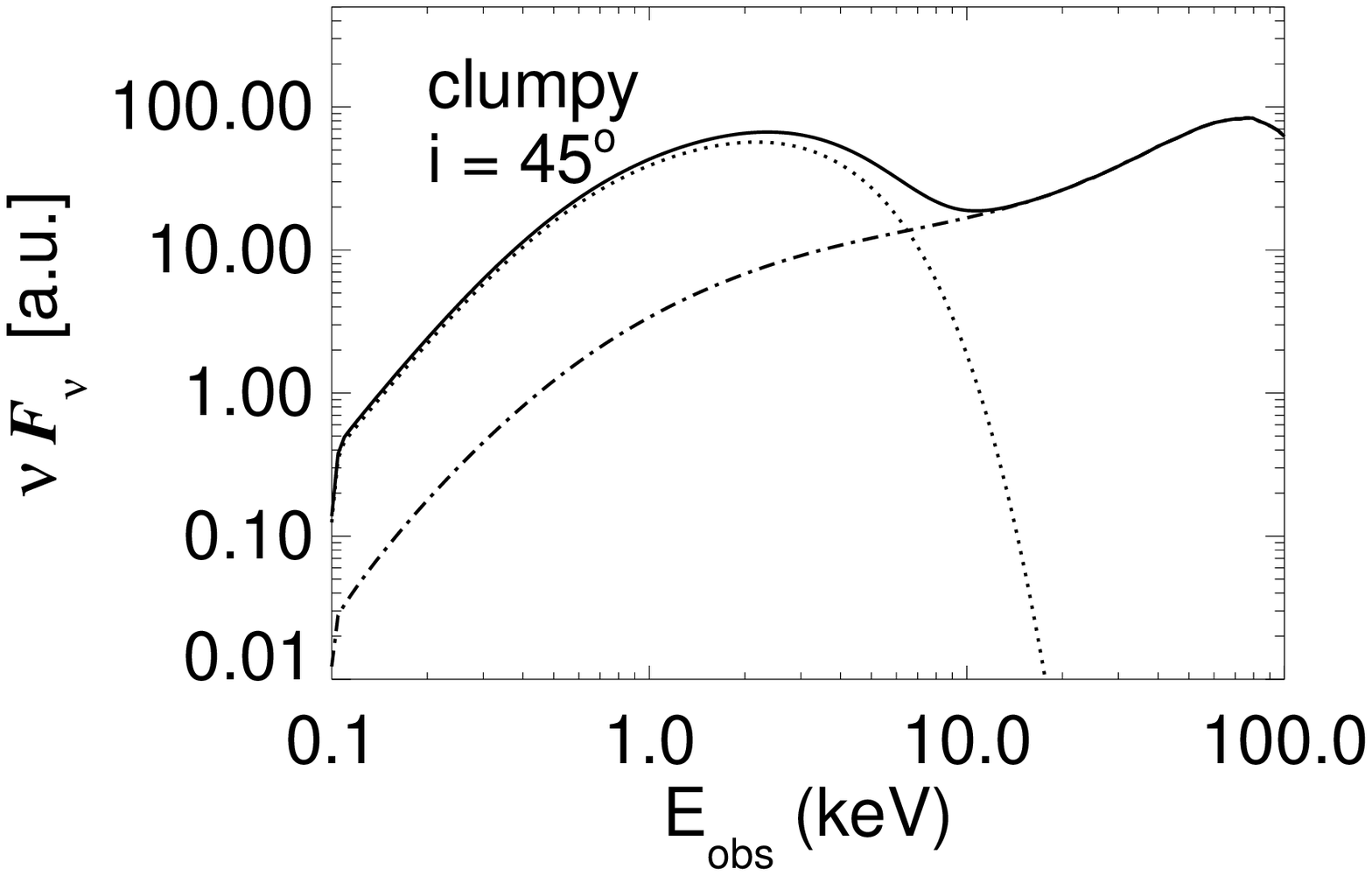}}
\scalebox{0.3}{\includegraphics{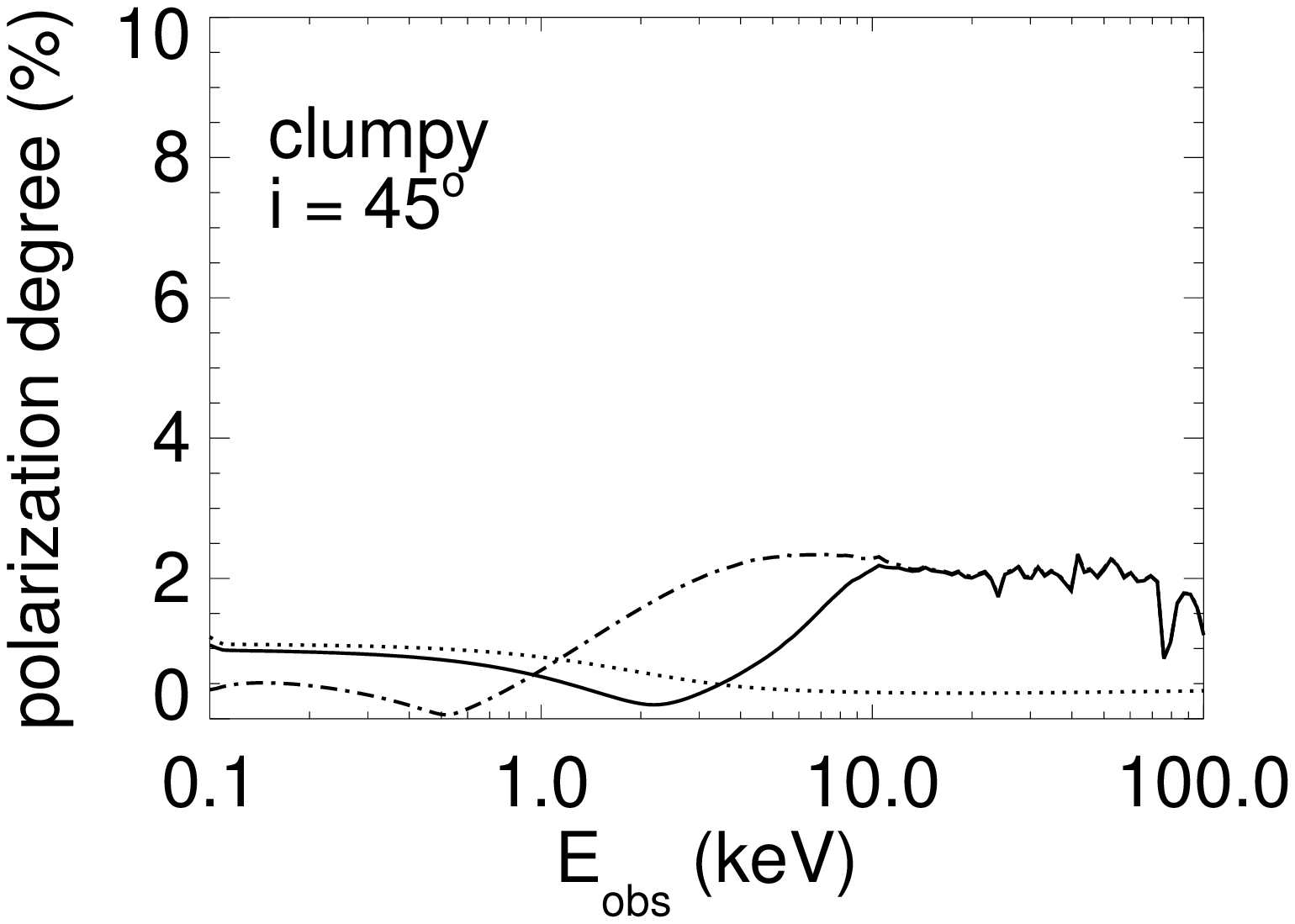}}
\scalebox{0.3}{\includegraphics{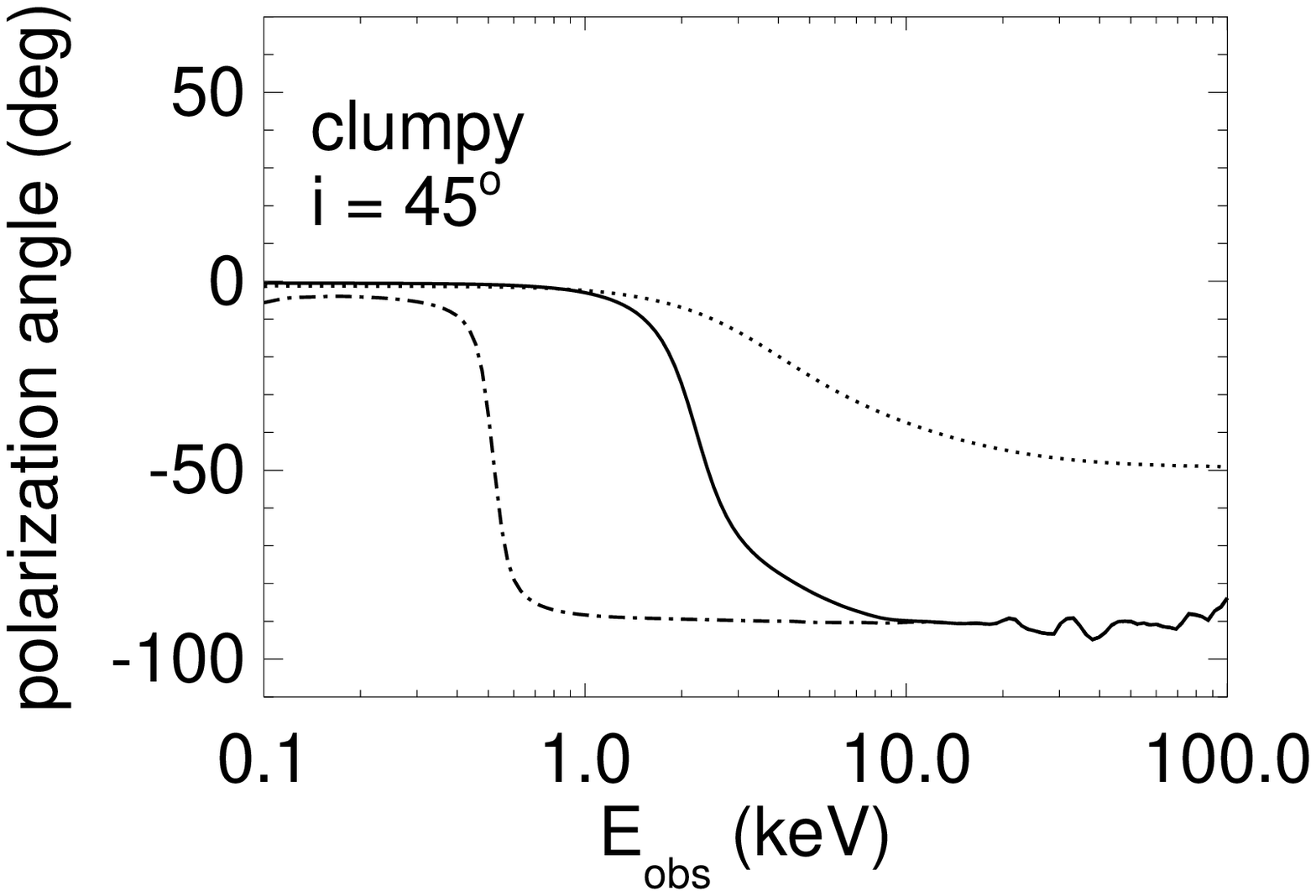}}\\
\scalebox{0.3}{\includegraphics{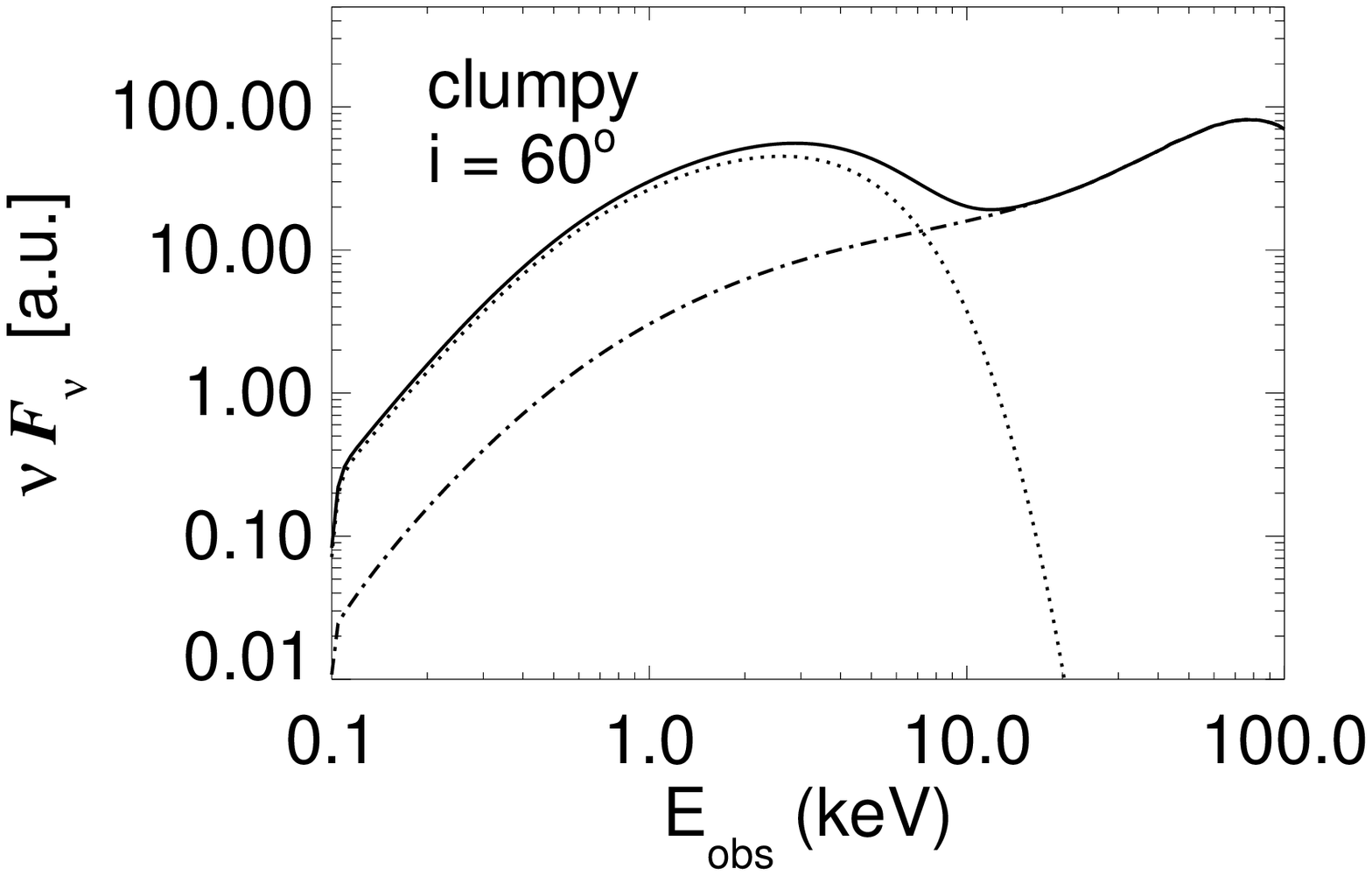}}
\scalebox{0.3}{\includegraphics{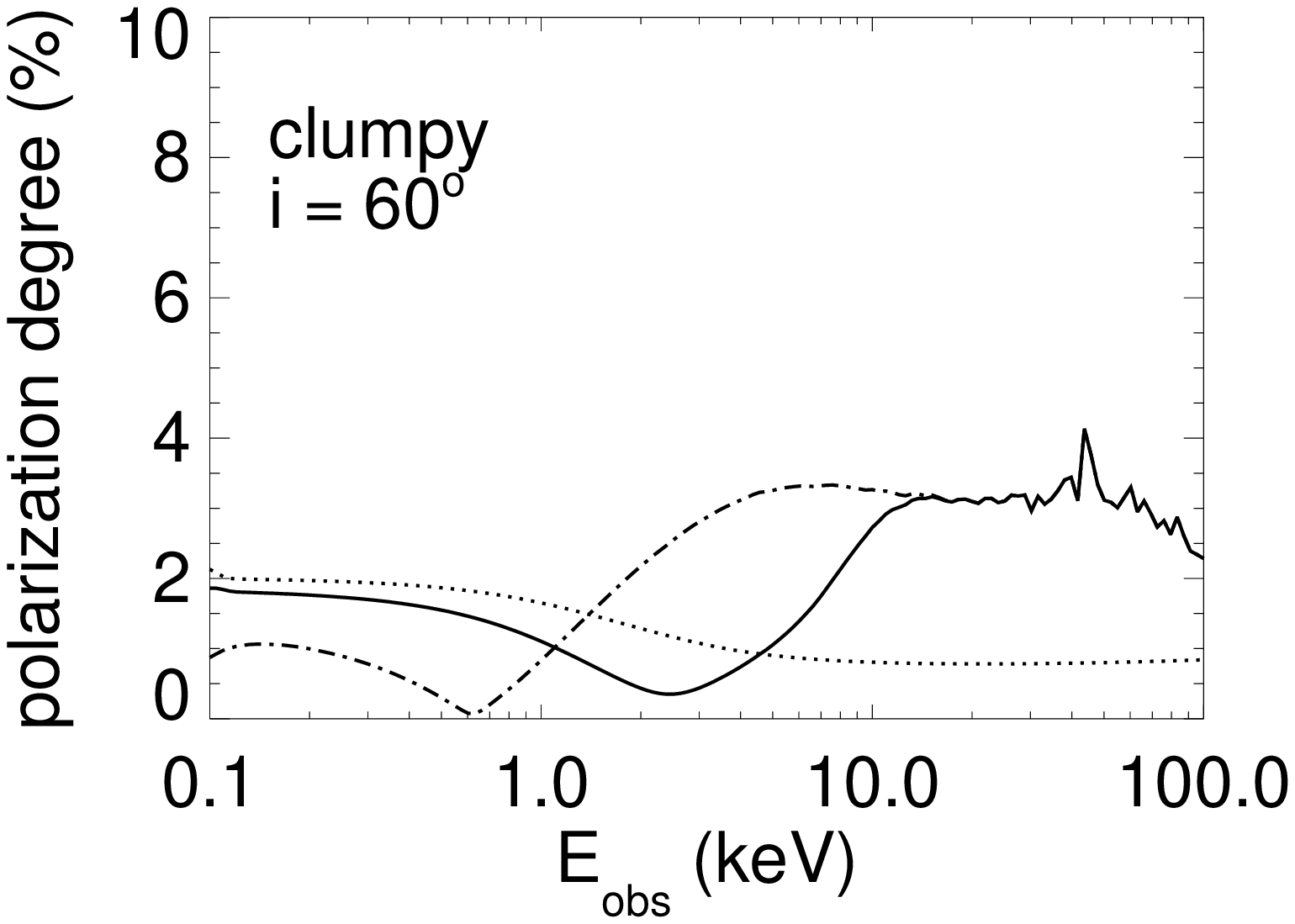}}
\scalebox{0.3}{\includegraphics{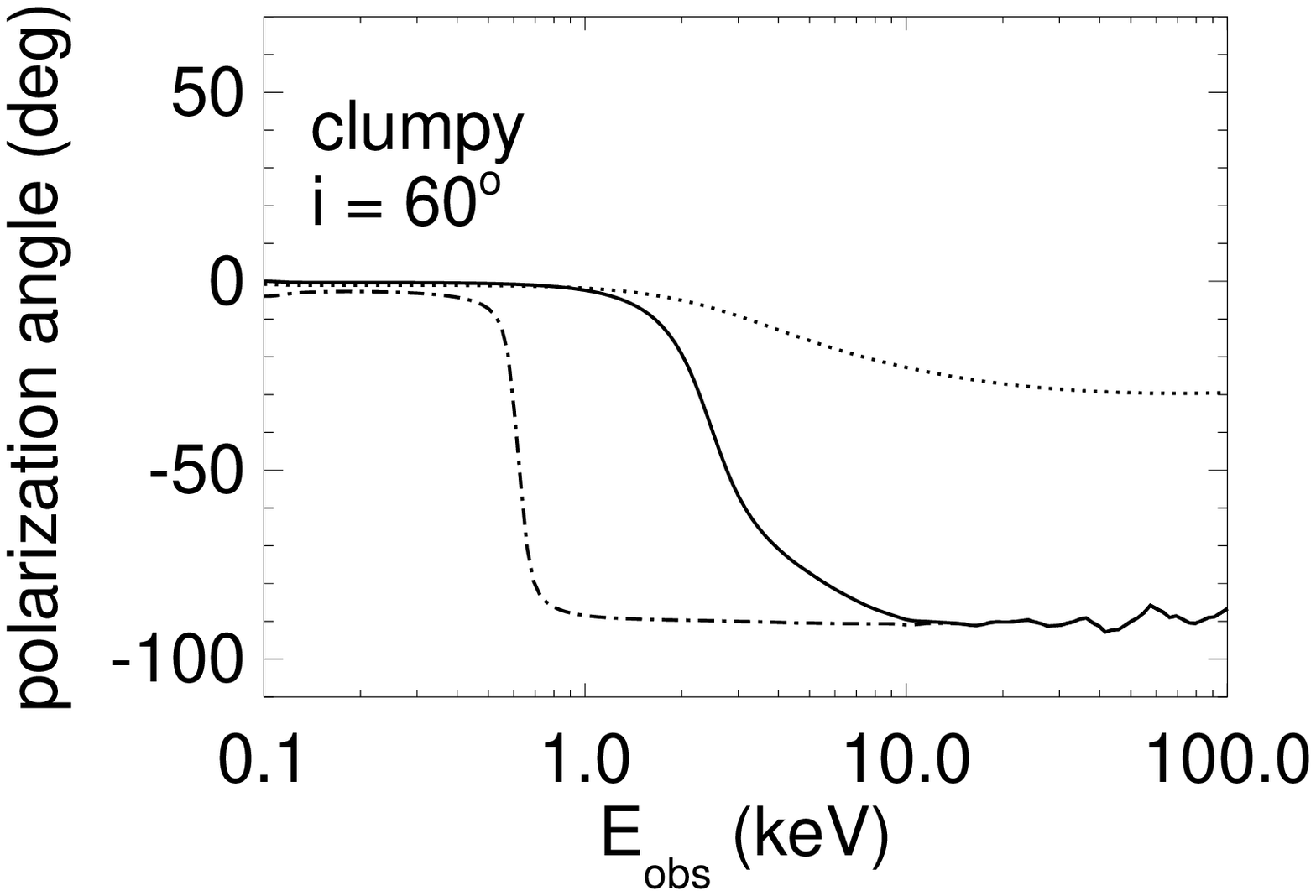}}\\
\scalebox{0.3}{\includegraphics{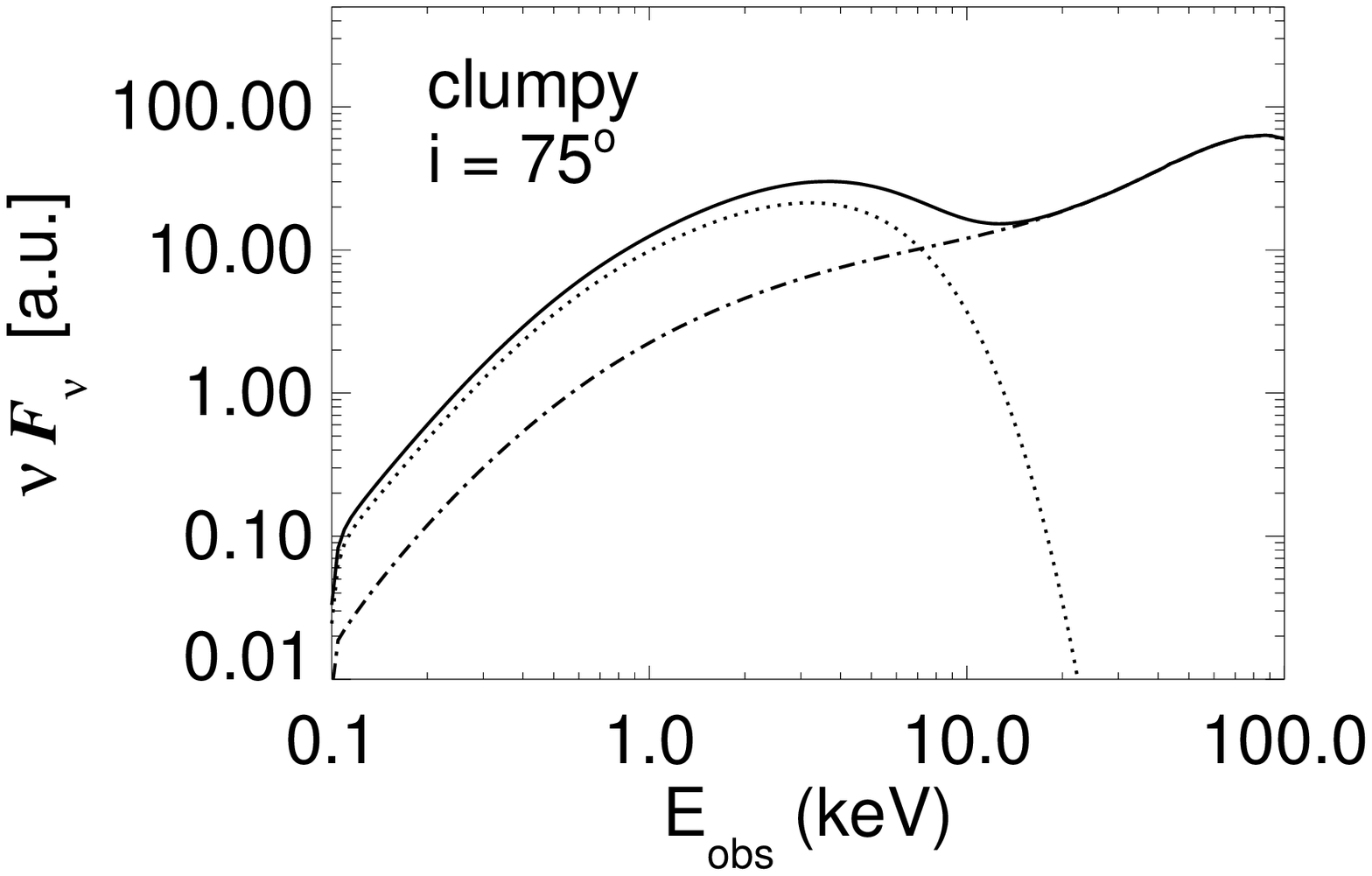}}
\scalebox{0.3}{\includegraphics{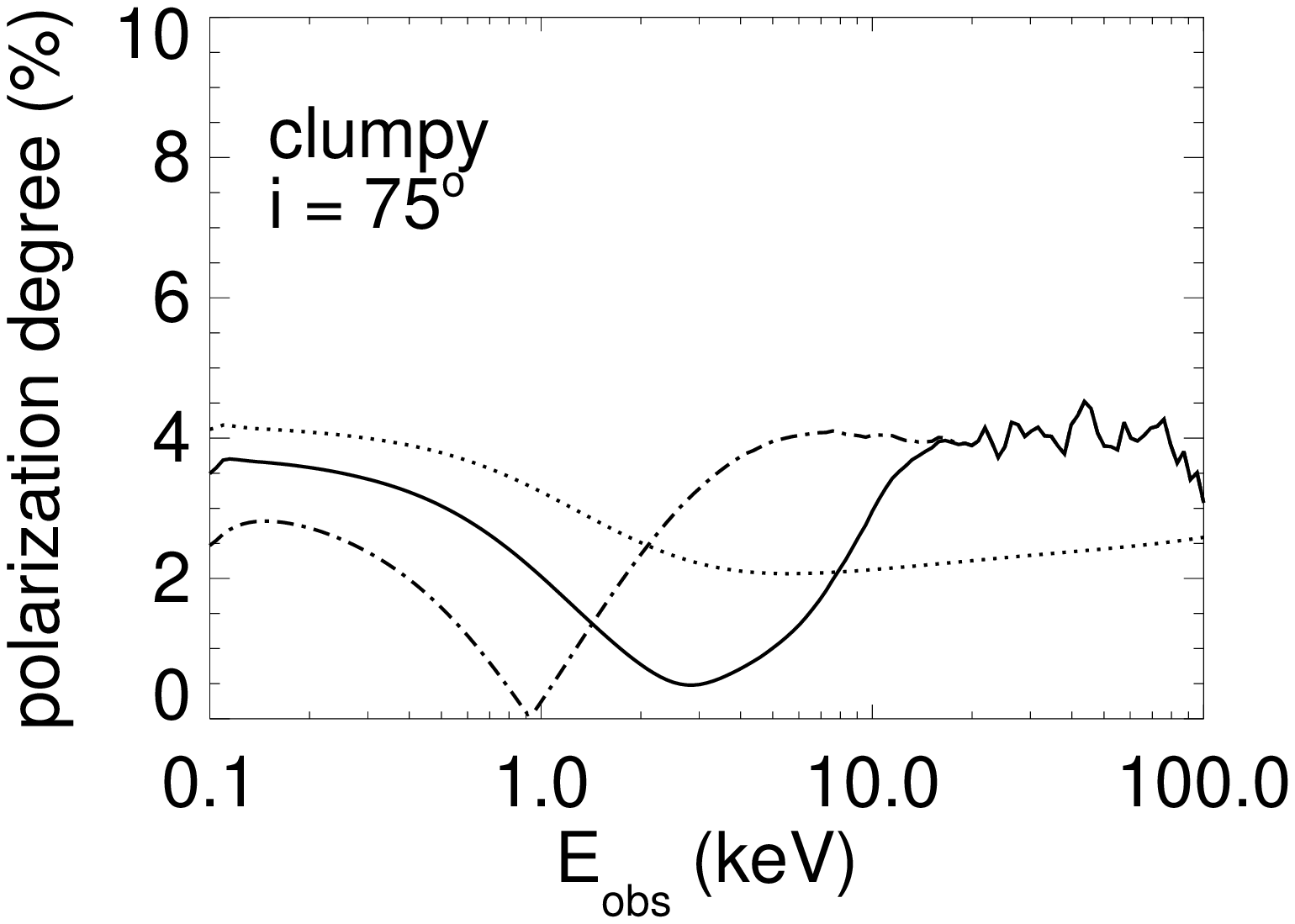}}
\scalebox{0.3}{\includegraphics{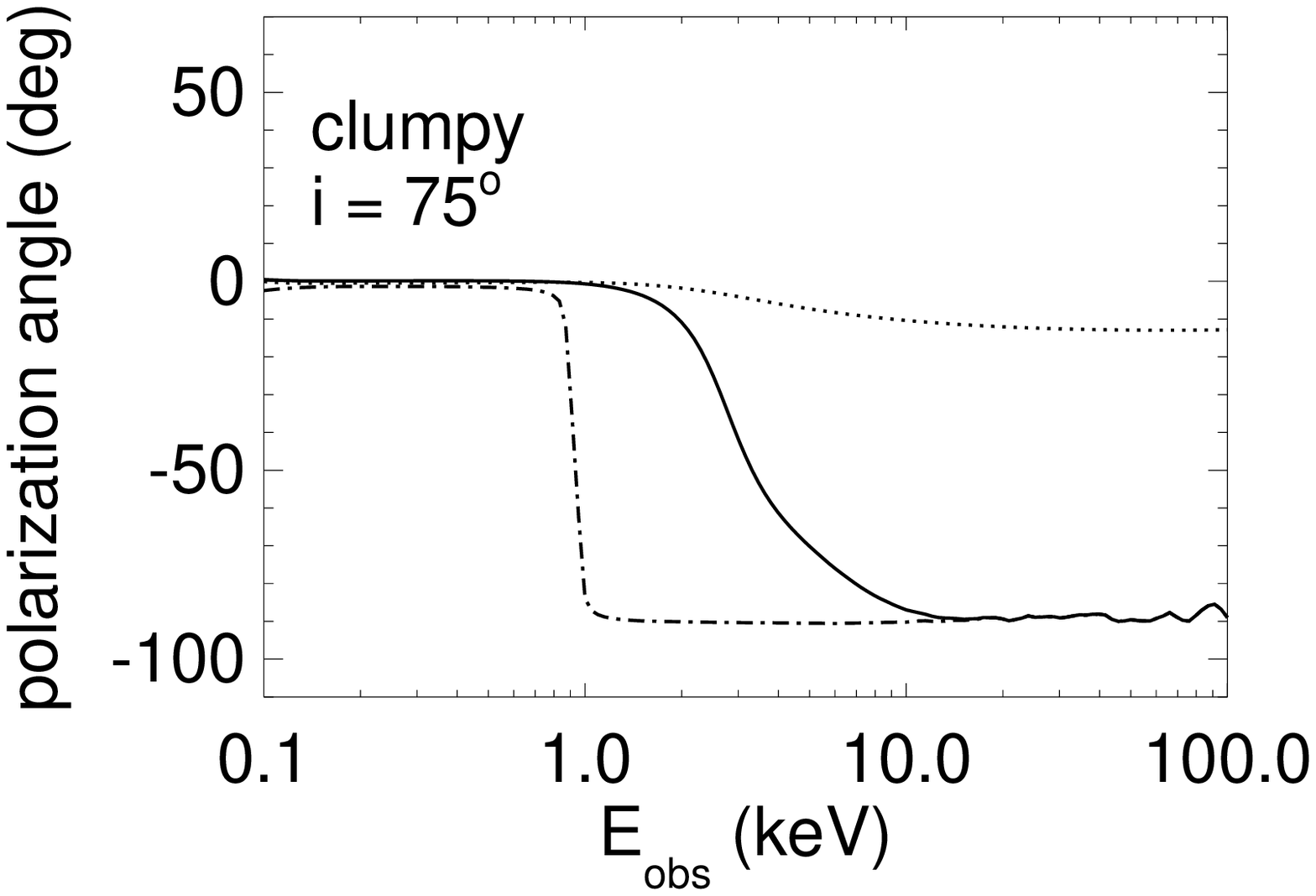}}
\end{center}
\end{figure}

\begin{figure}
\caption{\label{hotspot_rho} Degree and angle of polarization for a
  hot spot corona, varying the compactness of the coronal hot spots.}
\begin{center}
\scalebox{0.8}{\includegraphics{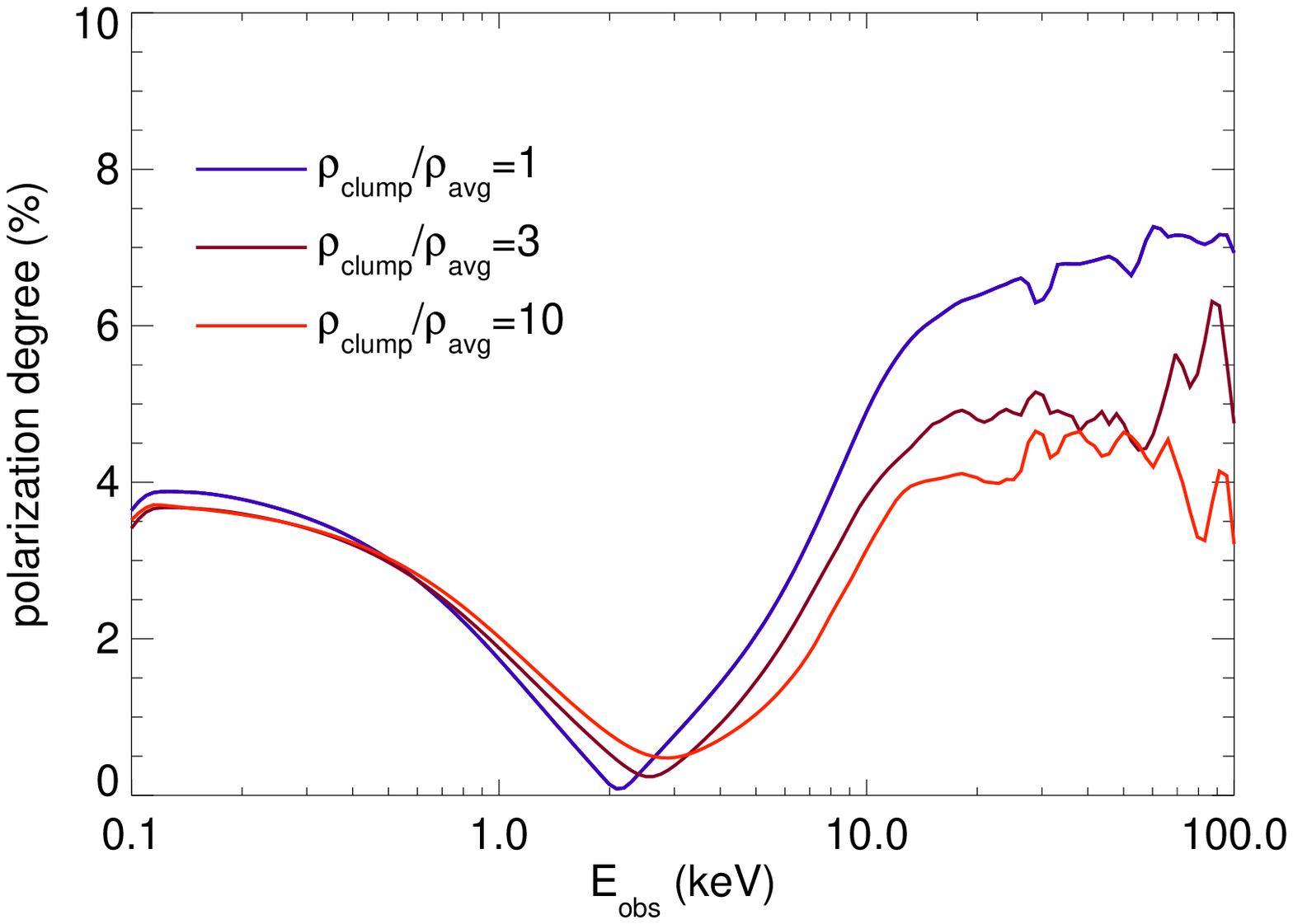}}\\
\scalebox{0.8}{\includegraphics{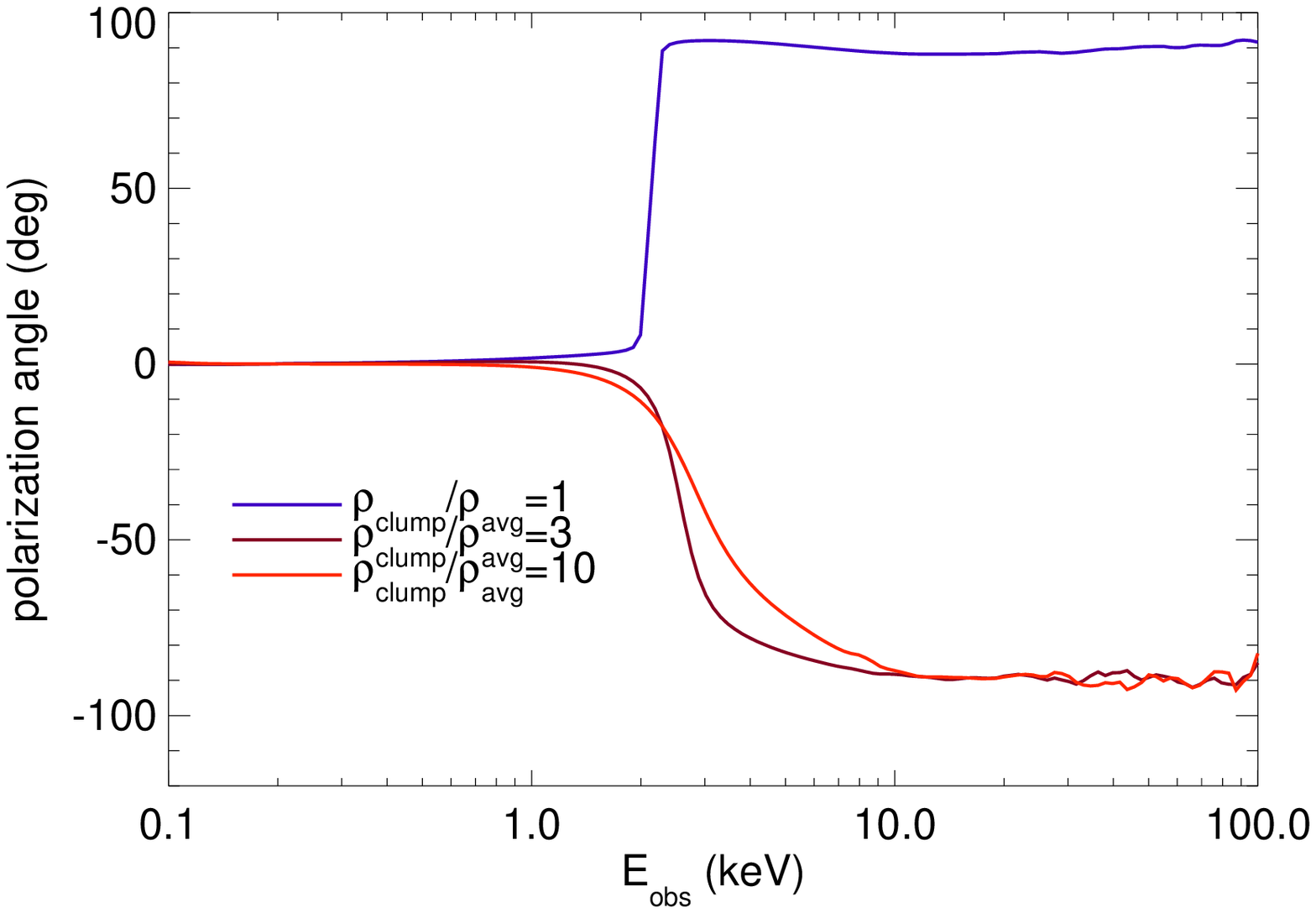}}
\end{center}
\end{figure}

\begin{figure}
\caption{\label{hotspot_Nc} Degree and angle of polarization for a
  hot spot corona, varying the number density of coronal hot spots.}
\begin{center}
\scalebox{0.8}{\includegraphics{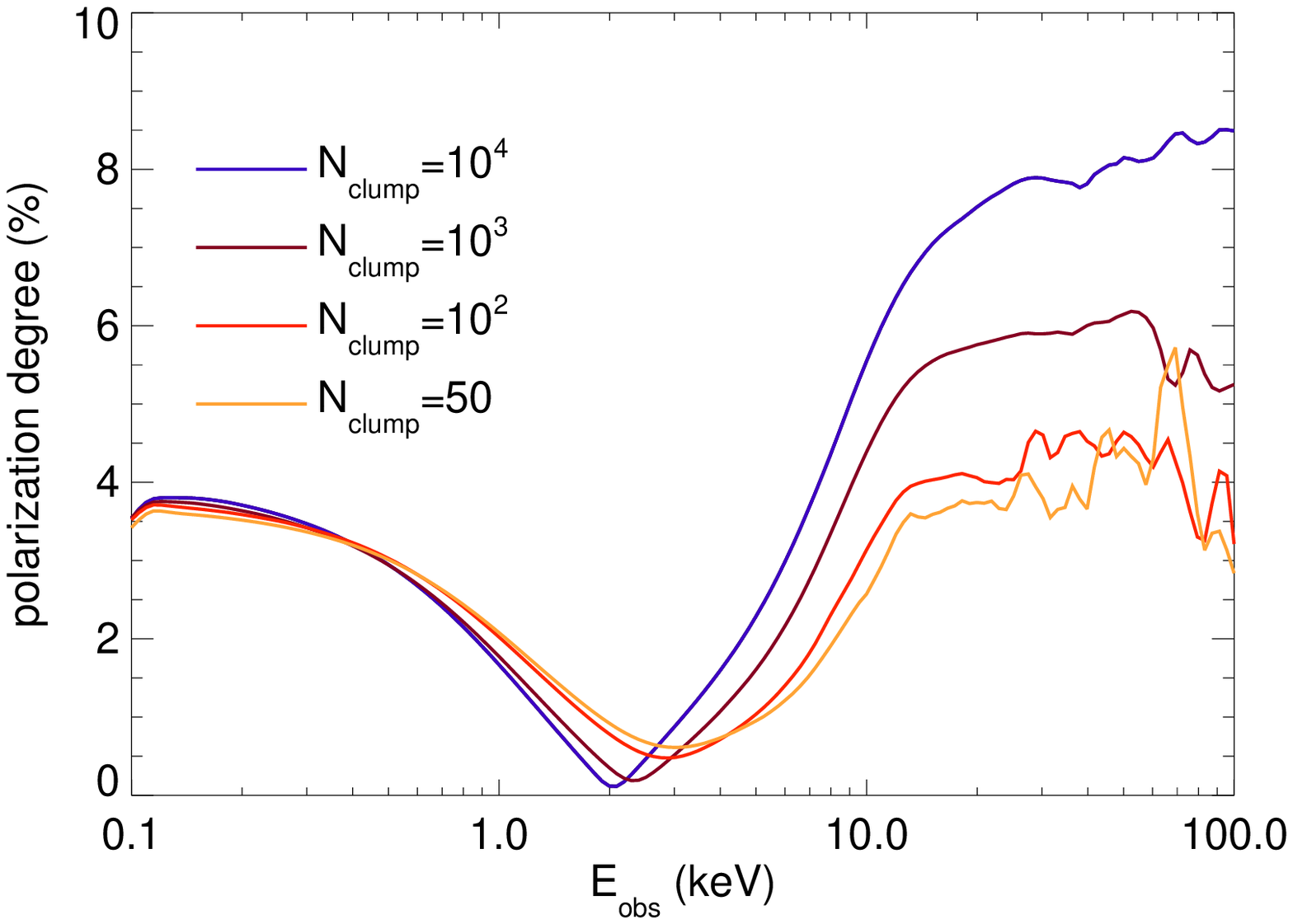}}\\
\scalebox{0.8}{\includegraphics{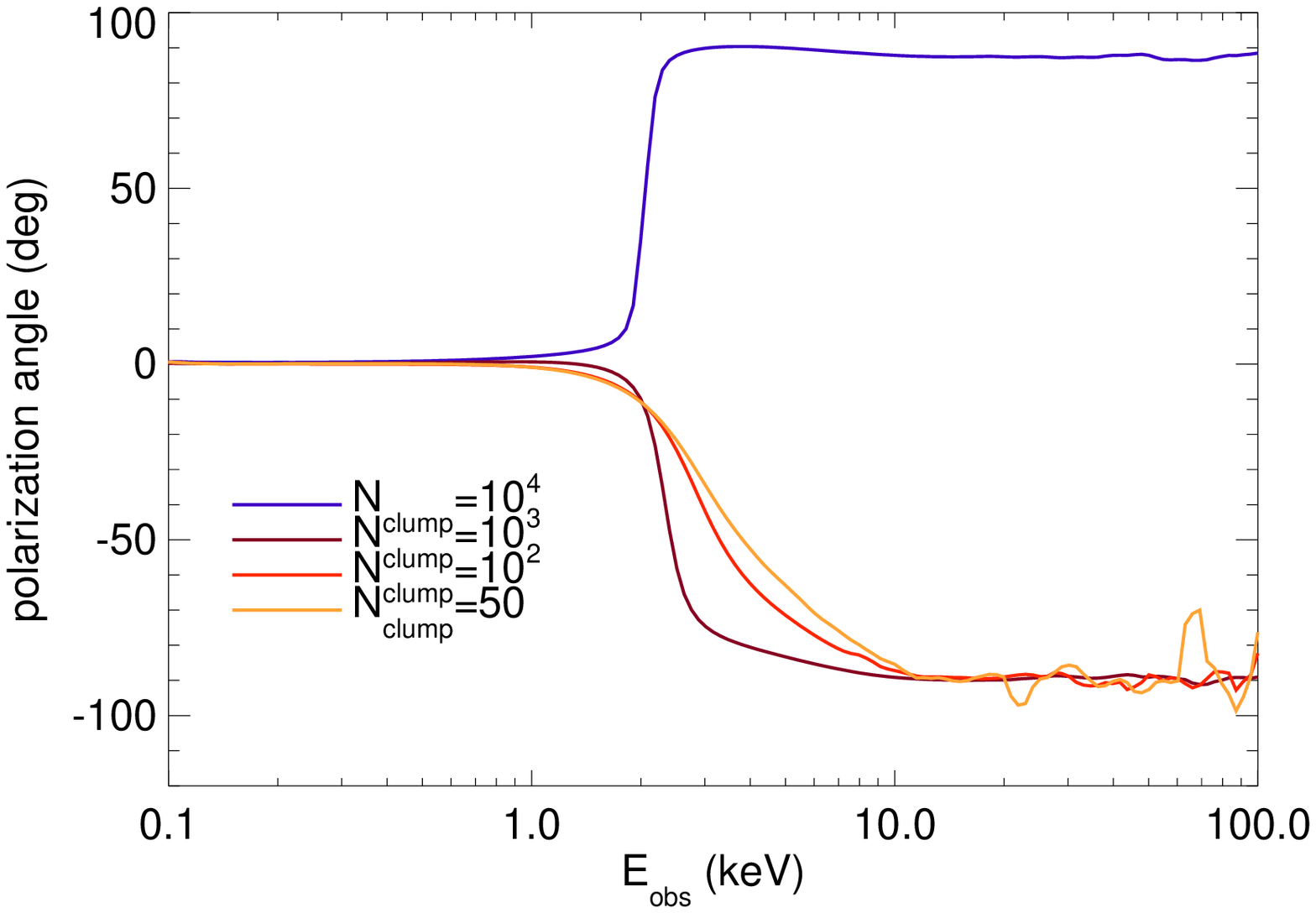}}
\end{center}
\end{figure}

\begin{figure}
\caption{\label{sphere_grid} Same as Figure \ref{wedge_grid}, but for the
  spherical corona geometry shown in Figure \ref{image_sphere}.}
\begin{center}
\scalebox{0.3}{\includegraphics{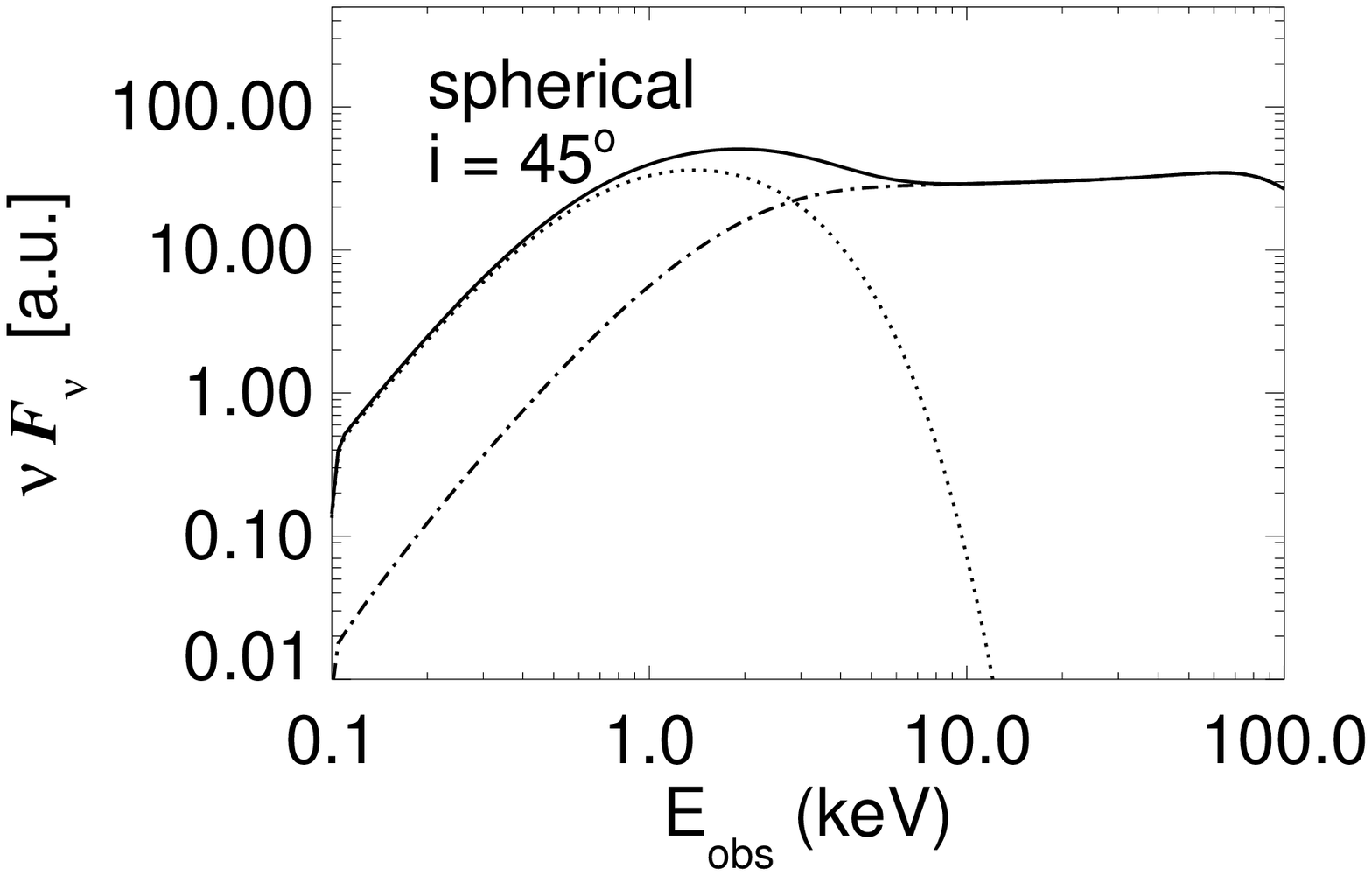}}
\scalebox{0.3}{\includegraphics{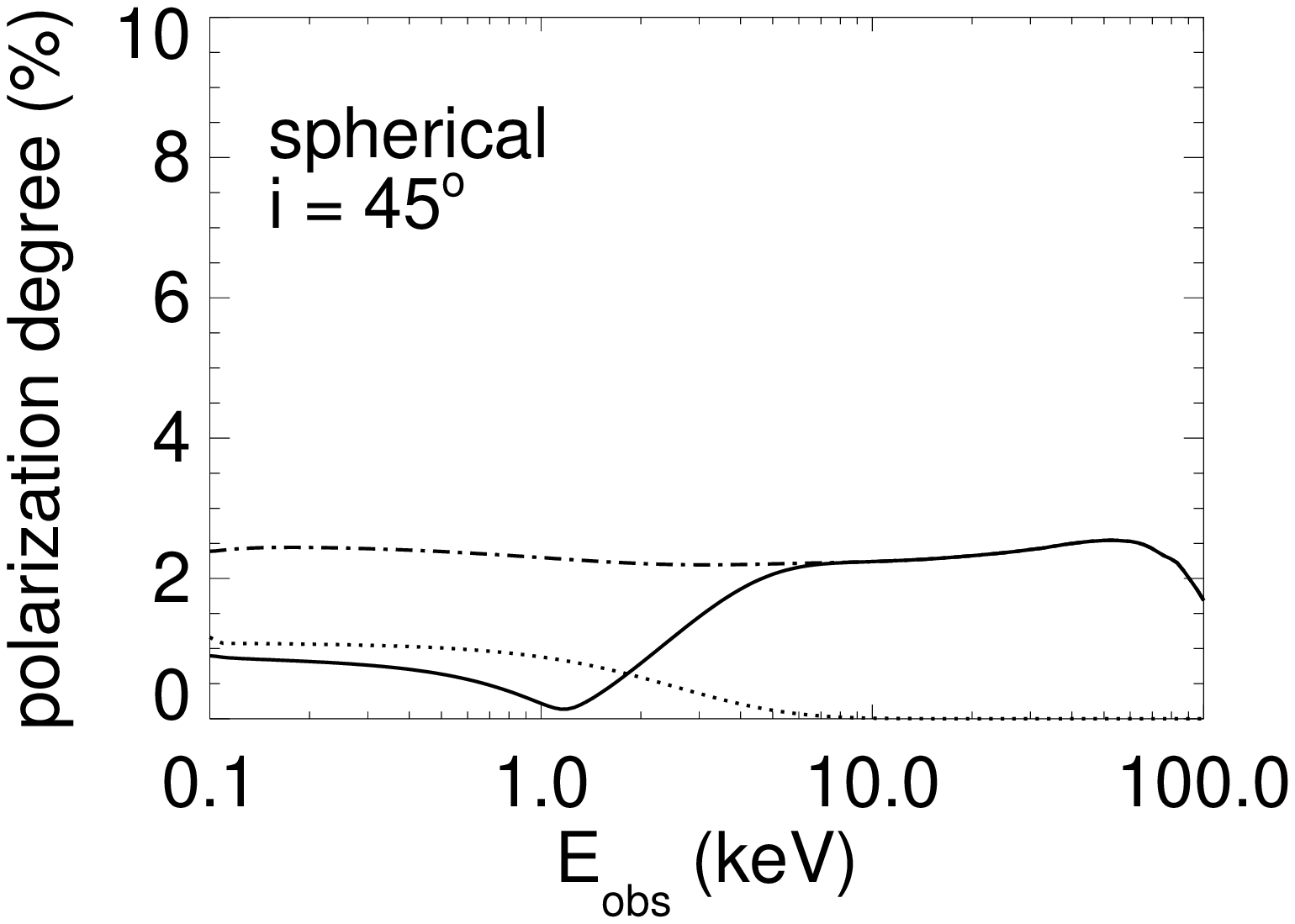}}
\scalebox{0.3}{\includegraphics{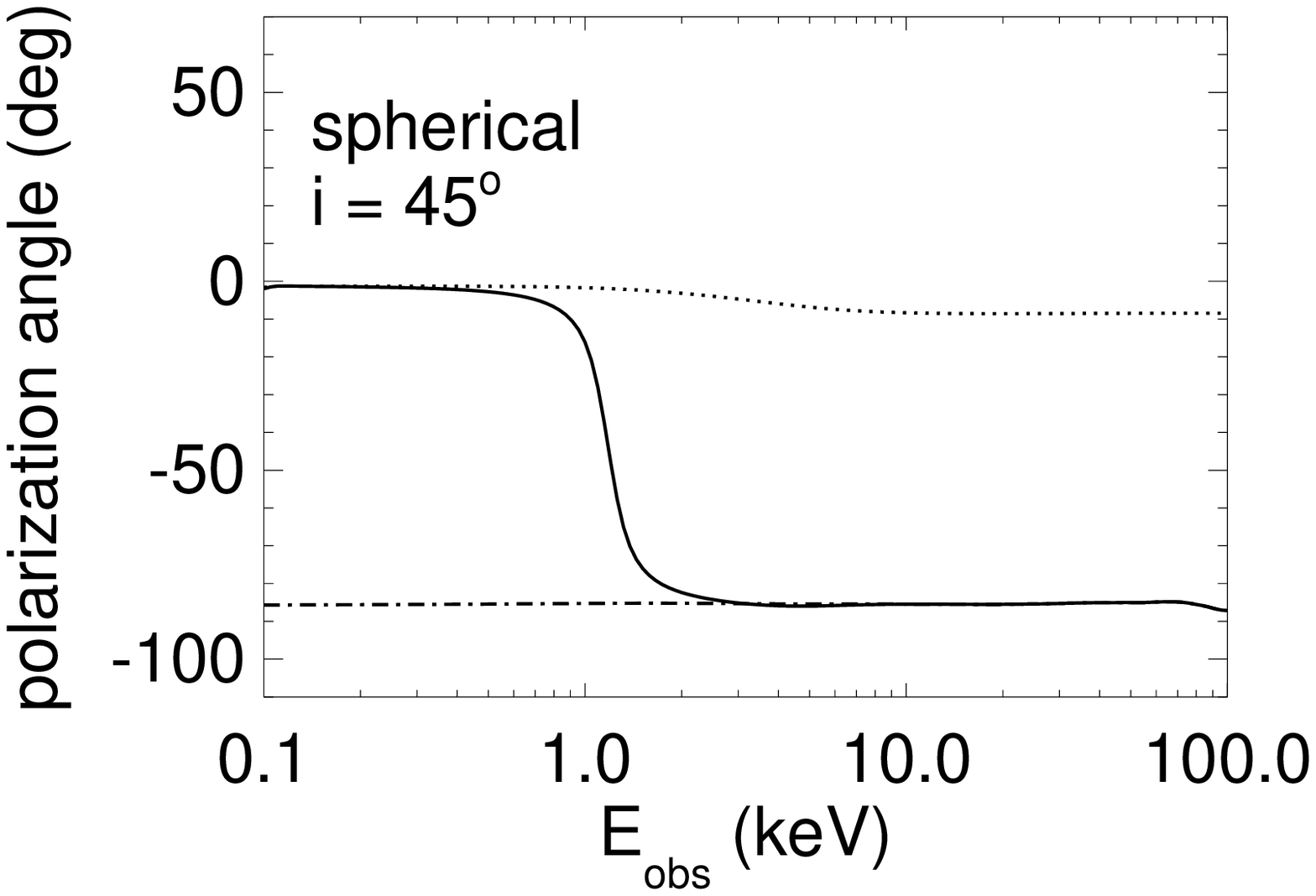}}\\
\scalebox{0.3}{\includegraphics{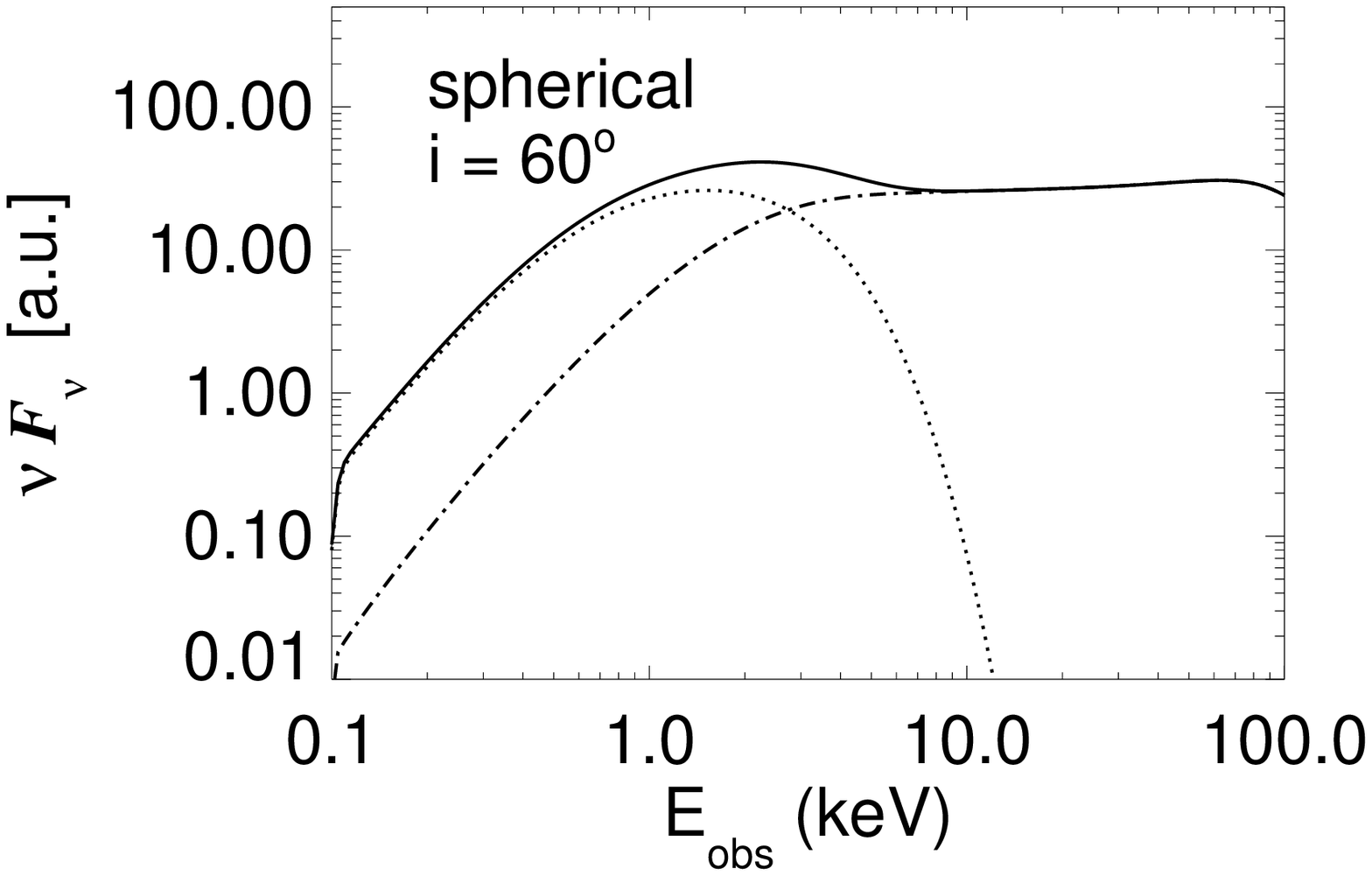}}
\scalebox{0.3}{\includegraphics{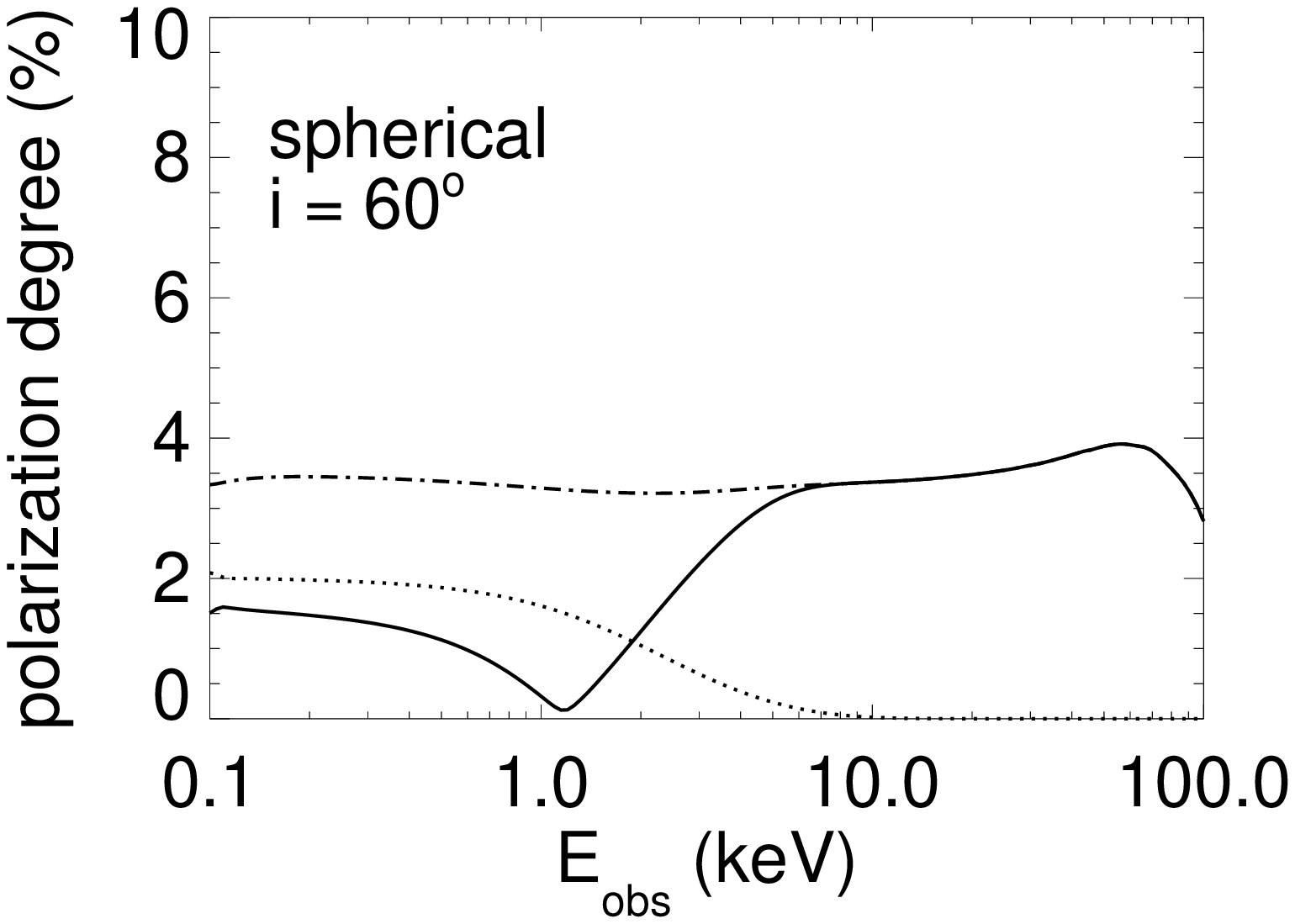}}
\scalebox{0.3}{\includegraphics{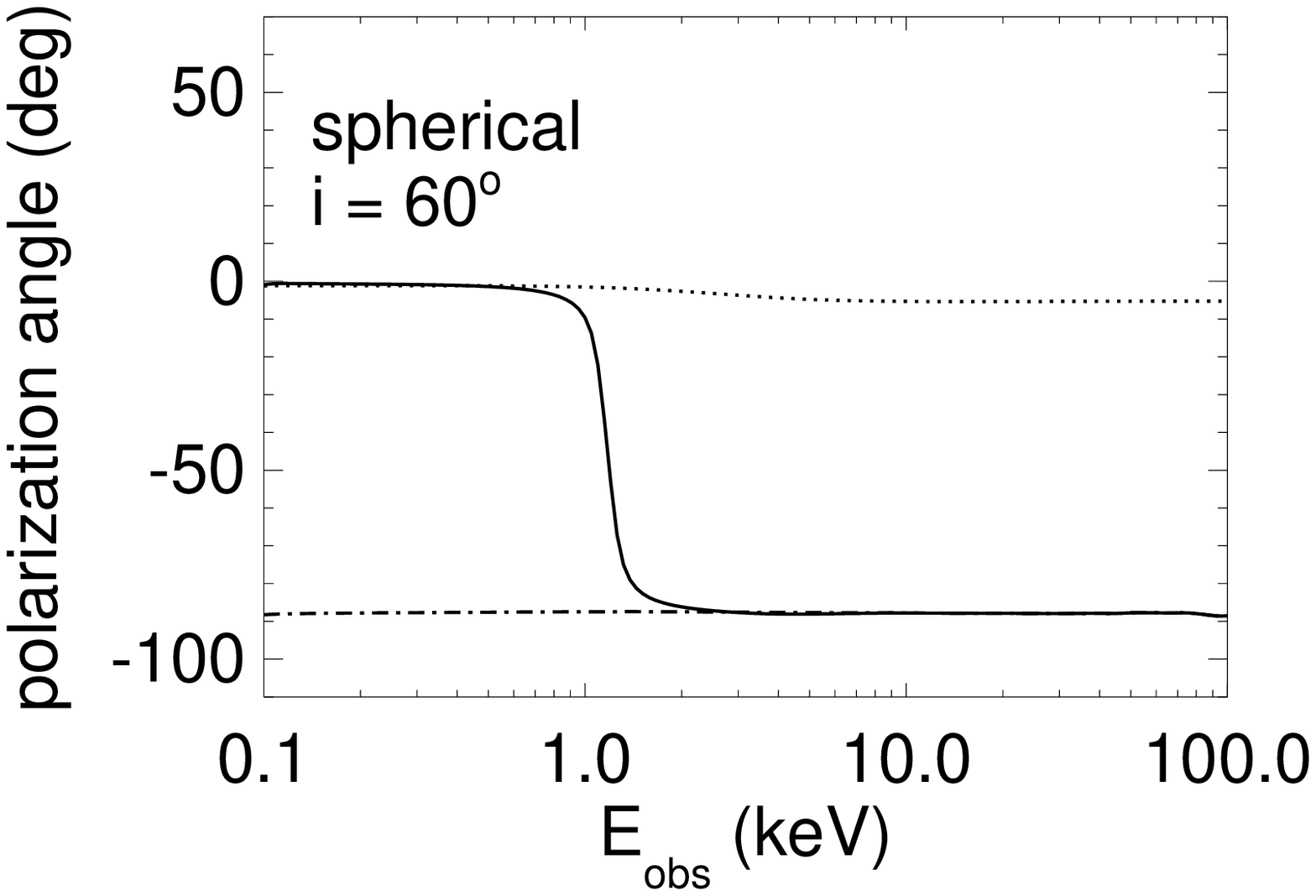}}\\
\scalebox{0.3}{\includegraphics{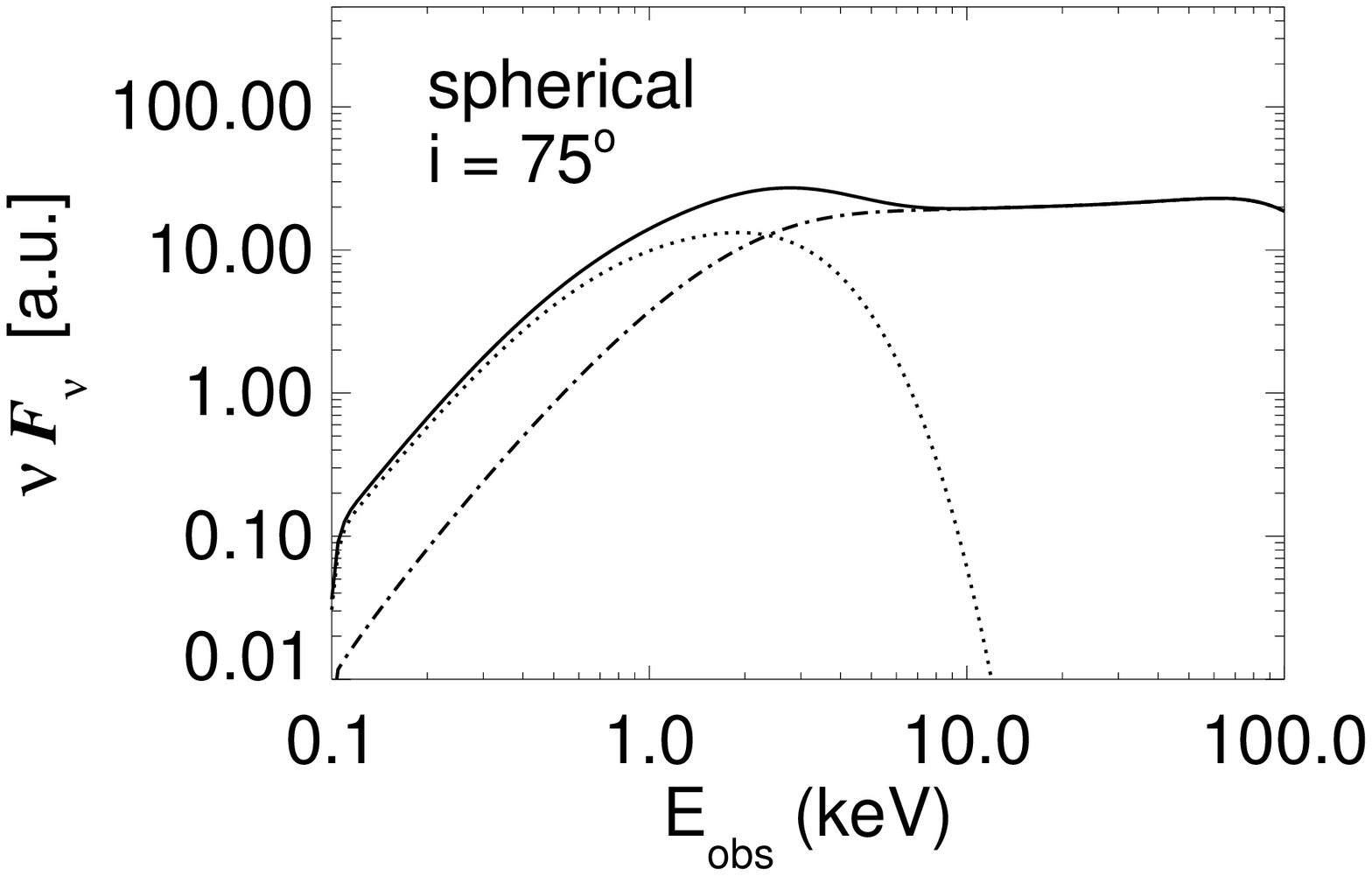}}
\scalebox{0.3}{\includegraphics{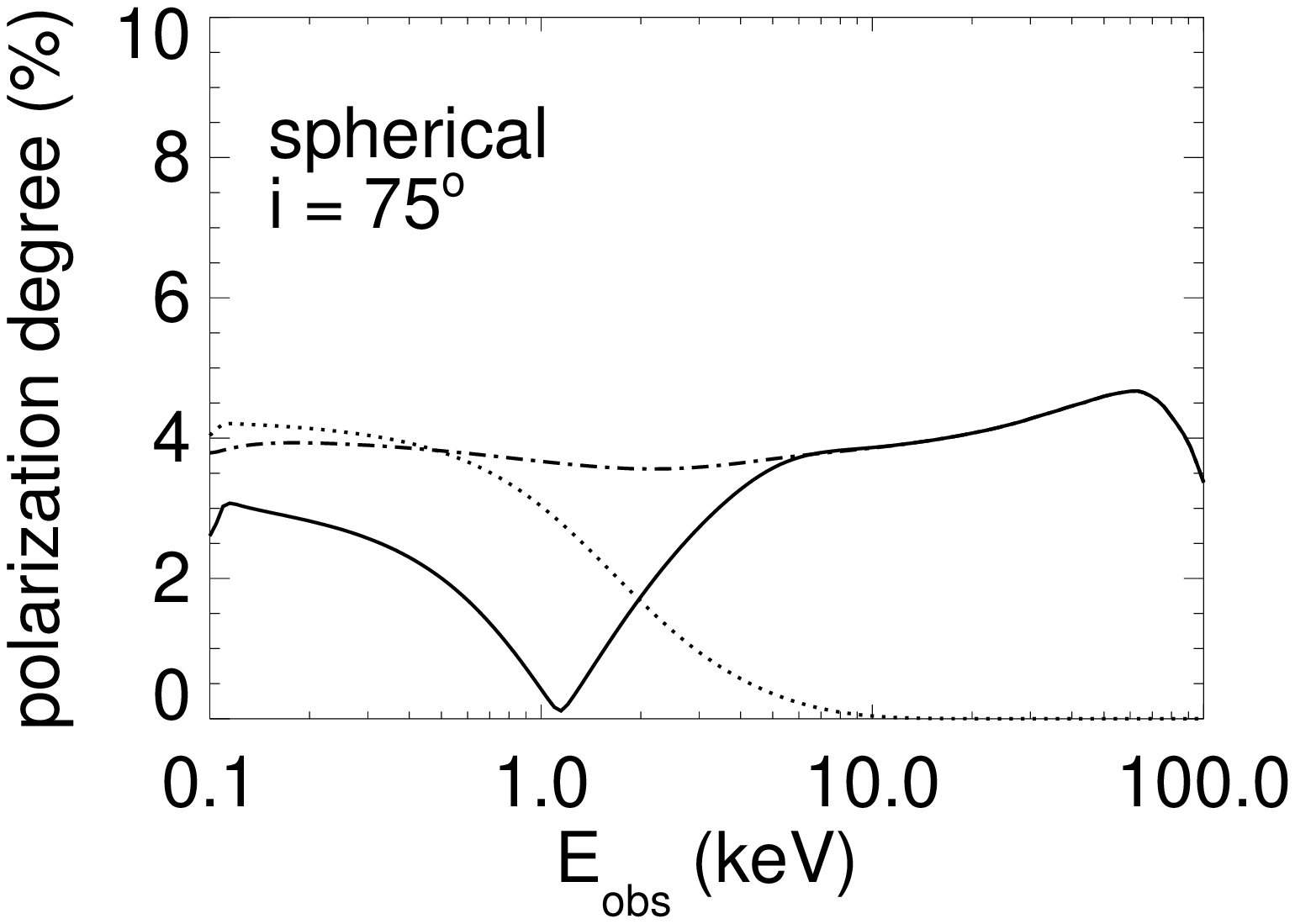}}
\scalebox{0.3}{\includegraphics{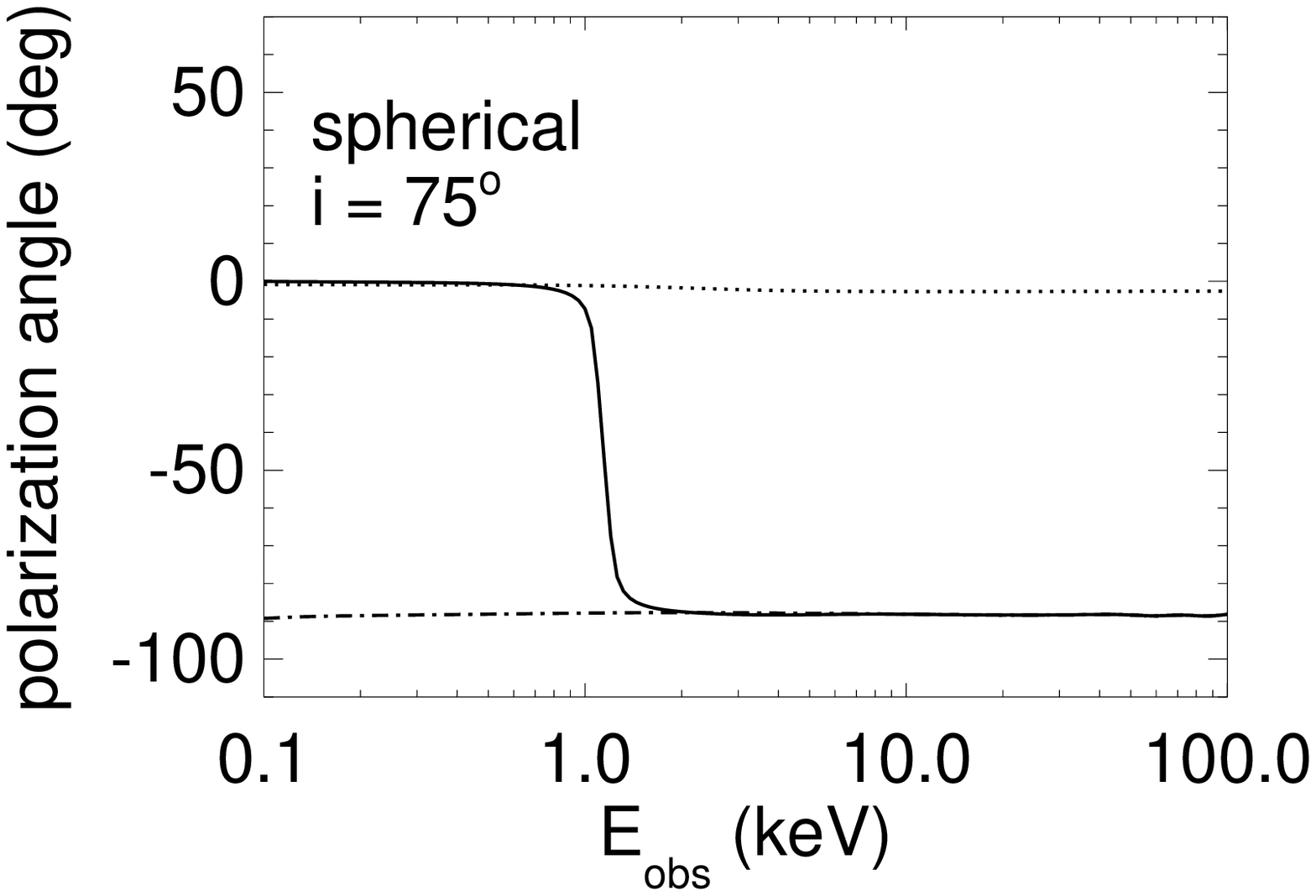}}
\end{center}
\end{figure}

As another
example, if the coronal properties are well-known (perhaps through more
detailed spectral observations over a greater range of energies, which
could allow an independent determination of the electron temperature),
the polarization could be used to determine the disk
inclination. While the inclination could in principle be measured
directly from the low-energy polarization \citep{connors:80,li:08},
first-generation polarimeters are likely not going to be very
sensitive below $\sim 1$ keV, where in any case, magnetic turbulence
in the disk may reduce the net polarization via Faraday rotation
\citep{davis:09}. Yet as we see from Figure \ref{wedge_grid}, a single
high-energy polarization measurement at $\sim 10$ keV could give the
inclination angle, assuming the other model parameters are known
reasonably well. In practice, it is more likely that we could
use some independent method (e.g.\ 
optical light curves, Fe line fitting, etc.) to determine the disk
inclination, and then use the amplitude of polarization in the $\sim
1-10$ keV band to determine the inhomogeneity of the corona. Coronae
with large-amplitude density fluctuations produce weaker
polarization at high energies and a more gradual transition from
horizontal to vertical orientation, as shown in Figures
\ref{hotspot_rho} and \ref{hotspot_Nc}.

\citet{poutanen:97} showed that, using basic physical arguments about
the covering fraction of the corona and the relative flux in the hard
and soft components, the smooth sandwich corona could
be distinguished from a 
patchy corona on the basis of the spectral properties alone. Yet in
Section \ref{sphere}, we have
shown that the spectrum from a truncated disk around a spherical
corona can be nearly indistinguishable from that of a
wedge corona above an extended disk. However, the polarization
signatures from these two models are quite distinguishable: the
spherical geometry leads to a horizontal-vertical transition at lower
energy and gives significantly weaker polarization at all energies.

\begin{figure}
\caption{\label{wedge_HR} Degree and angle of polarization for a
  sandwich corona, varying the scale height of the corona.}
\begin{center}
\scalebox{0.8}{\includegraphics{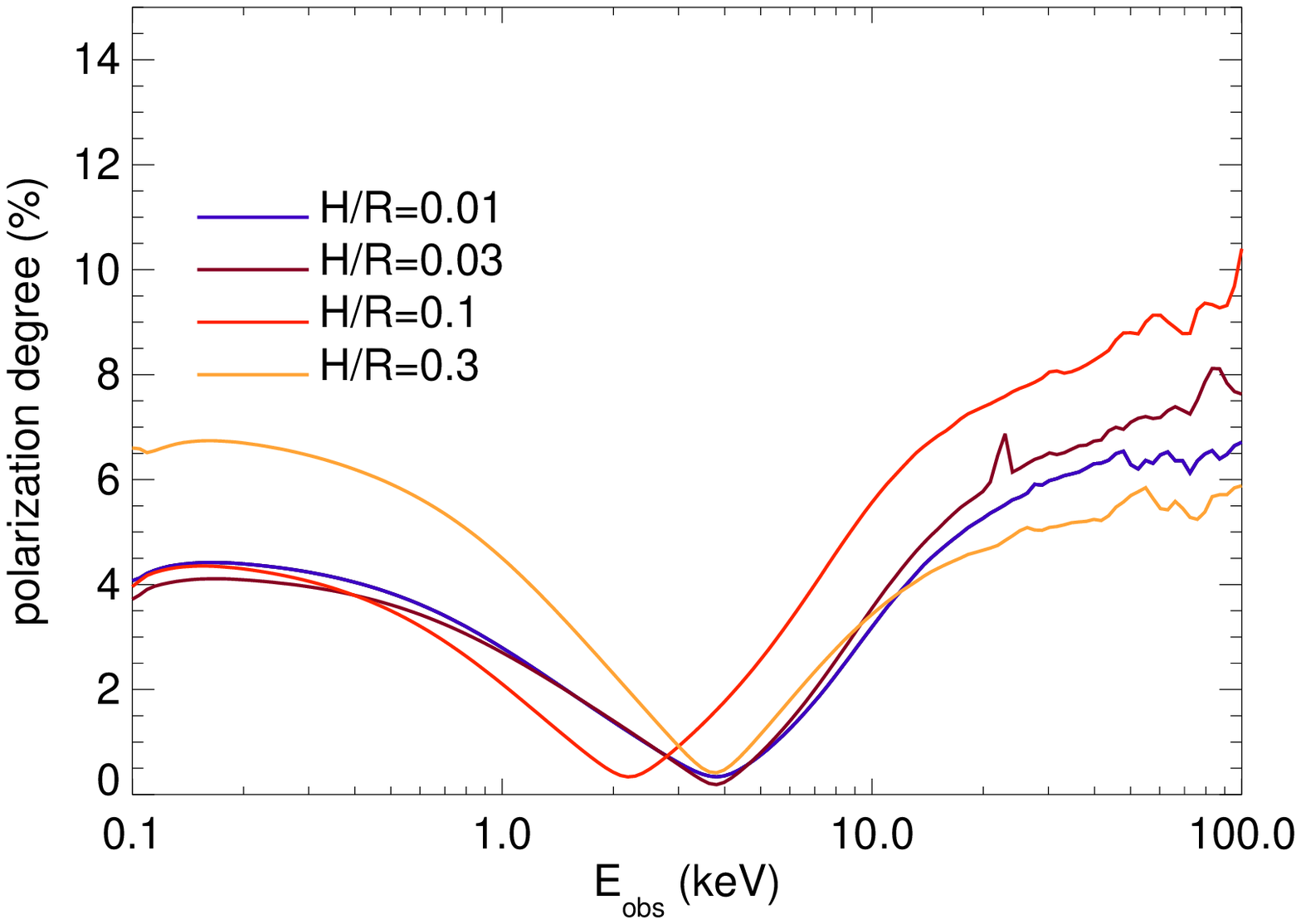}}\\
\scalebox{0.8}{\includegraphics{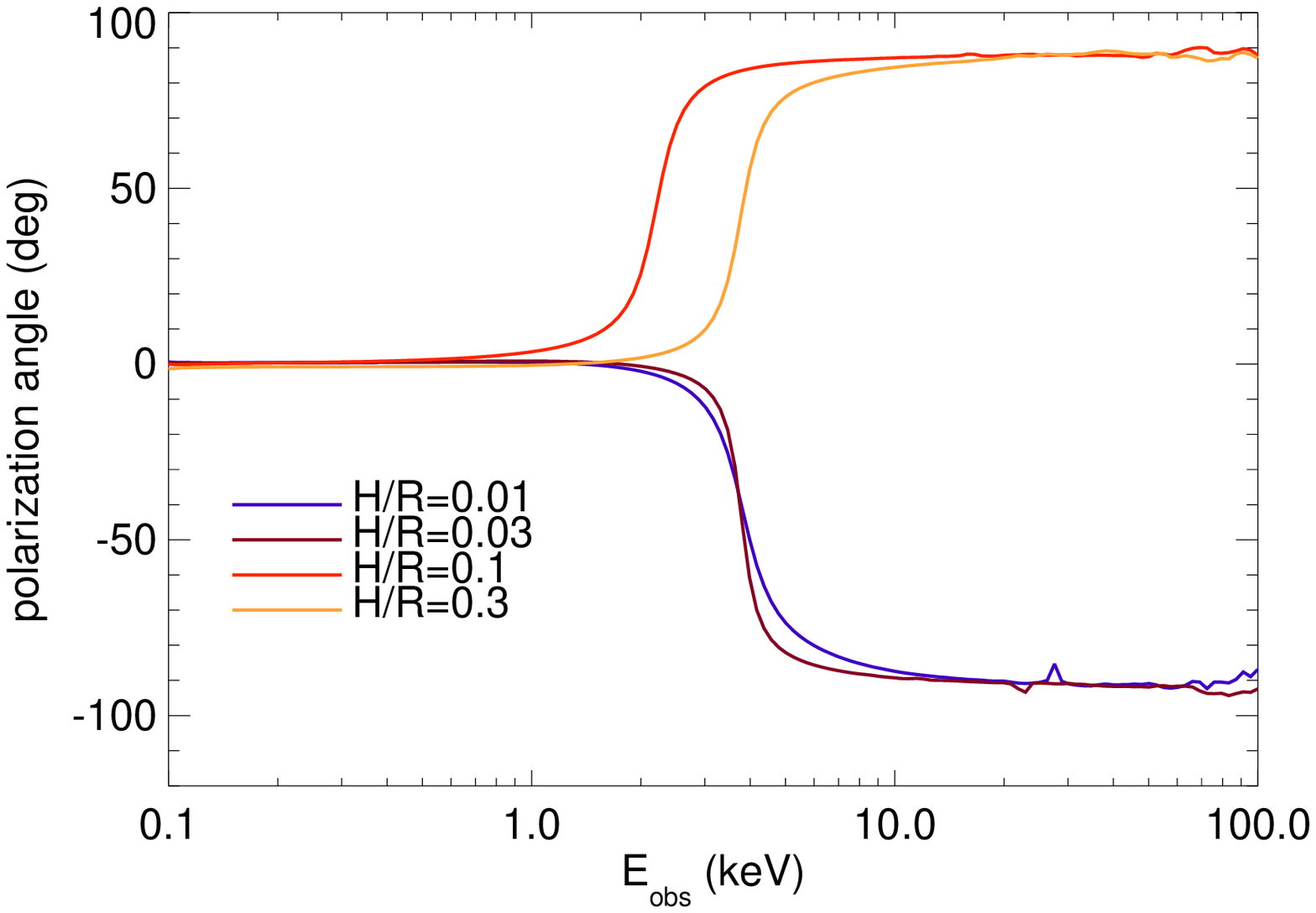}}
\end{center}
\end{figure}

As seen in Sections \ref{wedge} and \ref{hotspot},
the sense of rotation in the polarization transition (i.e., $\psi>0$
or $\psi <0$) contains
additional information that constrains the corona scale height and
level of inhomogeneity. The angle of polarization at low
energies defines the projection of the disk plane on the sky, but
not the direction of the BH spin (as in many binary systems,
the line of nodes is uniquely determined, but it is impossible to know
which is the ascending node). From observations of the thermal state,
we can use the
sense of the rotation toward vertical polarization to determine
the sense of orbital motion of the disk, thus defining the $\psi>0$
direction. Given
that result, the sign of $\psi$ in the coronal state can constrain
the scale height of a wedge corona (Fig.\ \ref{wedge_HR}) or the
covering fraction of a clumpy corona (Figs.\ \ref{hotspot_rho} and
\ref{hotspot_Nc}).

Additional constraints on clumping properties $\rho_c/\rho_0$ and
$n_c$ can come from the amplitude of polarization above the thermal
peak as follows: In stellar-mass systems, 
we might determine the inclination of the disk from radial velocity
observations of the companion. In AGN, the inclination may be inferred
from the blue edge
of a broad iron line \citep{reynolds:03}. In either stellar-mass
systems or AGN, the slope of the hard power-law tail of the X-ray
spectrum can give the covering fraction $f_c$, which is related to
$\rho_c/\rho_0$ and $n_c$ through equation (\ref{f_c}). For a given
$i$ and $f_c$, the degree of polarization at high energies can then be
used to determine the coronal properties $\rho_c/\rho_0$ and $n_c$. 

Furthermore, with timing observations we expect an inverse correlation
between the amplitude of hard X-ray luminosity fluctuations and
the number of coronal clumps, due
simply to Poisson statistics. In a related way, polarization could
provide valuable insight into the nature of poorly-understood timing
phenomena such as quasi-periodic oscillations (QPOs). The QPOs
observed in stellar-mass BHs are most prominent in the hard X-rays
($\gtrsim 6$~keV), even in the SPL state when most of the flux is
at lower energies \citep{remillard:06}, suggesting that the QPOs come
from the hot corona. By comparing the polarization signal with the
timing properties at different epochs, we should be able to constrain
the location, size, and coherence of the regions from which QPOs
originate. For example, if we find that periods of large amplitude
timing fluctuations correlate with low-amplitude polarization, then it
is quite likely that both features are caused by a relatively small
number of hot, dense clouds orbiting above the inner disk.

Lastly, in Section \ref{AGN}, we showed how the special properties of AGN
could be exploited through X-ray spectropolarimetry. Because the seed
photons in such systems are at much lower energies, more scattering events
are required to inverse Compton scatter the seed photons to X-ray
energies. Additionally, because the relatively cool
disk is largely opaque to absorption below $\sim 10$ keV, 
the range of available scattering angles is more restricted.
Together, these features particular to AGN systems lead to a
stronger ability to discriminate between different coronal covering
fractions, as well as the characteristic size and density of the
scattering clouds. Combining spectral and polarization information, it should
be possible to determine the coronal covering fraction and the disk
inclination independently with moderate precision. In contrast to the
galactic BHs, in the AGN case it is unlikely that we will be able to
observe a single system 
in multiple accretion states, yet the much longer time scales should allow
for more detailed comparisons between polarization and variability.

\vspace{0.35cm}\noindent We would like to thank Tim Kallman,
Jean Swank, and Shane Davis for helpful discussions and
comments. This work was supported by the Chandra Postdoctoral
Fellowship Program grant PF7-80051 (JDS) and NSF grant AST-0908336
(JHK).

\newpage

\clearpage


\newpage
\begin{thebibliography}{99}
\bibitem[Agol \& Krolik(2000)]{agol:00} Agol, E., \& Krolik, J.\
  H. 2000, ApJ, 528, 161
\bibitem[Bardeen et al.(1972)]{bardeen:72} Bardeen, J.\ M., Press, W.\
  H., \& Teukolsky, S.\ A. 1972, ApJ 178, 347
\bibitem[Bellazzini et al.(2006)]{bellazzini:06} Bellazzini, R., et
  al. 2006, Nucl.\ Instr.\ Meth., A560, 425
\bibitem[Black et al.(2003)]{black:03} Black, J.\ K., et al. 2003,
  Nucl.\ Instr.\ Meth., 513, 639
\bibitem[Blandford \& Begelman(2004)]{blandford:04} Blandford, R.\ D.,
  \& Begelman, M.\ C. 2004, MNRAS, 349, 68
\bibitem[Carter(1968)]{carter:68} Carter, B. 1968, Phys.\ Rev., 174,
  1559
\bibitem[Chandrasekhar(1960)]{chandra:60} Chandrasekhar, S.
  1960. {\it Radiative Transfer}, Dover, New York
\bibitem[Connors \& Stark(1977)]{connors:77} Connors, P.\ A., \&
  Stark, R.\ F. 1980, Nature, 269, 128
\bibitem[Connors et al.(1980)]{connors:80} Connors, P.\ A., Piran,
  T., \& Stark, R.\ F. 1980, ApJ, 235, 224
\bibitem[Costa et al.(2008)]{costa:08} Costa, E., et al. 2008, Proc.\
  SPIE, vol.\ 7011-15, [arXiv:0810.2700]
\bibitem[Cunningham(1976)]{cunningham:76} Cunningham, C.\ T. 1976,
  ApJ, 208, 534
\bibitem[Davis et al.(2009)]{davis:09} Davis, S.\ W., Blaes, O.\ M.,
  Hirose, S., Krolik, J.\ H. 2009, ApJ 703, 569
\bibitem[Dolence et al.(2009)]{dolence:09} Dolence, J.\ C., Gammie,
  C.\ F., Moscibrodzka, M., Leung, P.\ K. 2009, [arXiv:09090708]
\bibitem[Done \& Zycki(1999)]{done:99} Done, C., \& Zycki, P.\
  T. 1999, MNRAS, 305, 457
\bibitem[Done \& Trigo(2009)]{done:09} Done, C., \& Trigo, M.\
  D. 2009, MNRAS submitted, [arXiv:0911:3243]
\bibitem[Dovciak et al.(2004)]{dovciak:04} Dovciak, M., Karas, V.,
  \& Matt, G. 2004, MNRAS, 355, 1005
\bibitem[Dovciak et al.(2008)]{dovciak:08} Dovciak, M., Muleri, F.,
  Goodmann, R.\ W., Karas, V., \& Matt, G. 2008, MNRAS submitted,
  [arXiv:0809.0418]
\bibitem[Elvis et al.(1978)]{elvis:78} Elvis, M., et al. 1978, MNRAS,
  183, 129
\bibitem[Esin et al.(2001)]{esin:01} Esin, A.\ A., et al. 2001, ApJ,
  555, 483
\bibitem[Gierli\'{n}ski et al.(1997)]{gierlinski:97} Gierli\'{n}ski,
  M., et al. 1997, MNRAS, 288, 958
\bibitem[Gierli\'{n}ski \& Done(2005)]{gierlinski:05} Gierli\'{n}ski, M., \&
  Done, C. 2005, MNRAS, 347, 885
\bibitem[Haardt \& Maraschi(1993)]{haardt:93} Haardt, F., \& Maraschi,
  L. 1993, ApJ, 413, 507
\bibitem[Haardt et al.(1994)]{haardt:94} Haardt, F., Maraschi,
  L., \& Ghisellini, G. 1994, ApJ, 432, L95
\bibitem[Ingram et al.(2009)]{ingram:09} Ingram, A., Done, C., \&
  Fragile, P.\ C. 2009, MNRAS, 397, L101
\bibitem[Jahoda et al.(2007)]{jahoda:07} Jahoda, K., Black, K.,
  Deines-Jones, P., Hill, J.\ E., Kallman, T., Strohmayer, T., \&
  Swank, J. 2007, [arXiv:0701090]
\bibitem[Laor et al.(1990)]{laor:90} Laor, A., Netzer, H., \& Piran,
  T. 1990, MNRAS, 242, 560
\bibitem[Li et al.(2008)]{li:08} Li, L.-X., Narayan, R., \&
  McClintock, J.\ E. 2008, ApJ submitted, [arXiv:0809.0866]
\bibitem[Makishima et al.(2008)]{makishima:08} Makishima, K., et
  al. 2008, PASJ, 60, 585
\bibitem[Matt et al.(1993)]{matt:93} Matt, G., Fabian, A.\ C., \&
  Ross, R.\ R. 1993, MNRAS, 264, 839
\bibitem[McClintock et al.(2001)]{mcclintock:01} McClintock, J.\ E.,
  et al. 2001, ApJ, 555, 477
\bibitem[Miller et al.(2006)]{miller:06} Miller, J.\ M., et al. 2006,
  ApJ, 653, 525
\bibitem[Mushotzky et al.(1993)]{mushotzky:93} Mushotzky, R.\ F.,
  Done, C., \& Pounds, K.\ A. 1993, ARA\& A, 31, 717 
\bibitem[Nandra et al.(1991)]{nandra:91} Nandra, K., et al. 1991,
  MNRAS, 248, 760
\bibitem[Noble et al.(2008)]{noble:08} Noble, S.\ C., Krolik, J.\ H.,
  \& Hawley, J.\ F. 2008, ApJ submitted, [arXiv:0808.3140]
\bibitem[Novikov \& Thorne(1973)]{novikov:73} Novikov, I.\ D., \& Thorne,
  K.\ S. 1973, in \textit{Black Holes}, ed. C.\ DeWitt \& B.\ S.\
  DeWitt (New York: Gordon and Breach)
\bibitem[Pietrini \& Krolik(1995)]{PK95} Pietrini, P. \& Krolik, J.H.
  1995, ApJ, 447, 526
\bibitem[Poutanen et al.(1997)]{poutanen:97} Poutanen, J., Krolik, J.\
  H., \& Ryde, F. 1997, MNRAS, 292, L21
\bibitem[Poutanen \& Fabian(1999)]{poutanen:99} Poutanen, J., \&
  Fabian, A.\ C. 1999, MNRAS, 306, L31
\bibitem[Reis et al.(2009)]{reis:09} Reis, R.\ C., Fabian, A.\ C., \&
  Miller, J.\ M. 2009, MNRAS accepted, [arXiv:0911.1151]
\bibitem[Remillard \& McClintock(2006)]{remillard:06} Remillard, R.\
  A., \& McClintock, J.\ E. 2006, ARA\& A, 44, 49
\bibitem[Reynolds \& Nowak(2003)]{reynolds:03} Reynolds, C.\ S., \&
  Nowak, M.\ A. 2003, Phys.\ Reports, 377, 389
\bibitem[Rybicki \& Lightman(1979)]{rybicki:79} Rybicki, G.\ B., \&
  Lightman, A.\ P. 1979, \textit{Radiative Processes in Astrophysics}
  (New York: Wiley-Interscience)
\bibitem[Schnittman \& Krolik(2009)]{schnittman:09a} Schnittman, J.\
  D., \& Krolik, J.\ H. 2009, ApJ 701, 1175
\bibitem[Schnittman \& Krolik(2010)]{schnittman:09c} Schnittman, J.\
  D., \& Krolik, J.\ H. 2010, in preparation
\bibitem[Shakura \& Sunyaev(1973)]{shakura:73} Shakura, N.\ I., \& Sunyaev,
  R.\ A. 1973, A\&A, 24, 337
\bibitem[Shen et~al.(2006)]{shen:06} Shen, Shiyin, White, S.D.M., Mo, H.J.,
Voges, W., Kauffmann, G., Tremonti, C. \& Anderson, S.F. 2006, MNRAS 369, 1639
\bibitem[Shimura \& Takahara(1995)]{shimura:95} Shimura, T., \&
  Takahara, F. 1995, ApJ, 445, 780
\bibitem[Stark \& Connors(1977)]{stark:77} Stark, R.\ F., \&
  Connors, P.\ A. 1977, Nature, 266, 429
\bibitem[Stern et al.(1995)]{stern:95} Stern, B.\ E., Poutanen, J.,
  Svensson, R., Sikora, M., \& Begelman, M.\ C. 1995, ApJ, 449, L13
\bibitem[Sunyaev \& Titarchuk(1985)]{sunyaev:85} Sunyaev, R.\ A., \&
  Titarchuk, L.\ G. 1985, A\&A, 143, 374
\bibitem[Swank et al.(2009)]{swank:09} Swank, J., et al., in {\it
  X-ray Polarimetry:  A New Window in Astrophysics}, Rome 2009, 
  edited by R. Bellazzini, E. Costa, G. Matt and G. Tagliaferri
\bibitem[Titarchuk \& Shrader(2002)]{titarchuk:02} Titarchuk, L., \&
  Shrader, C. 2002, ApJ, 567, 1057
\bibitem[Turolla et al.(2002)]{turolla:02} Turolla, R., Zane, S., \&
  Titarchuk, L. 2002, ApJ, 576, 349
\bibitem[Walker \& Penrose(1970)]{walker:70} Walker, M., \& Penrose,
  R. 1970, Commun.\ Math.\ Phys., 18, 265
\bibitem[Zdziarski \& Gierli\'{n}ski(2004)]{zdziarski:04} Zdziarski, A.\
  A., \& Gierli\'{n}ski, M. 2004, Prog.\ Theor.\ Phys.\ Suppl., 144, 99
\bibitem[Zdziarski et al.(2005)]{zdziarski:05} Zdziarski, A.\
  A., Gierli\'{n}ski, M., Rao, A.\ R., Vadawale, S.\ V., \& Mikalojewska,
  J. 2005, MNRAS, 360, 825
\end{thebibliography}
\end{document}